\documentclass[
preprint,
superscriptaddress,
nofootinbib,
]{revtex4-2}

\usepackage{amsmath}
\usepackage{amssymb}
\usepackage[pageanchor=true,plainpages=false, pdfpagelabels, bookmarks,bookmarksnumbered,hidelinks]{hyperref}
\usepackage{times}
\usepackage{graphicx}
\usepackage[first=0,last=9]{lcg}

\usepackage{color, colortbl}
\usepackage{multirow}
\usepackage{amsmath}
\usepackage{esint}
\usepackage[table, svgnames, dvipsnames]{xcolor}

\newcommand{\mixedconn}{\mathbf{A}}
\newcommand{\gait}{\phi}

\newcommand{\fiber}{\boldsymbol{g}}

\newcommand{\beq}{\begin{equation}}
\newcommand{\eeq}{\end{equation}}

\newcommand{\fibercirc}{\boldsymbol{\dot{\fiber}^{b}}}
\newcommand{\fibercircnot}{\boldsymbol{\dot{\fiber}_{0}^{b}}}

\usepackage[single]{accents}

\newcommand{\fibercircave}{\boldsymbol{\dot{\fiber}^{b}_{ave}}}

\makeatletter
\renewcommand*\env@matrix[1][*\c@MaxMatrixCols c]{%
  \hskip -\arraycolsep
  \let\@ifnextchar\new@ifnextchar
  \array{#1}}
\makeatother

\begin{document}

\title{Geometric phase predicts locomotion performance in undulating living systems across scales}


\author{Jennifer M. Rieser\footnote{Equal contribution}}
\affiliation{Emory University, Atlanta, Georgia 30322, USA}
\affiliation{Georgia Institute of Technology, Atlanta, Georgia 30332, USA}%

\author{Baxi Chong$^\mathrm{a}$}%
\affiliation{Georgia Institute of Technology, Atlanta, Georgia 30332, USA}%

\author{Chaohui Gong}%
\affiliation{Bito Robotics, Pittsburgh, PA 15203, USA}%

\author{Henry C. Astley}%
\affiliation{Biomimicry Research and Innovation Center, Departments of Biology and Polymer Science, University of Akron, Akron, OH 44325, USA}%

\author{Perrin E. Schiebel}%
\affiliation{Georgia Institute of Technology, Atlanta, Georgia 30332, USA}%

\author{Kelimar Diaz}%
\affiliation{Georgia Institute of Technology, Atlanta, Georgia 30332, USA}%

\author{Christopher Pierce}%
\affiliation{Georgia Institute of Technology, Atlanta, Georgia 30332, USA}%

\author{Hang Lu}%
\affiliation{Georgia Institute of Technology, Atlanta, Georgia 30332, USA}%

\author{Ross L. Hatton}%
\affiliation{Oregon State University, Corvallis, OR  97331, USA}%

\author{Howie Choset}%
\affiliation{Carnegie Mellon University, Pittsburgh, PA 15213, USA}%

\author{Daniel I. Goldman}%
\affiliation{Georgia Institute of Technology, Atlanta, Georgia 30332, USA}%

\date{\today}

\begin{abstract}
Self-propelling organisms locomote via generation of patterns of self-deformation. Despite the diversity of body plans, internal actuation schemes and environments in limbless vertebrates and invertebrates, such organisms often use similar travelling waves of axial body bending for movement. Delineating how parameters (wave amplitudes, frequencies) lead to locomotor performance (e.g. speed, energy, turning capabilities) remains challenging. Here we show that a geometric framework that replaces laborious calculation with a diagrammatic scheme is well suited to discovery and comparison of optimal patterns of wave dynamics in diverse living systems. We focus on a regime of undulatory locomotion, that of highly damped environments, which is applicable not only to small organism movement in viscous fluids, but also larger animals moving in frictional fluids (sand) and on frictional ground. We find that the travelling wave dynamics used by mm-scale nematode worms and cm-scale desert dwelling snakes and lizards can be described by time series of the weights associated with two principal modes. The approximately circular closed path trajectories of mode weights in a self-deformation space enclose near-maximal surface integral (geometric phase) for organisms spanning two decades in body length. We hypothesize that such trajectories are targets of control (which we refer to as ``serpenoid templates"). Further, the geometric approach reveals how seemingly complex behaviors such as turning in worms and sidewinding snakes can be described as modulations of templates. Thus, the use of differential geometry in living systems can assist in the growth of comparative neuromechanics, allowing a common description of locomotion across taxa and providing hypotheses for function at lower levels of organization.
\end{abstract}


\maketitle
\onecolumngrid


\newpage

Locomotion (or self-propulsion) is an essential behavior in most living systems~\cite{alexander2003principles,childress2012natural} and important for engineered devices like robots~\cite{gravish2018robotics,kim2013soft,Aguilar:2016bq}. In organisms as diverse as jumping kangaroos, swimming eels, crawling nematodes, and spiraling bacteria, self-propulsion results from cyclic changes in body and/or appendage configuration. These configuration sequences are ultimately generated by numerous interacting and coordinated components coupled to environments of varying composition. A major challenge in locomotor biology is to discover general principles that govern how organisms generate and control fast, stable, or energetically efficient locomotion. At the organismal scale, these principles have long been discussed in the physiology and motor control literature~\cite{grillner1985neurobiological,loeb2012optimal} with the term ``neuromechanics'' used to indicate the importance of concomitant consideration of nervous, musculoskeletal, and biomechanical systems in explaining performance. 

Several approaches are used to develop neuromechanical control principles. One approach (``bottom up'') to addressing this question is to directly incorporate the non-linearly coupled nervous/musculoskeletal systems, and environments in full detail. While success has been achieved using this approach in ferreting out mechanisms of legged~\cite{Holmes:2006ku,schilling2013walknet} and undulatory locomotion~\cite{boyle2012gait,johnson2020neuromechanical,mcmillen2008nonlinear} the complexity of such models leads to challenges in discovering broad principles. Another approach (``top down'') ignores the complexity of organisms and seeks to discover broad (cross taxa) and relatively simple patterns of dynamics. These models are often referred to as templates~\cite{full1999templates,Holmes:2006ku}, defined as a behavior that ``contains the smallest number of variables and parameters that exhibits a behavior of interest''. This approach has the benefit of producing models which are analyzable, can be used to test lower level mechanisms and yield insight into features across organisms. The template approach has been useful in rationalizing locomotor performance and control across taxa in legged and undulatory systems~\cite{Holmes:2006ku, roth2014comparative,cowan2014feedback,sponberg2017emergent,srinivasan2006computer,ding2013emergence}. In addition to descriptive power, templates can also serve a prescriptive role, generating dynamics which are targets of control for the neuromechanical system that can yield beneficial locomotor properties (speed, energetics, stability, etc), enabling robots with performance approaching those of living systems~\cite{bhounsule2012design,Holmes:2006ku,seok2014design,astley2015modulation}. Finally, a top-down approach can offer insights into how locomotors adapt their templates to changing locomotion objectives (e.g., from forward locomotion to turning behaviors), locomotor morphology (e.g., lizard limblessness~\cite{chong2022coordinating}), and/or environments (e.g., from swimming to walking in amphibious salamander~\cite{ijspeert2007swimming}).


Despite the apparent simplicity in the template-based approach, a question arises: given all the ways an organism could organize its neuromechanical system to self-deform~\cite{berman2018measuring}---e.g., bounce across the ground like a pogo stick ~\cite{srinivasan2006computer} or send waves down its body from head to tail---how can one determine ``good'' ones? That is, does a general framework for locomotion exist that could provide a priori useful template dynamics which could then be used to evaluate performance and test optimality~\cite{loeb2012optimal} in terms of speed, energy use, or stability? Surprisingly, an answer to this question arose in the 1980s through the work of particle physicists and control theorists~\cite{shapere1987self,berry1990anticipations,berry1988geometric,montgomery1990,marsden1990reduction,krishnaprasad1994,kelly1995geometric,ostrowski1998geometric,lewis2000,Bloch2003,melli2006}. These researchers developed a scheme which now goes by the name ``geometric mechanics'' (and which we will refer to as the geometric phase approach) which first uses environmental models to link small self-deformations around each body configuration to small translations and rotations in world space. Line integrals over closed paths in a space of body configurations (gaits) link cyclic sequences of body/limb changes (``configuration changes'') can then lead to translation/rotation in the world. These translations or rotations can be expressed as ``geometric phases'' -- net changes in global quantities like position that depend on the shape of a cyclical path whose local parameters return to their initial values~\cite{shapere1987self,berry1990anticipations,montgomery1990,marsden1990reduction,berry1993geometric}. These geometric phases are independent of the specific temporal details, such as the speed at which the paths are executed. In other words, the geometric phase depends solely on the path's shape and is unaffected by the specific timing or speed of the movements. Geometric phases appear in diverse situations including the Ahranov-Bohm effect, parallel transport of vectors on curved manifolds, a swinging Foucault pendulum and polarization changes of light in coiled optical fibers~\footnote{We note that the use of the term geometric ``phase" initiated in analysis of wavefunction dynamics in quantum mechanics~\cite{berry1988geometric} but was later applied to translations and rotations by Marsden and collaborators~\cite{marsden1990reduction}}.

A major advance in the possibility of using geometric phase in realistic locomotor situations occurred with the introduction of the minimal perturbation coordinate~\cite{hatton2015EPJ,hatton2013geometric}. Such coordinates mitigate issues associated with the noncommutivity of translations and rotations in the plane and allow line integrals in the configuration space to be approximated by surface integrals over certain functions. These functions are referred to as ``constraint curvature functions" or ``height functions''. Critically, height functions replace potentially laborious calculations used in the line integral approach. For example, even in simple artificial systems ~\cite{purcell1977life,becker_koehler_stone_2003,tam2007optimal} to identify parameters that result in optimal performance requires considerable computational effort, comparing movements arising from an infinite combination of shape change sequences. Height functions instead enable a comparatively simple, diagrammatic approach. That is, their key utility is that they simplify the inverse problem: providing ready identification of gaits that maximize performance in diverse systems. Height functions also give a geometric rationalization of the marginal benefits that result from changing/adapting self-deformation patterns, without the need for significant calculation. Because of its utility, over the last decades, researchers have developed the theory so that it is applicable to a broad range of situations and applied the scheme to optimal control of artificial devices including satellite reorientation~\cite{wuhr1995satellite}, robot swimming~\cite{hatton2013geometric, dai2016rss}, sidewinding~\cite{chongmoving} and walking~\cite{Chong2019Coordination,chong2022coordinating} in granular and frictional environments.

A particular regime of self-propulsion which could be amenable to geometric analysis is that in which dissipative forces dominate inertia. Here, cyclic patterns of undulatory self-deformations solely dictate performance (provided the environment is uniform like in open fluid)-- unlike in inertia dominated systems where gliding (movement without shape changes) and stored/returned elastic energy can be utilized. This is typically thought of as the world of very small scales (nicely narrated in~\cite{purcell1977life} and the subject of much effort devoted to locomotion~\cite{lauga2009hydrodynamics}).  Surprisingly, in the last decade, our experimental and granular Resistive Force Theory (RFT, first introduced for microscopic organisms~\cite{gray1955propulsion}) modeling studies of sand-swimming organisms have revealed that the dynamics of such terrestrial {\em macroscopic} undulatory locomotors~\cite{hu2009mechanics,maladen2009undulatory,sharpe2015locomotor,zhang2014effectiveness} operate in a mechanically analogous regime, where rate-independent friction, as opposed to viscosity, dominates inertia. Even within these dissipative locomotor regimes, comparing undulatory locomotor performance for different dynamics is challenging due again to infinite combinations of self-deformation sequences, necessitating the use of the height function formulation of locomotor geometric phase to establish cross-system principles of locomotion. 

\begin{figure}[ht!]
	\centering
	\includegraphics[width=0.8\textwidth]{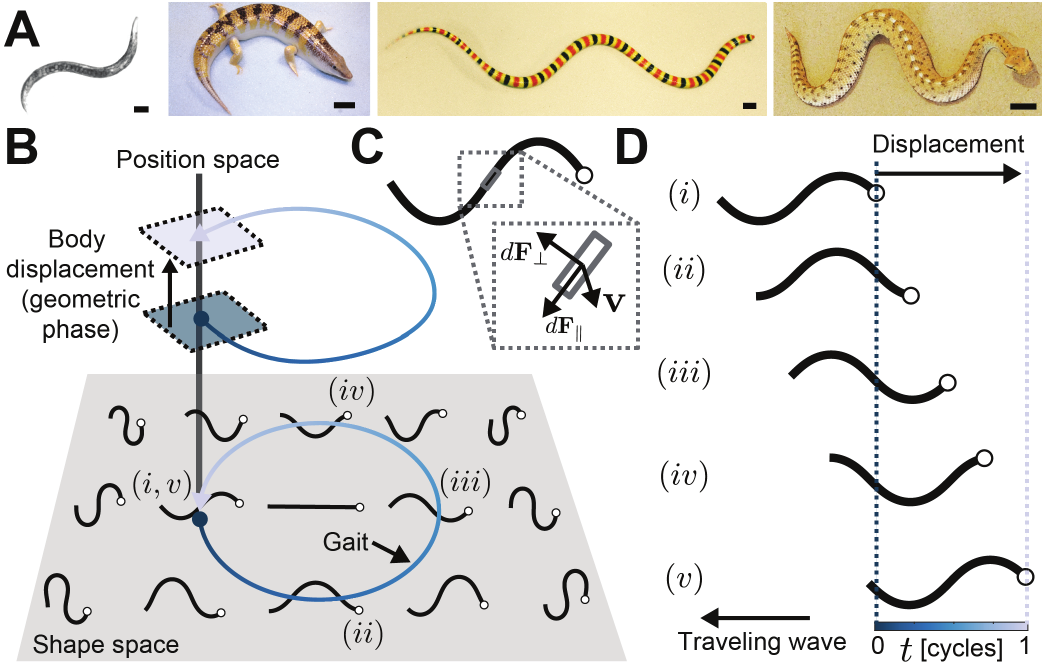}
	\caption{{\bf Undulatory locomotion as a geometric phase.} (A) Left to right: the nematode worm, \textit{Caenorhabditis elegans}; the sandfish lizard, \textit{Scincus scincus}; the shovelnose snake,  \textit{Chionactis occipitalis}; and the sidewinder rattlesnake, \textit{Crotalus cerastes}. The black scale bars denote  $0.1$ mm (left image) and 1 cm (all other images). (B) Depiction of a geometric framework that relates world-frame movements (in position space) to animal body deformations in shape space (adapted from~\cite{marsden1990reduction}). The circular path through shape space shows an example of a sequence of animal body configurations that produce a traveling wave of body curvature propagated along the body from head (shown as the white circle) to tail. (C) Environmental reaction forces on an infinitesimal body segment. (D)  Body shape changes coupled with the physics of the surrounding environment can give rise to net displacements.}
	\label{fig:fun}
\end{figure}

In this paper, we demonstrate that the geometric phase approach of locomotion provides a useful way to compare living organisms with seemingly very different and differently composed (e.g. exo- versus endoskeletal) locomotor systems across scales. We first show that diverse undulatory organisms (including microscopic nematodes and macroscopic snakes/lizards, Fig.~\ref{fig:fun}A) moving forward can be described using a planar wave template - a time series of two spatial basis functions. We notice that these time series exhibit similar patterns in locomotor systems across scales, suggesting the existence of a fundamental template for undulatory systems. Moreover, the time series emerge as circular paths in the two-dimensional spatial-basis space, which we refer to as serpenoid templates. Interestingly, we note that the parameters governing the animals' chosen serpenoid templates nearly maximize geometric phase, indicating that the animals are controlling their self-deformation patterns to achieve "good" locomotor performance (a notion we will discuss in more detail). Beyond forward crawling, serpenoid templates can be modulated (e.g., adding an offset to the center of the circle) to explain sidewinding and turning behaviors, such as the omega turn in nematodes and the differential turn in sidewinders.

\section*{Using a two mode description for planar undulation in dissipative environments}

Previous studies of highly damped locomotors using planar undulation revealed relative simplicity in wave shapes~\cite{stephens2008dimensionality,astley2020surprising}~\cite{fang2010biomechanical}. To describe the planar\footnote{Body undulation in these animals is planar but occurs dorsoventrally.} shapes used during \textit{C. elegans} locomotion, \cite{stephens2008dimensionality} used Principal Components Analysis (PCA) to diagonalize the covariance matrix of local body curvatures (determined from digitized midlines of animals during movement) to identify a set of orthonormal basis functions, referred to as Principal Components (PCs), whose weighted superposition can be used to capture observed animal body configurations.  Eigenvalues associated with each PC or ``eigenworm'' indicate the variance explained by each mode (and therefore the fraction of the variance explained by each mode is given by the eigenvalue divided by the sum of all eigenvalues). For steady forward crawling, \cite{stephens2008dimensionality} found that two PCs were sufficient to capture most of the shape variance, and the dynamics of movement could be represented within this two-dimensional space by projecting time traces of local body curvatures onto these PCs. A depiction of a ``shape space'' spanned by two undulatory PCs for forward movement is shown in Fig.~\ref{fig:fun}B, with an example of the dynamics of movement represented by the directed closed path within this space. Each point along this path corresponds to a body posture, and the direction indicates how postures change. The body shapes associated with five points identified along the path are shown in Fig.~\ref{fig:fun}D.

\begin{figure}[t!]
	\centering
	\includegraphics[width=0.8\textwidth]{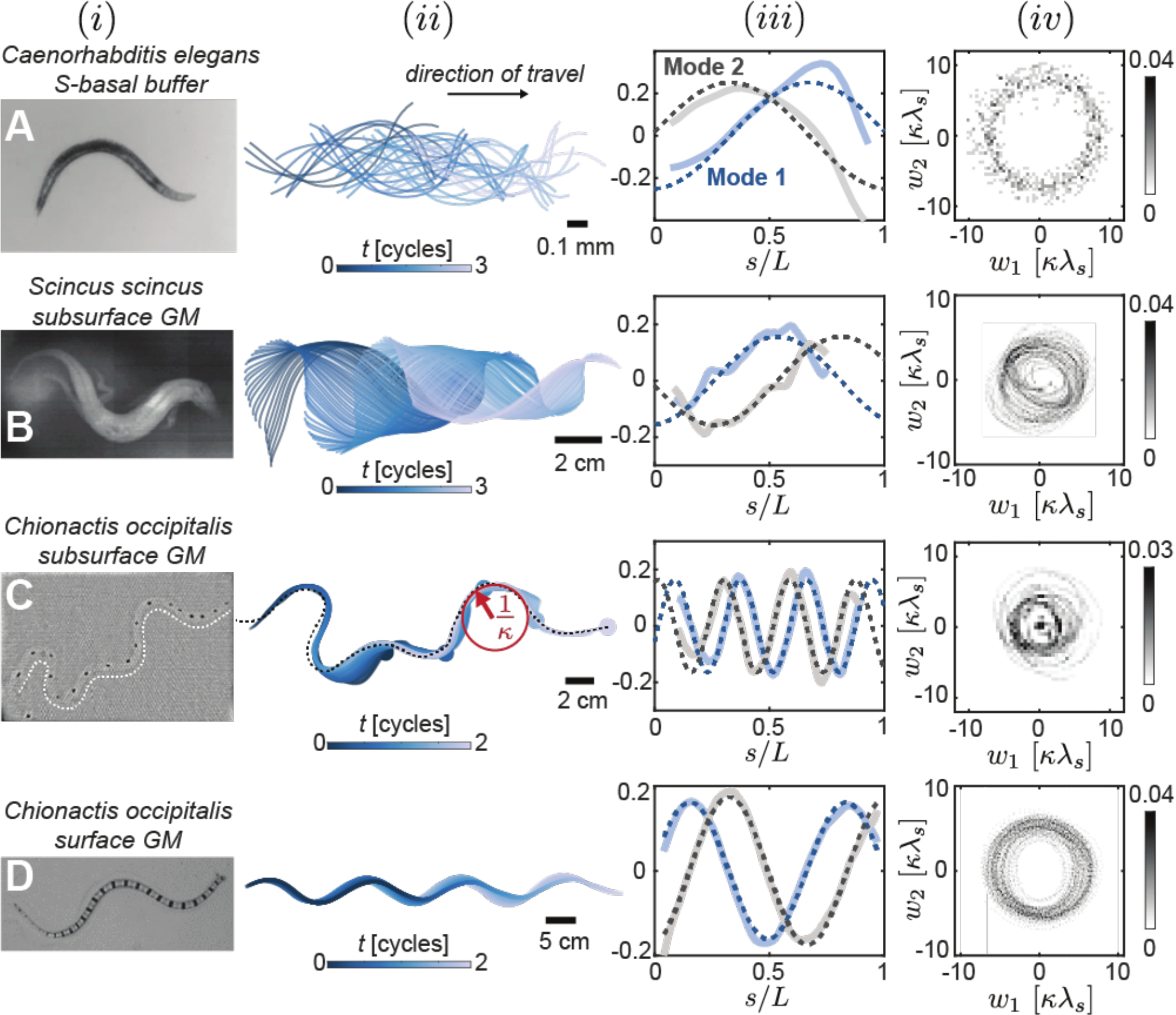}
	\caption{{\bf Low dimensional representation of animal movement through sand.} $(i)$ Photos of animals and $(ii)$ snapshots of animal body configurations colored by time (over two cycles) for (A) the nematode worm (\textit{Caenorhabditis elegans}) in S-basal buffer (B) the sandfish lizard (\textit{Scincus scincus}) $7.6$~cm below the surface of and fully immersed in $300$-$\mu$m glass beads, (C) the shovel-nosed snake (\textit{Chionactis occipitalis}) $7.6$~cm below the surface of and fully immersed in $300$-$\mu$m glass beads, and (D) \textit{Chionactis occipitalis} moving on the surface of $300$-$\mu$m glass beads. $(iii)$ Solid lines show the two dominant relative curvature ($\kappa \lambda_s$) PCA modes account for (A) $96.7\%$, (B) $94.7\%$, (C) $57.6\%$, (D) $90.7\%$ of the variation in observed body configurations. Dashed lines show best fits to $\sin$ and $\cos$ functions. $(iv)$ 2D probability density map of projections of curvatures onto the two PCA modes with the largest eigenvalues.
	}
	\label{fig:animals}
\end{figure}

Here, inspired by the similarity of the wave kinematics used by the sandfish lizard, \textit{Scincus scinus}~\cite{sharpe2015locomotor}, and the shovel-nosed snake, \textit{Chionactis occipitalis},~\cite{sharpe2015locomotor,schiebel2019mitigating} in sand to those used by low Reynolds number mm-scale locomotor \textit{C. elegans}, we investigated whether a low dimensional representation could capture the body postures and dynamics observed in these undulatory locomotors across scales. We collected new data on \textit{C. elegans} (Materials and Methods) and re-analyzed previously published data on lizards and snakes~\cite{sharpe2015locomotor,schiebel2019mitigating}. We started with the digitized animal midlines from high-speed kinematic data (Fig.~\ref{fig:animals}$(ii)$, SI Sec.~3), and characterized instantaneous body configuration using the relative curvature,  $\kappa(s,t)\lambda_s$, where $\kappa(s,t)$ is the local curvature   at position $s$ and time $t$ (Fig.~\ref{fig:animals}B-$(ii)$), $s$ is the position along the body, and $\lambda_s$ is the arc length of one wave 
(SI Sec.~3). $\kappa(s,t)\lambda_s$ is a non-dimensional and coordinate-invariant quantity that is measured as a function of position along the body for each moment in time.

PCA was applied to the entire data set of each species combining curvature measurements from all trials throughout all times~\cite{stephens2008dimensionality,schiebel2019mechanical}. We find that, for the forward movement of lizards and snakes, and nematodes in fluid, two PCs capture most of the variation in the body configurations of each species (SI Sec.~3). This observation further allows us to use the space spanned by the first two PCs as our low dimensional representation for each animal:

\begin{equation}
\kappa(s,t)\lambda_s = w_{1}(t) {v_1}(s) + w_{2}(t) {v_2}(s),
\label{eq:alpha}
\end{equation}

where ${v_1}(s)$ and ${v_2}(s)$ are principal components identified from PCA; $w_1(t)$ and $w_2(t)$ are the time series of weights associated with the corresponding principal components. We thus define the shape variable $\boldsymbol{\alpha(t)} = [ w_{1}(t), w_{2}(t)]$. Fig.~\ref{fig:GMnew}$(i)$ (inset) shows examples of the curvature basis functions for the corresponding undulatory locomotors.

Similar to the results from~\cite{stephens2008dimensionality}, we find that the two dominant PCs in the lizards, snakes, and nematodes were well-fit by $v_i(s)=\sin{(2\pi n s/L+\phi^s_i)},i\in{\{1,2\}}$ where  $n$, $\phi^s_i$ and $L$ are, respectively, the number of waves, spatial phase, and the body length (Fig.~\ref{fig:animals}$(iii)$ for modes and fits\footnote{For the subsurface movement of \textit{Ch. occipitalis}, we attribute the points near the origin to turning behavior that is not captured by the first two modes; however we will show that despite the more complex body postures and locomotion, we are still able to quantitatively describe forward locomotion.}). Note that to enforce the orthogonality of fitted basis functions, we assume $\phi^s_1-\phi^s_2=\pm\pi/2$. Specifically, in worms and subsurface snakes (Fig.~\ref{fig:animals}A,C), we have $\phi^s_1-\phi^s_2=\pi/2$, indicating that the phase of mode 2 is ahead of mode 1; in lizards and surface snakes (Fig.~\ref{fig:animals}B,D), we have $\phi^s_1-\phi^s_2=-\pi/2$, indicating that the phase of mode 1 is ahead of mode 2. The importance of relative phase relationship between modes will later be discussed in the context of turning\footnote{Because of the different phase relationship of mode 1 and mode 2, the paths propagate in $[w_1,w_2]$-space clockwise for worms and subsurface snakes (Fig.~\ref{fig:animals}A,C) and counterclockwise for lizards and surface snakes (Fig.~\ref{fig:animals}B,D).}.

The visualization of shape variables reveals that animals used nearly-circular templates (of radius $w = \sqrt{w_1^2+w_2^2}$ ) to transition through body configurations in $[w_1,w_2]$-space as they deform (Fig.~\ref{fig:animals}$(iv)$). Notably, the circle in shape space and thus the sinusoidal variation in curvature was first studied in the context of snake locomotion and limbless robot control by Hirose in his seminal work~\cite{Hirose:1993}. Following Hirose's terminology for such a wave, we will refer to the circular pattern in shape space as a ``serpenoid" template~\footnote{The serpenoid dynamics differ from characterization of animal body shapes as sinusoidal amplitude displacements of away from the midline of a straight animal during a posteriorly-traveling wave, previously utilized in~\cite{maladen2009undulatory,Maladen:2011es}.}

\section*{Resistive force theory to model environmental interactions}

To develop the geometric phase framework to rationalize such templates as well as to discover novel behaviors, we first require a model of body-environment interactions that connects a particular change in shape to center-of-mass translation and/or rotation in the environment.  Resistive force theory (RFT) modeling has had successes describing movement in fluids, for example in predicting forward swimming speeds of {\em C. elegans}~\cite{berri2009forward,sznitman2010propulsive}; however, in other situations (e.g., waves with higher curvatures), more elaborate schemes are often required to accurately capture the dynamics~\cite{RodenbornE338}. One of the key assumptions in RFT modeling is that environmental disturbances induced by the movement of a swimmer are sufficiently localized that forces and flow fields from neighboring body segments are completely decoupled. In the last decade, studies of locomotion of animals in moving in dry granular media, have revealed that the simplest form of RFT is remarkably successful in describing movement in such environments~\cite{maladen2009undulatory,zhang2014effectiveness,sharpe2015locomotor}. 

\newpage Further from the assumption of decoupled forces along a deforming body, a swimmer can be divided into many infinitesimal segments that can be treated independently.  In dissipation-dominated  environments, the net force on a body is zero at every moment in time, giving

\begin{equation}
\boldsymbol{F} = \int_{body} (d\boldsymbol{F}_\perp + d\boldsymbol{F}_\parallel) = 0. 
\label{eq:RFT}
\end{equation}

$d\boldsymbol{F}_\perp$ and $d\boldsymbol{F}_\parallel$ are the environmental reaction forces acting perpendicular and parallel to the surface of an infinitesimal segment of the body as it moves within the surrounding medium.

In the fluid-swimming nematodes analyzed here, locomotion occurs at sufficiently low Reynolds number (the ratio of inertial to viscous forces is approximately $0.1$) such that the assumption of zero net force (inertialess locomotion) is well justified. This has an important locomotor consequence: if a nematode stops self-deforming, its locomotory speed will decay to one half of its steady state speed within approximately $\approx$5 millisecond via viscous Stokes drag (see SI Sec. 4 and~\cite{berg1993random} for details on the calculation). We will refer to this as the coasting time, $\tau_{coast}$, and can form a non-dimensional parameter, the ``coasting number" $\mathcal{C}=2\tau_{coast}/\tau_{cycle}$, the ratio of this time to a typical undulatory timescale, $\tau_{cycle}$; for a nematode this is $\tau_{cycle} \approx 1$ seconds\footnote{Note that the coasting number can also be defined as the coasting distance (from stop self-deforming to complete stop) normalized by the characteristic length scale. The relationship between time-scale coasting number and length-scale coasting number is discussed in SI Sec. 4}.

We can extend this idea (and thus the inertialess locomotion assumption) to granular undulating systems. We justify our extension by estimating the ratio of inertial to frictional forces in Coulomb friction dominated systems (which are approximately rate independent): $\frac{mv_0/\tau_{cycle}}{\mu m g}$, where the numerator is the characteristic dynamic inertial forces with $m$, $v_0$, and $\tau_{cycle}$ are body mass, average speed, and temporal period respectively; the denominator is the characteristic frictional force where $\mu$ and $g$ are the friction coefficient and gravitational acceleration constants respectively. We can rewrite the above as $\frac{v_0/(\mu g)}{\tau_{cycle}}$, where the numerator can be interpreted as the time required to go from steady-state locomotion to a complete stop. Because force in a frictional fluid environment is approximately rate-independent, we have $\tau_{coast} = \frac{1}{2}v_0/(\mu g)$. In doing so, this ratio is then (in friction dominated systems) exactly $\mathcal{C}$. Thus, like in the viscous swimmer, in the macroscopic granular swimmers we have analyzed, $\mathcal{C}$ is sufficiently small (order $0.1$) such that we can neglect inertial effects in granular locomotion (Table S1 and S2).

\begin{figure}[t!]
	\centering
	\includegraphics[width=0.8\textwidth]{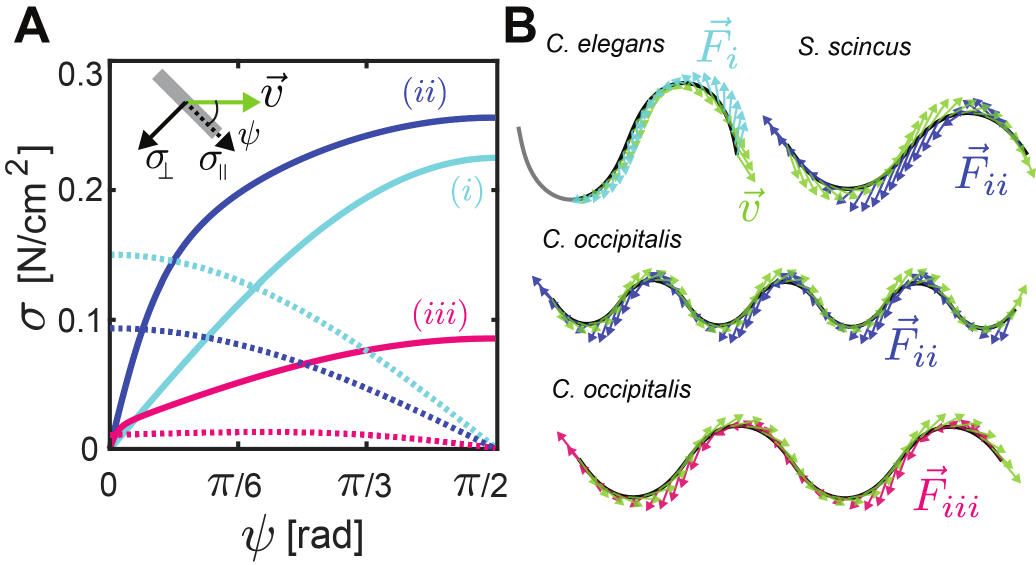}
	\caption{{\bf Resistive forces in frictional and viscous fluid environments.} (A) Curves $(ii)$ and $(iii)$ are fitted to environmental stresses (force divided by submerged area) on a thin plate moving through a granular substrate of $300$-$\mu$m glass particles (SI Sec.~4). Stresses parallel to (dashed lines) and perpendicular to (solid lines) are compared for $(ii)$ subsurface granular movement and $(iii)$ surface granular movement. We also illustrate the relationship between stresses and attack angle in $(i)$ a viscous fluid (e.g. buffer) assuming drag as in~\cite{sznitman2010propulsive,Backholm2015effects} . (B) Local velocity vectors (green) and segment velocities (colors correspond to (A)) for undulatory locomotors. 
	}
	\label{fig:RF}
\end{figure}

\newpage For forces acting on the body of a viscous drag dominated nematode we use Stokes drag from measurements from~\cite{sznitman2010propulsive,Backholm2015effects}. In the present study, these values likely represent an approximation of the drag forces, due to the presence of surface-induced hydrodynamic effects arising when swimming near a glass substrate.  Regarding forces on the granular swimmers in the locomotion systems we studied, elemental forces were determined from previously developed empirical force relations~\cite{Maladen:2011es,sharpe2015locomotor}, with modifications made (SI Sec.~4) to account for contact dynamics at granular surfaces (Fig.~\ref{fig:RF}A). Knowing the elemental forces, the instantaneous swimming speed, $\fibercirc(t)$, satisfying the force balance in Eq.~\ref{eq:RFT}, can be numerically determined (SI Sec. 4 and 5). Notably, $\fibercirc = [\xi_x, \xi_y, \xi_\theta]$, where each element denotes instantaneous velocity component in the forward, lateral, and rotational directions expressed in the time-varying local frame of the locomotor (which does not always align with the laboratory frame of reference). It is worth noting that the evaluation of instantaneous velocity is independent of the position of the locomotor with respect to the laboratory frame of reference ($\boldsymbol{g}=[x,y,\theta]$). The net displacement over a cycle can be numerically obtained by solving the ordinary differential equation:

\begin{equation}
    \dot{\boldsymbol{g}} = \begin{bmatrix}
    \dot{x} (t) \\
    \dot{y}(t) \\
    \dot{\theta}(t) \\
    \end{bmatrix} 
    = \begin{bmatrix} 
    \cos{\big(\theta(t)\big)} & -\sin{\big(\theta(t)\big)} & 0 \\
    \sin{\big(\theta(t)\big)} & \cos{\big(\theta(t)\big)} & 0 \\
    0 & 0 & 1 \\
    \end{bmatrix}\fibercirc(t) \label{eq:fullRFT}
\end{equation}

Our previous work~\cite{maladen2009undulatory,sharpe2015locomotor,schiebel2019mitigating} (reproduced as the dashed curves in Fig.~\ref{fig:pred1}($ii$)) demonstrated that RFT agrees with experimentally measured forward speeds (displayed here as body lengths per undulation cycle) of the sandfish lizard and the shovel-nosed snake. RFT calculations predict that intermediate body curvatures result in the largest displacements, and previous studies demonstrated that animals move using body curvatures that nearly coincide with the RFT-predicted speed-maximizing shapes (animal data reproduced from~\cite{sharpe2015locomotor,schiebel2019mitigating} as crosses in Fig.~\ref{fig:pred1}($ii$)). Given the success of RFT, we will model environmental forces with this approach.

\section*{Introducing the geometric framework}

RFT does not immediately facilitate ready understanding of how variation in paths in the configuration space result in different amounts of displacement (e.g. what happens to displacement when we go from circle paths to  ellipse paths or vary parameters like radius of circle), nor does it permit rapid understanding of why certain paths could be better than others~\cite{hatton2013geometric,Chong2019Coordination}. That is, beyond merely rationalizing observed gaits, what are the principles by which we can predict paths for other living (and ultimately non-living systems like robots) to achieve optimal performance without having to do  laborious calculations? While we will not fully address these questions in this paper, we now illustrate how progress can be made using the geometric phase approach.

\begin{figure}[t!]
	\centering
	\includegraphics[width=0.77\textwidth]{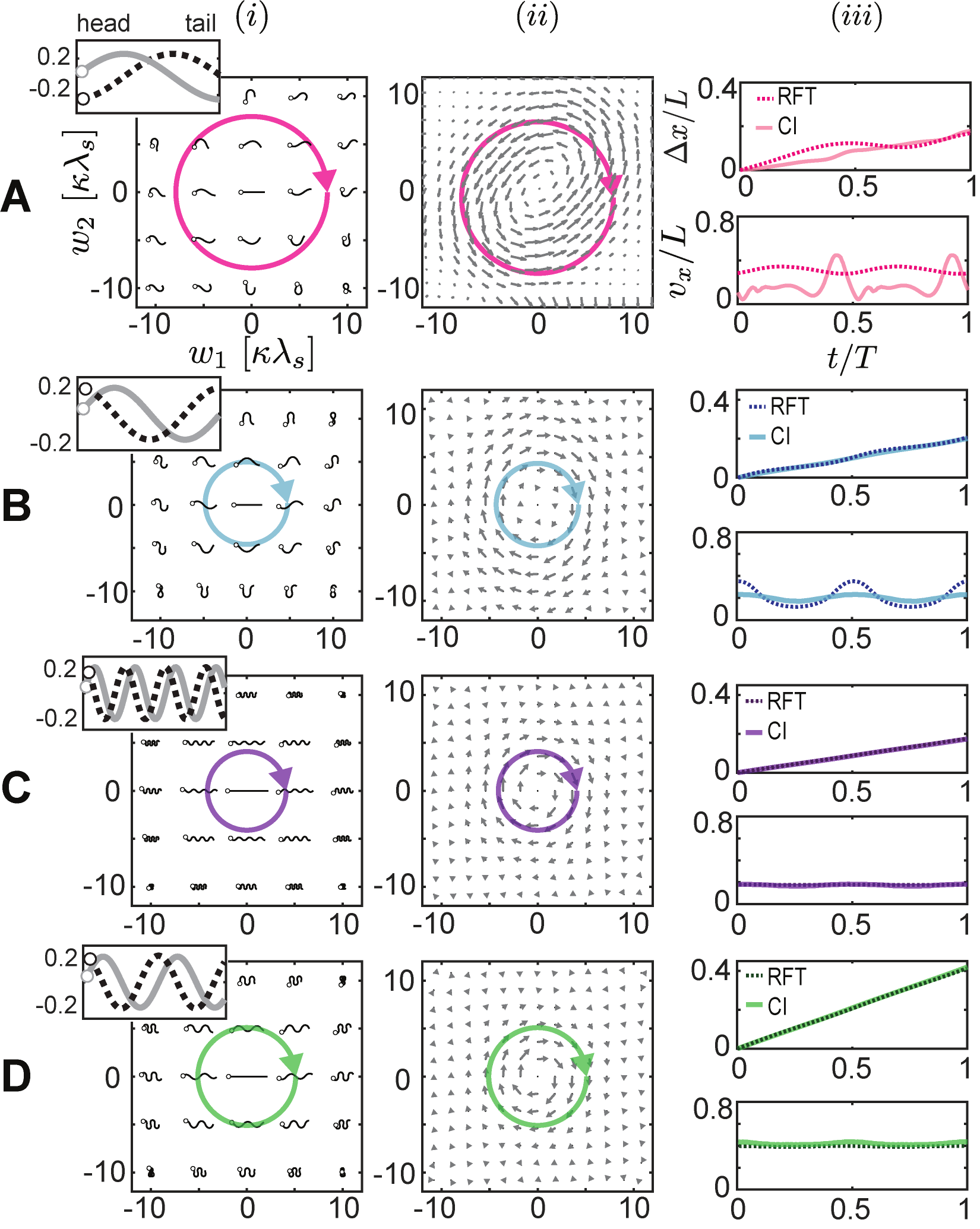}
	\caption{{\bf Geometric phase approach in  undulatory locomotors.} Row show calculations for (A) the nematode (B) the sandfish, (C) the shovelnose snake within sand, and (D) the shovelnose snake on the surface of sand. $(i)$ Space of shapes spanned by the two modes. The directed circle in on each plot represents a particular gait. Inset shows dominant two modes of body curvatures determined from PCA (with the open circles representing the head of the locomotor). $(ii)$ Connection vector fields determined using corresponding resistive forces shown in Fig.~\ref{fig:RF}A. Each arrow represents an infinitesimal displacement, and the directed circle again represents a gait. $(iii)$ Displacements throughout a gait cycle, determined via a contour integral (CI) along the gait specified in $(ii)$, are in agreement with RFT predictions.
	}
	\label{fig:GMnew}
\end{figure}

The first step in applying the geometric framework to our systems is to think of movement as a series of small displacements (translations or rotations) each introduced by small body configuration changes (``self-deformations'').   We seek a mapping that relates changes in body configuration space to changes in real world space. To construct such a map (which will prove valuable in subsequent sections as we build-up the machinery of the geometric approach), we next introduce a fundamental assumption in the geometric theory. That is, we assume that the small changes in displacement (small body velocities) are {\em linearly related} to small shape changes (small ``shape'' velocities) via the following: 

\begin{equation}
\fibercirc = \mixedconn(\boldsymbol{\alpha})\dot{\boldsymbol{\alpha}},
\label{eq:xi}
\end{equation}
where $\boldsymbol{\alpha}=[ w_{1}, w_{2}]^{T}$  is the \emph{shape} of the system;  $\dot{\boldsymbol{\alpha}}$ is the \emph{shape velocity}, the speed with which the body curvatures are changing; and $\mixedconn(\boldsymbol{\alpha})$ is the 
\textit{local connection}, which encodes environmental constraint forces that relate changes in body shape to the changes in position that they induce~\cite{krishnaprasad1994,kelly1995geometric,ostrowski1998geometric,lewis2000,Bloch2003}.  Each row of the local connection $\mixedconn(\boldsymbol{\alpha})$ can the be visualized as a vector field in the shape space (Fig.~\ref{fig:GMnew}.ii).

In our previous work on geometric methods in locomotion, we  showed the validity of the linearity in Eq.~\ref{eq:xi} of a granular Purcell's three-link swimmer~\cite{hatton2013geometric},  which only has two joints (internal shape Dof = 2). Here, we consider a 2-dimensional shape space identified by PCA as discussed earlier. We numerically obtained the local connection matrix  $\mixedconn$  using the same approach as discussed in~\cite{hatton2013geometric}. The validity of the linearity assumption for the two-dimensional reduced shape space representation is shown in Figs. S1-S4.

Given the local connection, we aim to identify relationships between the geometries of gaits and the displacements (net body translations or rotations) they produce (the accumulated geometric phase). 
Notably, Eq.~\ref{eq:fullRFT} is nonlinear and time-dependent, causing additional challenge to directly calculate the geometric phase. To simplify our analysis, we can approximate Eq.~\ref{eq:fullRFT} as:

\begin{equation}
    \Delta \approx \int_{0}^{T}\mixedconn(\boldsymbol{\alpha})\dot{\boldsymbol{\alpha}}dt, \label{eq:linearRFT}.
\end{equation}

\noindent where $\Delta = [\Delta_x,\ \Delta_y,\ \Delta_{\theta}]$ is the net displacement per cycle. Notably, without the time-varying rotation matrix in Eq.~\ref{eq:fullRFT}, Eq.~\ref{eq:linearRFT} is now time-invariant. Thus, the net displacement $\Delta$ is uniquely determined by the trajectory of shape change. The establishment of such approximation can allow us to evaluate the geometric phase as the linear, time-independent line integral over the local connection vector field. To further visualize and analyze the geometric phase, it can be convenient to turn the line integral into a surface integral via Stokes' theorem~\cite{shammas2007geometric,hatton2015EPJ}. Thus, the net displacement can be approximated by taking the integral of the \emph{curvature} of $\mixedconn$ over the region of the shape space enclosed by a gait, 

\begin{equation}
\Delta
\approx \iint_{\gait} \bigl( \nabla\times\boldsymbol{A}\bigr)\ d\boldsymbol{\alpha}.
\label{eq:stokes}
\end{equation}

Taking the individual components of $\nabla\times\boldsymbol{A}$ as \textit{height functions} yields gait-independent signed scalar maps that provide an intuitive visual depiction of how shape changes relate to net motions~\cite{hatton2011geometric,hatton2013geometric,hatton2015EPJ}. Surface integrals over the height functions predict displacements caused by the cyclic sequences of self-deformation described by the boundary of the enclosed area. As a result, the height function provides a way to identify gaits which produce no displacements (i.e., enclose no net curvature) as well as gaits that yield large displacements (i.e., enclose significant net signed curvature).

The primary source of error in our approximation Eq.\ref{eq:linearRFT} is the fact that matrix multiplication is not commutative~\cite{hatton2015EPJ}. This effect can be seen in the context of ``parallel parking'': a car cannot move sideways, but the interplay of forward and rotational velocities can generate an emergent lateral displacement if the forward and rotational velocities are properly sequenced. Hatton et al. \cite{hatton2011geometric} introduced the notion of systematically selecting the system gauge (the choice of body frame) such that the parallel-parking effect is suppressed. In the chosen gauge (which is generally close to, but not exactly, the center-of-mass and mean-orientation of the body elements).


\begin{figure}[ht!]
    \centering
	\includegraphics[scale=0.3]{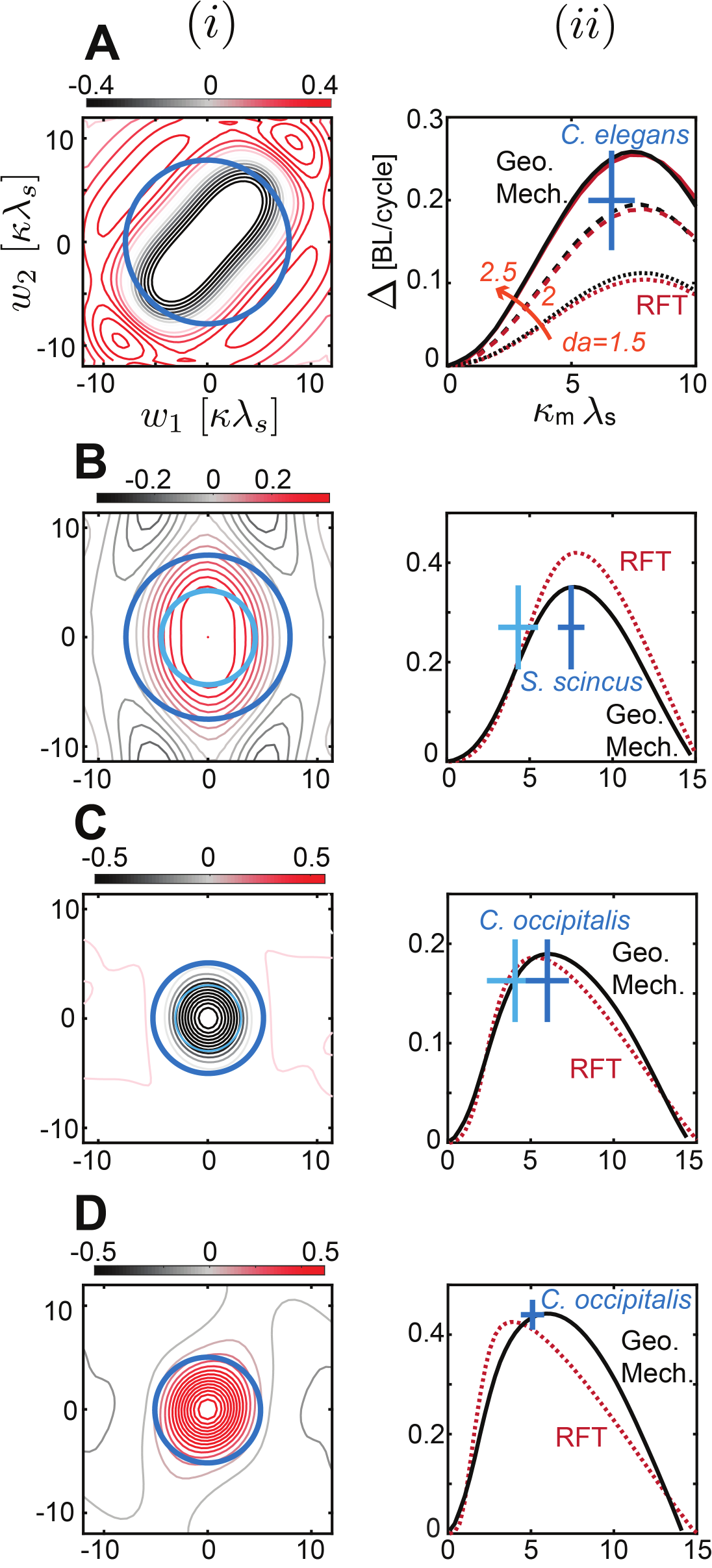}
	\caption{{\bf Results of geometric phase approach in planar undulation.}  $(i)$ Height functions (contours; here, color scale multiplied by $100$) and average animal gait (blue circles); and $(ii)$ Animal performance (average $\pm$ standard deviation, represented by the blue crosses), geometric phase (solid black curve), and RFT (dashed red curve) predictions for (A) \textit{C. elegans} in S-basal buffer, (B) \textit{S. scincus} in $300$-$\mu$m glass beads, (C) \textit{Ch. occipitalis} in $300$-$\mu$m glass beads, and (D) \textit{Ch. occipitalis} on the surface of $300$-$\mu$m glass beads. Crosses in A-$(ii)$ show average animal gaits from different data sets. In \cite{maladen2009undulatory}, curvatures were measured manually once per cycle (dark blue cross); in \cite{sharpe2015locomotor}, curvatures were measured throughout time ~\cite{sharpe2013environmental} (light blue cross). The light blue cross in B-$(ii)$ represents the postural dynamics as defined by projections onto the two dominant PCs, and the dark blue cross shows postural dynamics measured directly from kinematic data (as reported in~\cite{sharpe2015locomotor}, SI Sec. 3).	}
	\label{fig:pred1}
\end{figure}

\section*{Locomotion with continuous environmental contact}

We used Eq.~\ref{eq:stokes} to obtain the height functions shown in the contour plots of Fig.~\ref{fig:pred1}$(i)$ with average animal gaits overlaid in blue. Integrating the height function over circles of different radii provides predictions of how displacements per gait cycle, $\Delta$, depend on the maximum local curvature along the body of the animal, $\kappa_m \lambda_s$ (Fig.~\ref{fig:pred1}$(ii)$). 

\begin{figure}[ht]
    \centering
	\includegraphics[width=0.48\textwidth]{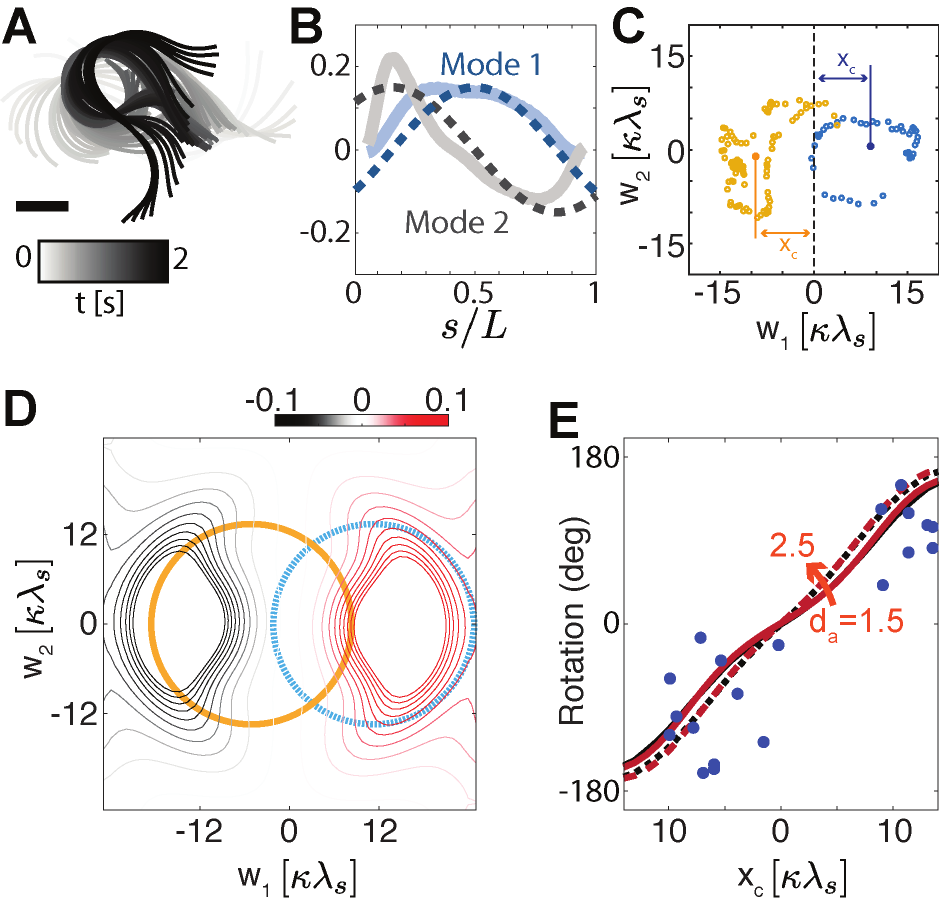}
	\caption{{\bf Geometric phase approach for turning in nematodes.}  One cycle of tracked midlines of a turning behavior in \textit{C. elegans} colored by time. Scale bar corresponds to 80 $\mu$m. (B) Two relative curvature modes (determined from PCA) account for 67.2\% and 17.2\% of variance observed in in-plane body configurations of 20 animals throughout 20 trials.  (C) Two examples of projections of body curvature onto a shape space identified by two PC modes. Blue points denote a typical CW turn and yellow points denote a typical CCW turn. (D) $\theta$ height functions for movement with $n = 0.7$ waves along the body. Blue and yellow circles denote the standard circle with off-origin center. (E) The comparison between GM predictions (surface integral over the height function, black curves), RFT calculation (red curves) and animal data (blue dots). Because of uncertainty in hydrodynamic effects relevant to the worm near a solid boundary, the GM prediction and RFT calculations are also generated for two other drag anisotropies $d_a=1.5,2.5$ (solid, dashed).
	}
	\label{fig:wormturning}
\end{figure}

For the subsurface movement of the sandfish lizard \textit{S. scincus} in a frictional fluid (Movie S1, Fig.~3.A.ii) and for swimming of \textit{C. elegans} in a true fluid (Movie S2, Fig.~3.A.i), comparisons of animal performance, direct RFT simulations, and height function surface integrals reveal that displacements per cycle are close to predictions (Fig.~\ref{fig:pred1}A,B). In the case of the sandfish, biological data from previous studies \cite{maladen2009undulatory,sharpe2013environmental,sharpe2015locomotor} provide a range of local curvatures (SI Sec.~3)
that are predicted to yield near-maximal displacements per cycle. This is in accord with previous muscle activation measurements that identified these template parameters as targets for the neuromechanical controller ~\cite{sharpe2013environmental}. In the case of the nematode, animal performance is in agreement with RFT and height function integral predictions. It is important to note here that because of power limitations in living (or synthetic) systems, displacement per cycle is not necessarily equivalent to speed. Since power generation capabilities of a swimmer are finite (e.g., muscles are not infinitely strong), larger shape changes (and therefore larger amplitude cycles) require more time to execute ~\cite{hatton2017cartography}. As a result, in the case of the sandfish, the peak power-limited speed occurs at a slightly smaller amplitude. In contrast, in low viscosity regimes \textit{C. elegans} is not power limited ~\cite{berri2009forward}. Power limits arise in higher viscosity regimes (e.g., agar), where the nematode uses greater muscle power to deform both its body and the surrounding fluid ~\cite{berri2009forward}.


The shovel nosed snake \textit{Chionactis occipitalis} used different waves for locomotion on the surface of sand and within sand (Fig.~\ref{fig:animals}B and C). For subsurface movement of \textit{Ch. occipitalis}, the first two PCs only capture $57.6\%$ of the variation in body configurations (Movie S3). One possible explanation for the apparent increase in postural complexity in subsurface movement may result from the increase in spatial frequency relative to surface crawling. As the number of waves along the body increases, travelling waves may become localized to such a degree that the coherence across the entire body is lost. Another possible mechanism may arise from the finite muscle torque provided by the organism in the increased stiffness subsurface environment. Torque limits may cause localized failures to realize a target waveform which may act as a random perturbation causing wave decoherence. 

Despite the lack of clean circularity in the configuration space, previous work showed that RFT and experiment are in good agreement~\cite{sharpe2015locomotor} which we attribute to the locality of granular resistive forces and the low slip locomotion. To test the efficacy of the geometric scheme in such a situation, we constructed the height function using the shape space spanned by the two dominant PCs and the fully-immersed environmental stresses (Fig.~\ref{fig:RF}A) is shown in Fig.~\ref{fig:pred1}B$(i)$. Predicted displacements from direct RFT simulations and height function surface integrals are in agreement with animal performance (Fig.~\ref{fig:pred1}B$(i)$, SI Sec. 4 and 5)
and reveal that animals use postural dynamics predicted to yield near-maximal displacement per cycle. 

The surface waveform used by \textit{Chionactis occipitalis}, which has fewer waves and lower curvatures, produces low-slip movement that leaves behind a well-defined track of depth $\approx 5$~mm~\cite{schiebel2019mitigating} (Movie S4). Given that RFT measurements for movement at the surface used a flat plate intruder, we added an additional term to the measured RFT relations to account for the kinetic Coulomb friction drag that opposes the motion of the local segment (SI. Sec.~4). 
Predictions from direct RFT simulation and height function surface integrals are in agreement with animal performance (Fig.~\ref{fig:pred1}C$(ii)$). We can rationalize the difference in the subsurface and surface waveforms by considering that granular drag force increases with intruder depth. For a fixed depth, increasing wave curvature and/or wavenumber decreases the amount of torque joints must produce \cite{schiebel2019mitigating}. Thus, when moving subsurface the snake can contend with the increased environmental forces by adjusting its waveform to reduce the torque required.

\section*{In place turns in limbless locomotors}

To navigate in complex terrain, effective turning (reorientation combined with or without translation) is as important as translation~\cite{astley2015modulation,stephens2008dimensionality,chevallier2008organisation}. 
However, unlike in forward motion, both the wavelength and the amplitude of the worm body wave become time dependent~\cite{stephens2008dimensionality}, leading to additional challenges to reconstruct the turning behavior. It was hypothesized that additional PCs (eigenworms) are need to fully describe the turning behaviors~\cite{stephens2008dimensionality}. Here, we hypothesize that turning and forward behaviors may share some common dynamics. Specifically, we posited that turning is a modulation of the serpenoid template in the shape space that can be rationalized by the geometric phase approach\footnote{Note that to fully describe worm turning behaviors on agar, we require 4 PC modes, as discussed in~\cite{wang2020omega,wang2022generalized}. However, in this study, we demonstrate that 2 PC modes are sufficient to characterize worm turning in buffer fluid.}.




The nematode exhibits extraordinary maneuverability in part because of its ability to perform \textit{omega turns}~\cite{srivastava2009temporal}. The turning motion is called an omega turn because during the course of turning the anterior end of the body (head) sweeps near the posterior end of the body (tail), inscribing an ``Omega" ($\Omega$) shape (Fig.~\ref{fig:wormturning}a). In Fig.~\ref{fig:wormturning}.A, we illustrate an example of an omega turn in \textit{C. elegans}. It is worth noting that the worm body in $\Omega$ shape cannot be readily prescribed by a sinusoidal function.

During an omega turn, the worm body orientation experiences significant (typically over $60^\circ$) rotation with negligible translation (typically less than 0.1 BL/cycle, see Fig.~\ref{fig:pred1}.A). We refer to this type of turning behavior as an \textit{in-place turn}. The efficacy of such turns have made them targets for control of turning in robot robots~\cite{wang2020omega,wang2022generalized}.

We perform PCA analysis to explore the modes that arise during Omega turns and notice that the modes of \textit{C. elegans} forward motion and turning motion are similar. Specifically, the two standard sinusoidal modes identified in \textit{C. elegans} forward motion (Fig.~\ref{fig:animals}A.iii) can explain over 80\% of the total variation of curvatures in turning motion. In Fig.~\ref{fig:wormturning}.B, we compare the PC calculated from turning motion (solid thick curves) and the standard sinusoidal basis identified by forward motion (dashed thin curves, same as Fig.~\ref{fig:animals}A.iii\footnote{Note that we shift the phase of the standard sinusoidal basis functions to align with the turning PC.}). The similar PC modes in forward and turning motions then suggests a conserved shape space for the variety of behaviors. 

Unlike in forward motion, for turning behaviors we notice that the trajectories in the shape space are centered off-origin along the $w_1$ axis. Notably, clockwise (CW) turns are typically associated with positive $w_1$ offset and counterclockwise (CCW) with a negative offset. For each trial (20 individuals, one trial per individual), we calculate the distance of the center of the recorded space shape trajectory from the origin ($x_c$) and measure the net rotation in the position space. Fig.~\ref{fig:wormturning}.E shows a clear correlation correlation between $x_c$ and the body rotation (each trial measurement is represented as a blue dot).

In the 2D plane, there are three connection vector fields (corresponding to forward, lateral and turning dynamics, Eq. 3). We can use the same methods narrated above to generate height functions for each. Performing these calculations (SI. Sec.~5E), the rotational height function reveals two distinct positive and negative regions. To quantify the turning modulation, we perform the surface integral over standard circular templates (with radius of $\kappa_m\lambda_s=8$) subject to different offsets in $w_1$ axis. We bound the uncertainty in effective drag anisotropy by computing a series of $\theta$ height function with different drag anisotropy ($d_a$). We compare the geometric phase approach predictions on displacement and rotation with the empirically measured data in Fig.~\ref{fig:wormturning}.E and observe good agreement. Thus, a shape space made with a common set of modes can describe both worm forward swimming and turning in fluids, simply by applying the appropriate height function. This shows that near-optimal turns can be achieved with the modes from forward crawling through amplitude modulation that serves to offset circular templates gaits from the origin. This suggests that to turn, organisms may capitalize on common neural means of coordination associated with forward locomotion.

\begin{figure}[t!]
	\centering
	\includegraphics[width=0.48\textwidth]{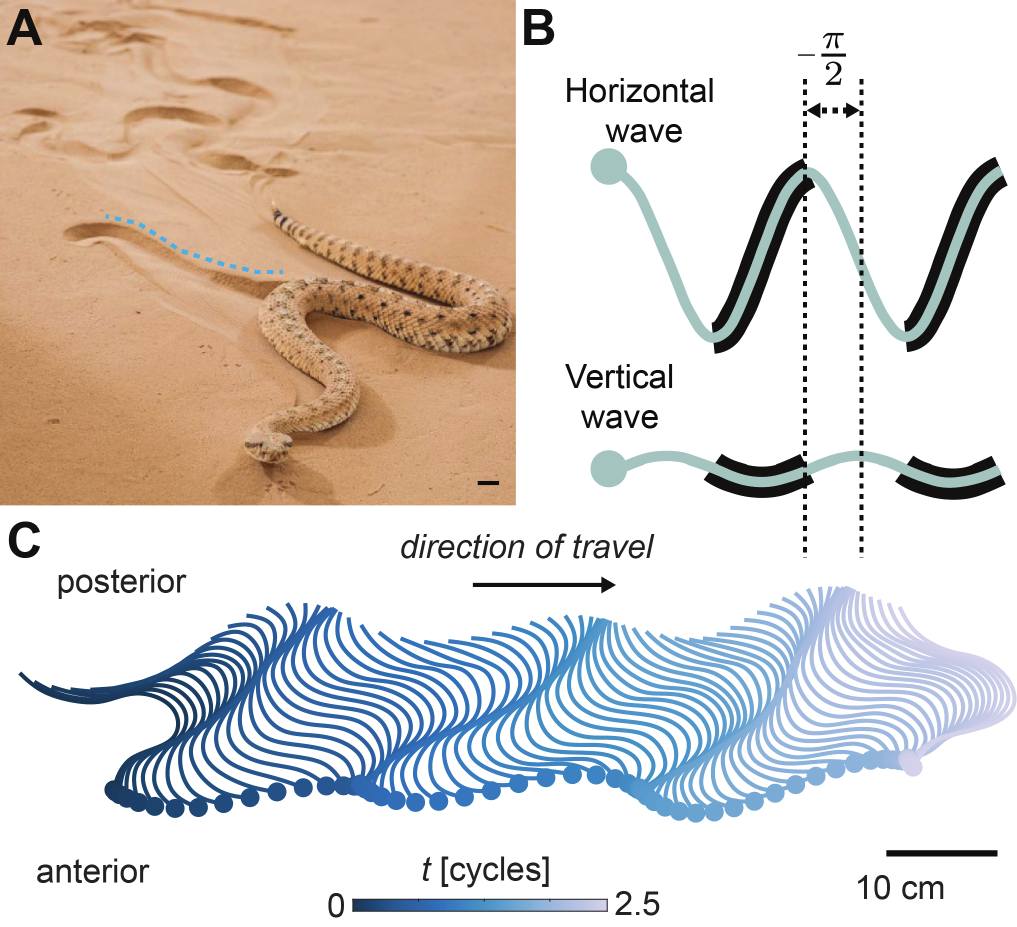}
	\caption{{\bf  Sidewinder rattlesnake locomotion on sand.} (A) Photo of \textit{Cr. cerastes} on sand. Black bar denotes 1 cm. (B) Depiction of coupling between in-plane and vertical waves (adapted from~\cite{astley2015modulation}). (C) Time-resolved kinematics are obtained from high-speed cameras.}
	\label{fig:swKinematics}
\end{figure}

 \section*{Sidewinding: Locomotion with changing environmental contact}

Thus far in the paper, the systems studied maintain continuous full-body contact with the environment during self-propulsion. However, many animals lift limbs or body portions as they move, changing their contact state throughout a gait cycle \cite{hildebrand1965symmetrical,jayne1986kinematics}. We therefore sought to build on our previous success in applying the geometric framework to such situations in robophysical models~\cite{mcinroe2016tail,chong2021frequency,chong2021coordination}.Following the other examples of undulatory behavior in this work, we chose an organism that modulates environmental contact within a flowable resistive environment with vertical waves, the rattlesnake, \textit{Crotalus cerastes} (Fig.~\ref{fig:swKinematics}A). This organism encounters sandy substrates in its native North American desert habitat and moves by sidewinding. Sidewinders locomote on homogeneous substrates ~\cite{Secor1994,Astley2020} by propagating a wave of planar body undulation coupled to an offset wave of body lifting, resulting in each body segment being cyclically lifted clear of the substrate, moved forwards, placed into a nearly static contact, then lifted again, with a slight phase offset between successive segments~\cite{Mosauer:1930,Burdick:1993,Hatton:2010ICRA:Sidewinding,jayne1986kinematics} (Movie S5). Thus, the snake generates multiple head-to-tail propagating regions of lifted movement and nearly static ground contact, and moves at a non-zero angle relative to the overall head-to-tail body axis (Fig.~\ref{fig:swKinematics}C)~\cite{Mosauer:1930,Burdick:1993,Hatton:2010ICRA:Sidewinding,jayne1986kinematics,chong2021frequency}. 

Despite the apparent complexity of these movements, our previous work indicated that the self-deformation pattern of \textit{Cr. cerastes} could be characterized as a template consisting of a superposition of a planar and vertical traveling wave~\cite{astley2015modulation}, with a phase shift of $\pm\pi/2$ between them (Fig.~\ref{fig:swKinematics}B). The modulation (e.g., changes in the maximum amplitude) of these waves can lead to diverse behaviors ~\cite{marvi2014sidewinding,astley2015modulation,chong2021frequency}. However, sidewinders tend to use relatively consistent horizontal waves during forward motion and are typically thought to regulate forward speed using temporal frequency changes of the wave~\cite{Secor1994,jayne1986kinematics,marvi2014sidewinding}.

\begin{figure}[t!]
	\centering
	\includegraphics[width=0.48\textwidth]{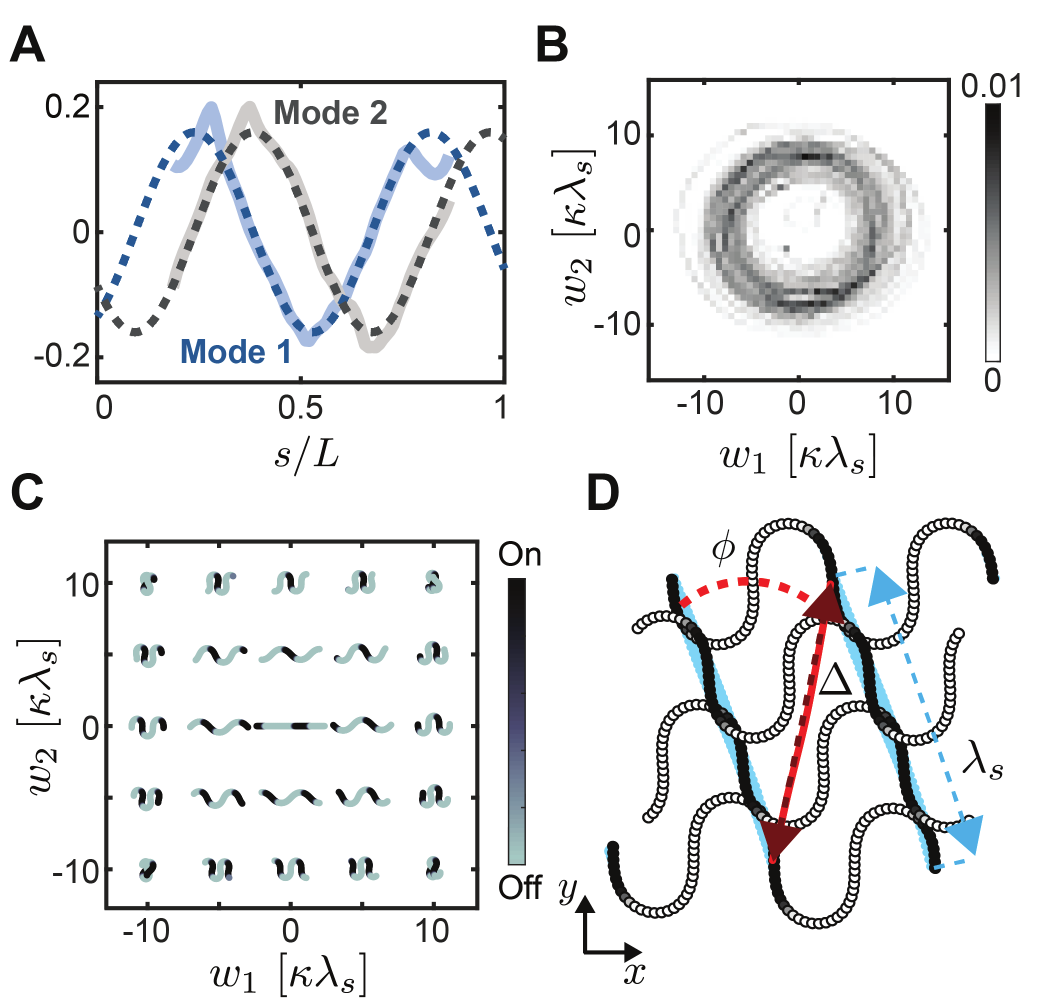}
	\caption{{\bf  Sidewinder rattlesnake substrate contact.} (A) Two relative curvature modes (determined from PCA) account for $42.4\%$ and $37.3\%$ of variance observed in in-plane body configurations of $4$ animals throughout $18$ trials. (B) 2D probability density map of animal data projected onto two dominant curvature principal components. (C) Shape space showing body configuration-dependent animal-environment contact model for an animal with $1.5$ waves along its body. (D) Schematic illustrating sidewinding locomotion. If the animal does not slip, displacements, $\Delta = \lambda_s \cos \phi$, can be predicted from geometry. }
	\label{fig:swContact}
\end{figure}

Indeed, when we applied PCA analysis to horizontal wave dynamics of previously collected \textit{Cr. cerastes} data, we discovered that across trials, dynamics of the horizontal wave consists of a circular path in a two sinuous mode configuration space (Fig.~\ref{fig:swContact}A-B)) of a characteristic radius (and therefore maximal body curvature). We thus posit that this circle forms a control template enabling these animals to move rapidly over loose granular surfaces.

Although the vertical body dynamics have not been carefully experimentally resolved ~\cite{marvi2014sidewinding}, they are assumed to be a traveling wave (and thus described by two modes) that sets the periodic contact pattern (Fig.~\ref{fig:swContact}A-B). On level granular media~\cite{marvi2014sidewinding,astley2015modulation}, parameters describing the vertical template remained approximately constant. Therefore, to model the vertical wave interaction, as in \cite{astley2015modulation,rieser2021functional,chong2021frequency} we introduced a weighting prefactor, $c$, that specified how much of the environmental force each infinitesimal segment experienced\footnote{We also modified our force balance to ensure that the total weight of remains unchanged (SI Sec.~4).}. Specifically, we modify the resistive force balance in Eq.~\ref{eq:RFT} as:

\begin{equation}
\boldsymbol{F} = \int_{body} \Big(c d\boldsymbol{F}_\perp(w_1,w_2,\fibercirc) +  c d\boldsymbol{F}_\parallel(w_1,w_2,\fibercirc)\Big) = 0. 
\label{eq:modRFT}
\end{equation}

Previous work~\cite{astley2015modulation} revealed that the three-dimensional pose of \textit{Crotalus cerastes} could be represented by a horizontal wave (characterized by $w_1,w_2$) coupled to a phase-shifted vertical wave that sets the environmental contact condition. To properly couple the contact function to the in-plane shape, we introduced the vertical wave: $\delta(s) = a \sin [2\pi n s/L + \tan^{-1}\left( w_2/w_1 \right) -\pi/2 ]$, where $a$ is the amplitude of the vertical wave, $w_1$ and $w_2$ describe the in-plane wave shape. To set the contact using the vertical wave description, $\delta$, we defined the smoothly-varying function $c(s) = 1/\big(1+\exp[\delta(s) + b]\big)$, where $c \in [0,1]$ sets the local fraction of the environmental force experienced as a function of position along the body, and $b$ sets contact width. To be consistent with previous observations, $a = 15$ and $b = 0.5$ are chosen so that approximately $34\%$ of the animal's body is on the ground~\cite{marvi2014sidewinding}. Fig.~\ref{fig:swKinematics}B shows how environmental contact couples to an in-plane shape, and Fig.~\ref{fig:swContact}C shows how this contact varies throughout the in-plane shape space for an animal with $n = 1.5$ waves along its body.

\begin{figure}[ht!]
	\centering
	\includegraphics[width=0.48\textwidth]{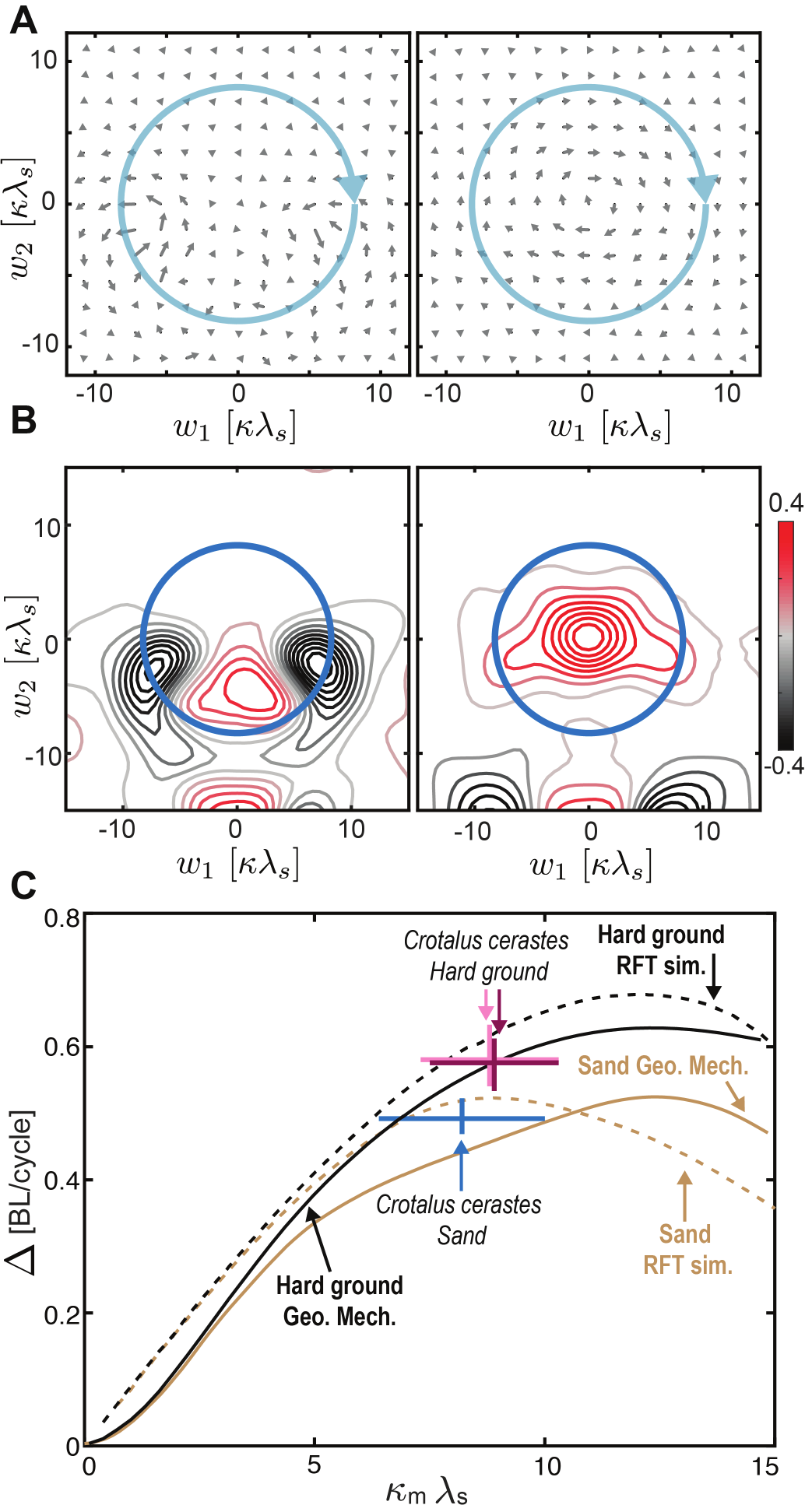}
	\caption{{\bf  Geometric phase approach in sidewinding.}  (A) Connection vector fields and (B) $x$- and $y$ height functions (shown here as contour plots with color scales multiplied by $100$) for movement on sand with $n=1.5$ waves along the body
	The blue circle shows average animal performance of \textit{Cr. cerastes} on sand. (C) Comparisons of RFT simulations (dashed tan curve) and geometric phase approach calculations (solid tan curve) for movement on sand. Biological data: \textit{Cr. cerastes} on a $7.6$-cm layer of sand (blue); \emph{N. fasciata} on a $5$-cm layer of sand (dark purple). \textit{Cr. cerastes} on an oak board roughened by a layer of adhered glass beads (dark magenta); \textit{Cr. cerastes} on a smooth oak board (light magenta); \emph{Cr. cerastes} on a $1.5$-cm layer of sand (light purple)~\cite{jayne1986kinematics}.}
	\label{fig:swCCFs}
\end{figure}

Fig.~\ref{fig:swContact}D shows four RFT simulation snapshots throughout one undulation cycle. Contact patches (dark regions) originate near the head and are propagated toward the tail. Given the experimentally observed oblique direction of travel (relative to the head-to-tail body axis), we expect the  kinematics in our modeling to produce significant displacements in both the $x$- (forward) and $y$ (lateral) directions. We therefore numerically computed height functions to visually and intuitively prescribe motions along both the $x$ and $y$ directions  (Fig.~\ref{fig:swCCFs}A). We define the total predicted displacement is given as $\Delta = (\Delta_x^2 + \Delta_y^2)^{1/2}$, where $\Delta_x$ and $\Delta_y$ are displacements predicted from $x$ and $y$ height functions, respectively.

Fig.~\ref{fig:swCCFs}B shows that, for movement on granular media, direct RFT simulations (dashed tan curve) and geometric computations (solid tan curve) predict similar maximal displacements. 
The RFT gait amplitudes predicted to yield peak performance differ slightly from those predicted to from the ``cartoon model'' presumably because at large amplitudes slip in the primary direction of motion occurs.
Despite the differences in predicted gait amplitude, the displacement curves predicted are not highly sensitive to gait amplitude variation over a broad range.
Note that the discrepancy between RFT simulation and geometric phase approach for sidewinding on sand can be a result of the non-commutativity of body velocities. As shown in Fig.~\ref{fig:swCCFs}a, the body velocity in $x-$ and $y-$ directions have comparable magnitudes, which can lead to relatively large non-commutativity effect in body velocities~\cite{hatton2015EPJ}.

\subsection*{Differential turns in sidewinding}

As with the worms, the geometric phase approach can also help rationalize the spectrum of``differential turns" observed in \textit{Cr. cerastes}~\cite{astley2015modulation} by appropriate template modulation. Such turns are interesting because those sidewinders can modulate the net translational displacement associated with the a particular turning angle. That is, sharp differential turns (e.g., $\sim90^\circ$ per cycle) are often accompanied by reduced translational displacement, and gradual differential turns (e.g., less than $20^\circ$ per cycle) by large translation. Therefore, unlike in straight sidewinding where the animals use relatively consistent horizontal waves (e.g., consistent wave amplitude and propagation speed), animals exhibit varying horizontal wave dynamics during differential turns.

 \begin{figure}[ht!]
	\centering
	\includegraphics[width=0.48\textwidth]{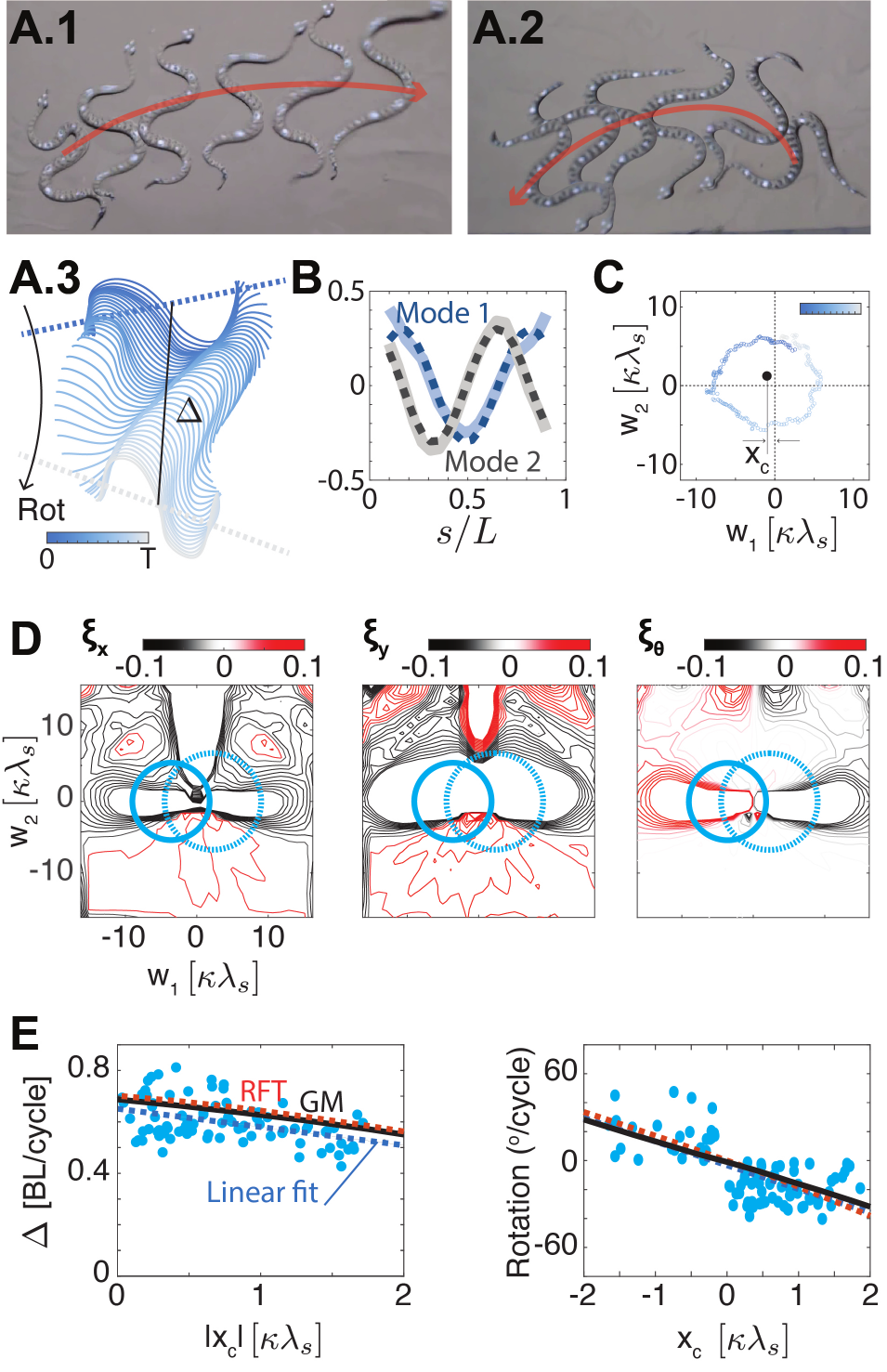}
	\caption{{\bf Geometric phase approach for the ``differential turn" in sidewinding snakes}  Snapshots of \textit{Cr. cerastes} body configurations performing (A.1) gradual (A.2) sharp differential. (A.3) One cycle of tracked midline of differential turn colored by time. Rotation (Rot) and displacement are labelled. (B) Two relative curvature modes (determined from PCA) account for 44.7\% and 24.5\% of variance observed in in-plane body configurations of 4 animals throughout 47 trials. (C) A typical projection of body curvature onto two dominant curvature principal components colored by time. (D) $x$, $y$, and $\theta$ height functions (shown here as contour plots, color scales in $x$, $y$ height function are multiplied by 100) for movement with n = 1.2 waves along the body. The blue circle typical CW and CCW differential turn in \textit{Cr. cerastes} (E) The comparison between direct RFT, GM (surface integral in height function), and animal data (light blue dots) with its linear fit ($p<0.001$).
	}
	\label{fig:diff}
\end{figure}

The differential turning mode, first analyzed by~\cite{astley2015modulation}, can be characterized by amplitude modulations on the horizontal wave (Fig.~\ref{fig:diff}a). Depending on the magnitude of amplitude modulation, animals can control the degree of turning during translation. Notably, in prior work, the amplitude modulation refers to the modulation of elemental velocity distribution (from anterior to posterior)~\cite{astley2015modulation}. It is yet not clear how internal curvature should be adapted to facilitate such amplitude modulation on spatial velocity distribution.  

We used PCA to find the curvature modes during differential turns. We noticed that the first two principal components can account for over 69.1\% of the variance (Fig.~\ref{fig:diff}b); and the two modes for straight sidewinding are almost identical to those for differential turns with two subtle differences: (1) the spatial frequency changes from 1.5 in straight sidewinding to 1.2 in differential turn, and (2) the phasing relationship between two modes changes from mode 1 ahead of mode 2 in straight sidewinding to mode 2 ahead of mode 1 in differential turn. We suspect that these subtle changes emerged from the variation in the horizontal wave.

We project body curvatures during differential turns onto the first two PC modes. We notice that the trajectory in PC space emerged as a circle, with its center offset from the origin. As we observed with worms, we posit that the offset of the center from the origin can serve as an indicator of the degree of turning and translation. We then measure the rotation and translation for each cycle (95 cycles over 47 trials, a trial might include multiple cycles). The rotation (Fig.~\ref{fig:diff}E, left) and translation in each cycle (represented as a blue dot) are plotted as a function of $w_1$-offset of the trajectory center (arithmetic average) from the origin. We run a linear regression between $w_1$-offset and rotation, and observe significant relationships (translation: $r^2=0.20$, $p<0.0001$; rotation: $r^2=0.50$, $p<0.0001$).

We use geometric phase approach to explain the observed correlation. We numerically compute the height functions on granular media using the same contact function and RFT relationships as straight sidewinding. We observe that the clusters of positive and negative volumes are distributed along the axis of $w_2=0$ in $\theta$ height function, indicating that the introduction of offset in $w_1$ direction can indeed lead to body rotation.

We then perform surface integrals over circular gaits on the height functions. Specifically, we used the following equations to prescribe off-centered circles in the calculation:

\begin{align}
    w_1(t) &= (\kappa_m\lambda_s-|x_c|)\sin{(t)}+x_c\nonumber \\
    w_2(t) &= (\kappa_m\lambda_s-|x_c|)\cos{(t)} \label{eq:ellipse}.
\end{align}

\noindent We compare the surface integral with RFT calculation (integrating Eq.~\ref{eq:ellipse} over a cycle $t\in [0\ 2\pi)$) and the fitted linear regression, and observed good agreement~\footnote{Note that forward and lateral height functions in Fig.~\ref{fig:diff}D have opposite sign to those in Fig.~\ref{fig:swCCFs} because of the changes in the relative phasing relationship between mode 1 and 2.}.

Our analysis illustrates that the seemingly distinct behaviors of straight sidewinding and differential turn (through amplitude modulation on spatial velocity distribution) share the same shape space (PC modes). Moreover, the simple modulation scheme (with a single variable: $w_1$-offset) reconstructs the complicated spectrum of behaviors with different degrees of displacement and rotation, which further facilitates a relatively simple understanding of seemingly complex sidewinder locomotion.

\section*{Discussion and Conclusions}
In this paper, we presented the first biological experimental application of a geometric framework of locomotion proposed in the 1980s by physicists~\cite{shapere1987self} to describe movement and developed during the last three decades for robotic applications by control theorists ~\cite{marsden1990reduction,kelly1995geometric,hatton2015EPJ}. Application of this framework to organisms across scales and levels of complexity revealed that self-deformation kinematics were well described by serpenoid templates (near circular paths in low-dimensional body configuration spaces). Further, observed animal self-deformation kinematics and locomotor performance coincided with predictions that nearly maximized the surface integral over this curve in a diagram called a height function, which corresponds to nearly maximizing a geometric phase in the space of animal body configurations. We can thus posit that a emergent guiding principle for control of limbless  undulatory locomotion in highly damped environments is: maximize geometric phase using a template which is approximately a circle in a two dimensional configuration space. This is true both in situations with continuous and variable environmental contact -- for cm scale lizards and snakes in flowable (granular) and frictional environments (both of which are ubiquitous in natural organism environments) as well as a tiny nematode worm in a fluid environment. Finally, modulations to serpenoid templates can explain the turning behaviors (omega turn and differential turn), which further indicates the generality of our proposed templates regardless of the underlying physiological, neural, and biomechanical mechanisms responsible for the self-deformation patterns.

Why is this seemingly abstract approach, requiring mathematical tools not traditionally represented in the fields of bio- and neuromechanics, of value in the study of locomotion? First, recognition that the dimensionality reduction scheme proposed in~\cite{stephens2008dimensionality} applies to diverse and significantly more complex organisms than nematode worms is a useful step in template recognition. That is, to evaluate performance of animals, we can initiate a search for principles starting with simplicity rather than complexity. Second, the diagrammatic approach simplifies search for candidate paths in configuration spaces; typically such a search requires brute force computation of all allowed paths but with height functions in hand, optimal self-deformation patterns for translation (or rotation~\cite{wang2020omega}) becomes relatively straightforward to hypothesize. Third, the scheme can be readily adapted~\cite{hatton2017cartography} to imposing biologically relevant constraints like internal force, power or energetic requirements. 

Amplifying on this last point: to demonstrate our method, in this paper all computations were performed in the space of shape changes, where all shape changes are equally easy to execute and the geometric phase nearly maximized displacement per cycle. But recent work~\cite{hatton2017cartography} has demonstrated connection vector fields can be computed within modified metric spaces (such as weighting shape changes by internal force and power requirements). Modifying the underlying metric to account for physiologically-relevant constraints could provide insights into the goals and limitations associated with a broader range of behavior and inherent biological variability. With such theoretical tools in concert with new experimental tools we can now address how lower level physiological, neural, and biomechanical mechanisms (``anchors'' in the parlance of~\cite{full1999templates}) conspire to generate the template dynamics~\cite{sharpe2013environmental,ding2013emergence}. In large scale organisms, electromyographic and neural recording tools at the macroscale have been used for decades to assay neuromechanical control across taxa~\cite{loeb1986electromyography}. Relevant to animals studied here, such tools proved useful to test a hypothesis of shape control in sandfish locomotion~\cite{sharpe2013environmental}. However, such tools provide crude assays relative to those available in model organisms (like \textit{C. elegans}). The optical transparency and genetic mutability of these worms provide experimental opportunities to connect templates to underlying anchors through genetic circuit manipulation, optogenetics and calcium imaging. Using the geometric scheme to develop templates for diverse environments, we can, for example, test the hypothesis that nematodes control for force and thus shapes emergently arise from neuromechanical feedback~\cite{johnson2021neuromechanical,boyle2012gait}.




Finally, the results in this paper used reductions of the kinematics of animal movements to simple patterns of self-deformation and used RFT to describe environmental interactions. We posit that this is because we focused on relatively simple tasks, such as escape and steady transit in homogeneous media. The extent to which more complex locomotor behaviors (e.g.~\cite{gart2019snakes,astley2009arboreal}) could be amenable to such simplification remains an open question. Complexity may arise from unusual gait paths in the shape spaces or from additional modes. While \textit{C. elegans} uses a circular gait in agar~\cite{stephens2008dimensionality}, changes in the viscosity of the environment could lead to a spectrum of gaits, ranging from circular to elliptical gaits. Such gait spectrum observed in the continuum of lizard body elongation and limb reduction, where elliptical gaits emerge as limb size reduces~\cite{chong2022coordinating}. But again, generalizations to more complex descriptions (e.g., more modes) are straightforward--geometric methods can handle higher dimension but the visualization of height functions becomes more difficult~\cite{ramasamy2016soap,chong2019hierarchical}. It will be interesting to explore other, potentially more complex, locomotor behaviors not described by planar traveling waves of body bends (e.g., rectilinear motion in snakes and peristalsis in worms, walking and crutching of mudskippers~\cite{mcinroe2016tail}).

\section*{Materials and Methods}

For all experiments, high-speed videos (excluding \textit{C. elegans}) were recorded as animals moved on various substrates, and animal postures were obtained by identifying tracking features located along the midline of the dorsal side (lateral side for \textit{C. elegans}) of the animal in each frame. For subsurface animals, lead markers glued to the animal were visible in x-ray images and extracted as a part of a previous study~\cite{sharpe2015locomotor}. For surface animals, either markers or features on the snakes were tracked though time, and again these points were interpolated using a cubic spline. For \textit{C. elegans}, animals were placed in 10 $\mu$L of S-basal on a glass slide. 3D movement was constrained by a glass coverslip to a 50 $\mu$m height, with the use of tape (Kapton). Worms were imaged using a brightfield microscope (Leica ATC 2000). For forward crawling data, midlines were obtained using custom MATLAB code. Binary masks of the organism were created via thresholding. The binary masks were then skeletonized and splined. For turning data, DeepLabCut~\cite{mathis2018deeplabcut} was used to track the self-occluding postures where binarization and skeletonization fails. For \textit{Chionactis occipitalis} on the surface of sand, the positions of naturally occurring evenly-spaced black bands were identified in each video image frame as part of a previous study~\cite{schiebel2019mitigating}. For \textit{Crotalus cerastes}, infrared-reflective markers were placed along the animal, and a Natural Point Optitrack Flex 13 camera system automatically identified and recorded marker positions at 120 frames per second.  Further biological information and experimental details for \textit{Scincus scincus} and \textit{Chionactis occipitalis} are given in Table S1 and for \textit{Crotalus cerastes} in Table S2. Details on numerical analysis and geometric phase approach are provided in SI Sec. 3-5.

\acknowledgements{The authors thank Gordon Berman, Paul Umbanhowar, Alfred Shapere, Sarah Sharpe and Zeb Rocklin for helpful discussions and Joseph Mendelson for assistance collecting biological data. This work was supported by NSF PoLS PHY-1205878, PHY-1150760, CMMI-1361778, CMMI-1653220, ARO W911NF-11-1-0514, U.S. DoD, NDSEG 32 CFR 168a (PES), and the DARPA YFA.}


%


\section*{Supplementary Information}

\subsection*{Figures}

\begin{figure}[h]
	\centering
	\includegraphics[width=0.8\textwidth]{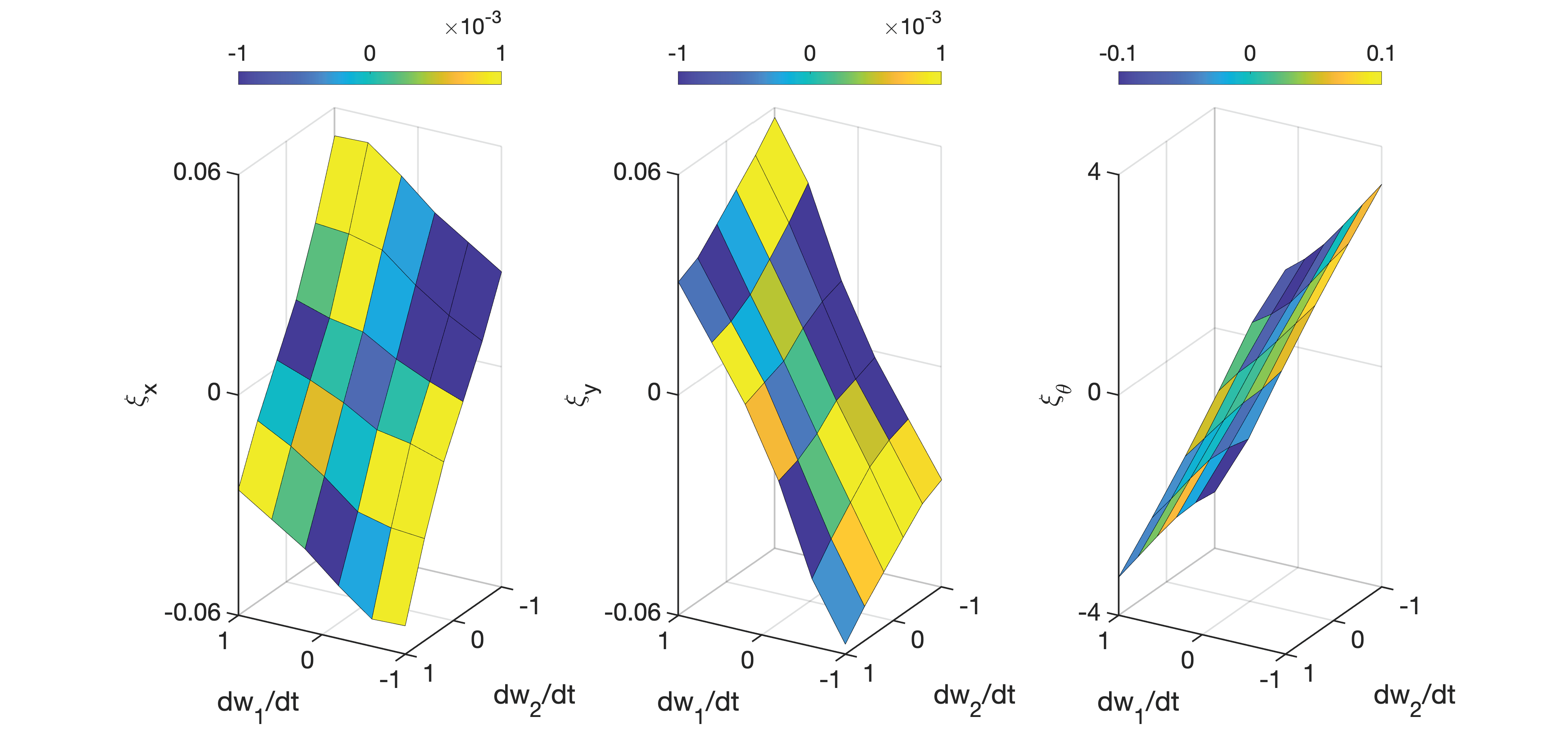}
	\caption{{\bf Testing the validity of the linearity ansatz of the local connection in a 2-dimensional reduced shape space representation of sandfish locomotion}  The mapping between shape velocity ($\dot{w_1}$ and $\dot{w_2}$) and body velocity ($\xi_x$, $\xi_y$, and $\xi_\theta$) evaluated at shape variable ($w_1= -0.7388$, $w_2=-0.7388$) are illustrated. Color on surface denotes the deviation from actual body velocity from a linear plane fitting. The mapping evaluated at other shape variables are available online.
	}
	\label{linearity1}
\end{figure}

\begin{figure}[h]
	\centering
	\includegraphics[width=0.8\textwidth]{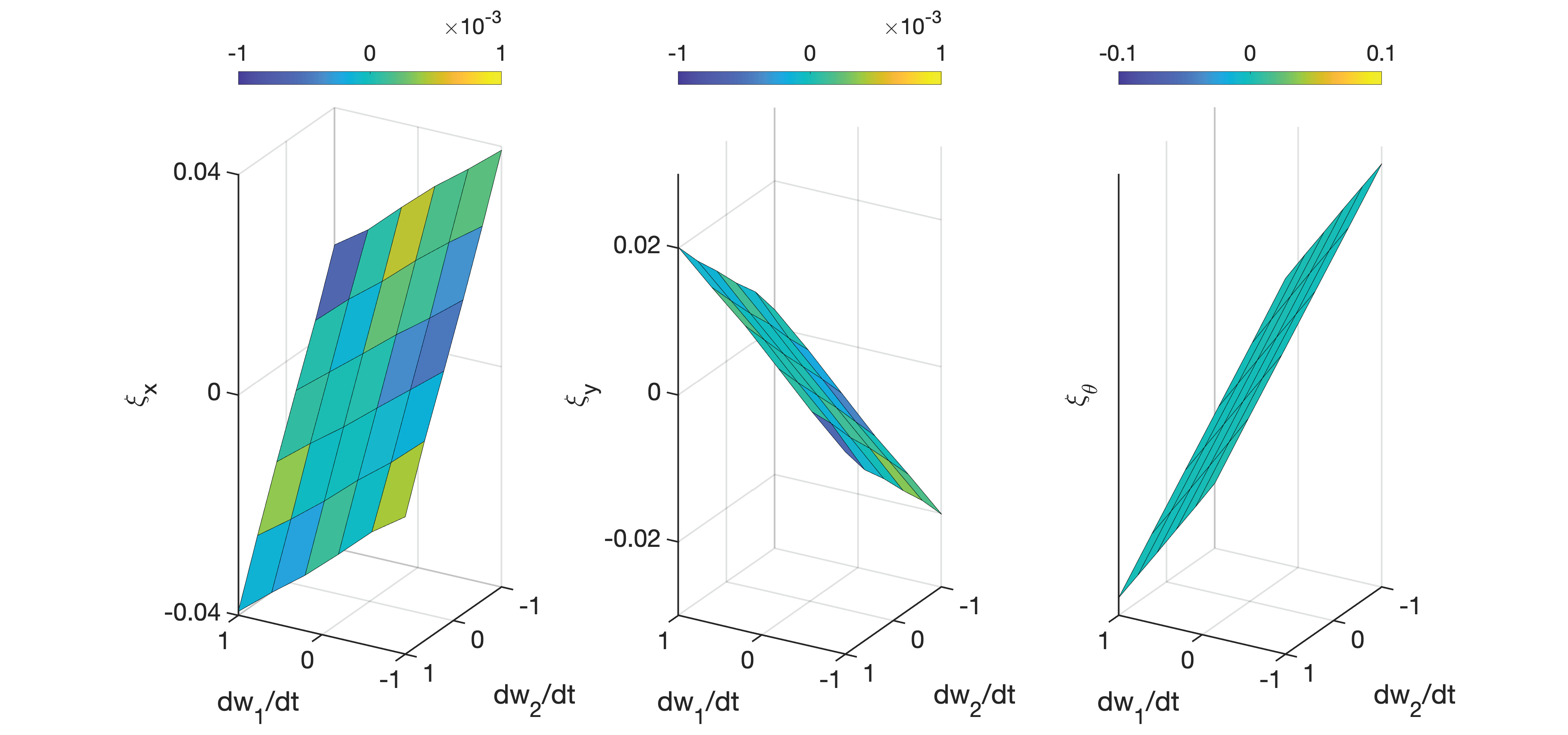}
	\caption{{\bf Testing the validity of the linearity ansatz of the local connection in a 2-dimensional reduced shape space representation of shovelnose snake within sand}  The mapping between shape velocity ($\dot{w_1}$ and $\dot{w_2}$) and body velocity ($\xi_x$, $\xi_y$, and $\xi_\theta$) evaluated at shape variable ($w_1= -0.7388$, $w_2=-0.7388$) are illustrated. Color on surface denotes the deviation from actual body velocity from a linear plane fitting. The mapping evaluated at other shape variables are available online.
	}
	\label{linearity2}
\end{figure}

\begin{figure}[h]
	\centering
	\includegraphics[width=0.8\textwidth]{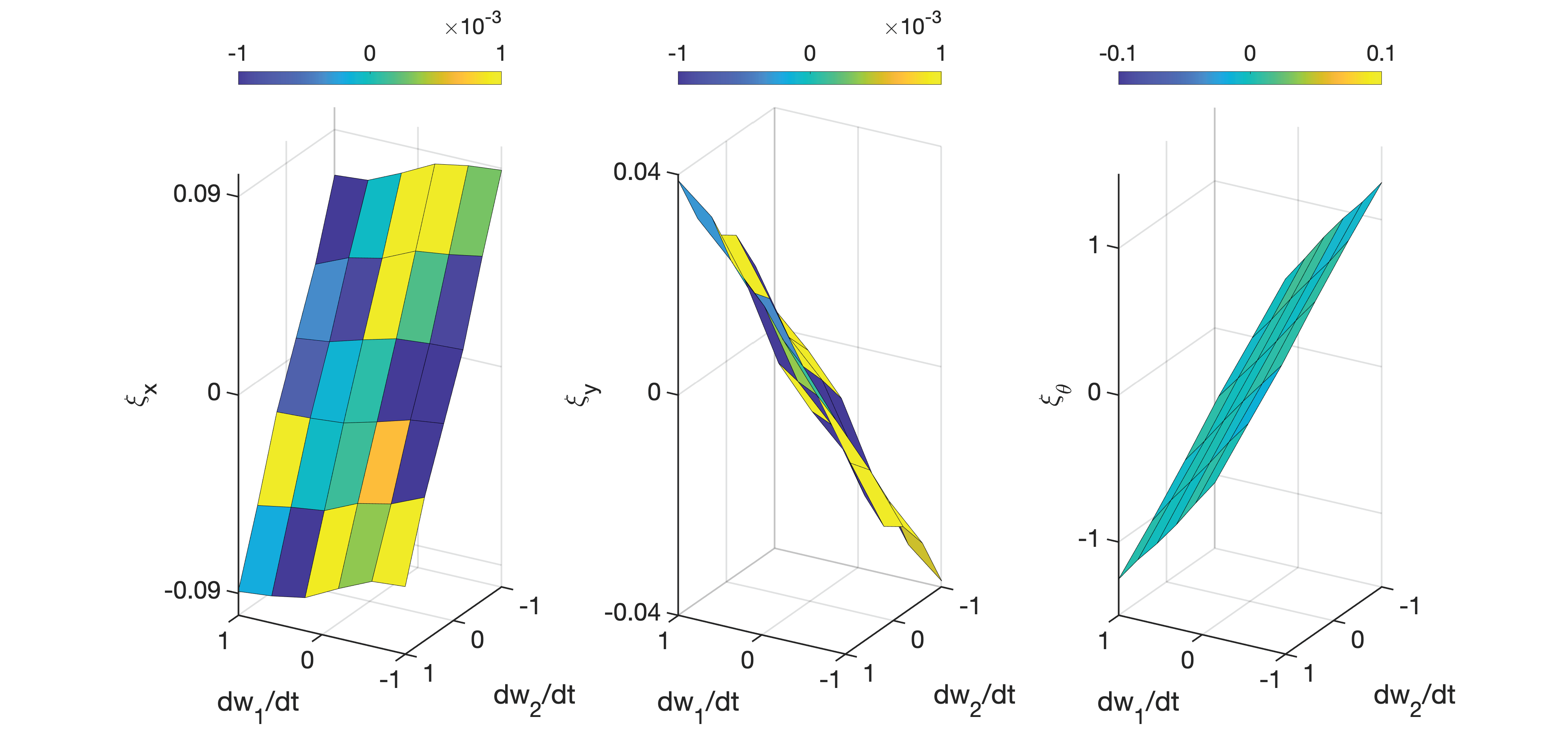}
	\caption{{\bf Testing the validity of the linearity ansatz of the local connection in a 2-dimensional reduced shape space representation of  shovelnose snake on the surface of sand}  The mapping between shape velocity ($\dot{w_1}$ and $\dot{w_2}$) and body velocity ($\xi_x$, $\xi_y$, and $\xi_\theta$) evaluated at shape variable ($w_1= -0.7388$, $w_2=-0.7388$) are illustrated. Color on surface denotes the deviation from actual body velocity from a linear plane fitting. The mapping evaluated at other shape variables are available online.
	}
	\label{linearity3}
\end{figure}

\begin{figure}[h]
	\centering
	\includegraphics[width=0.8\textwidth]{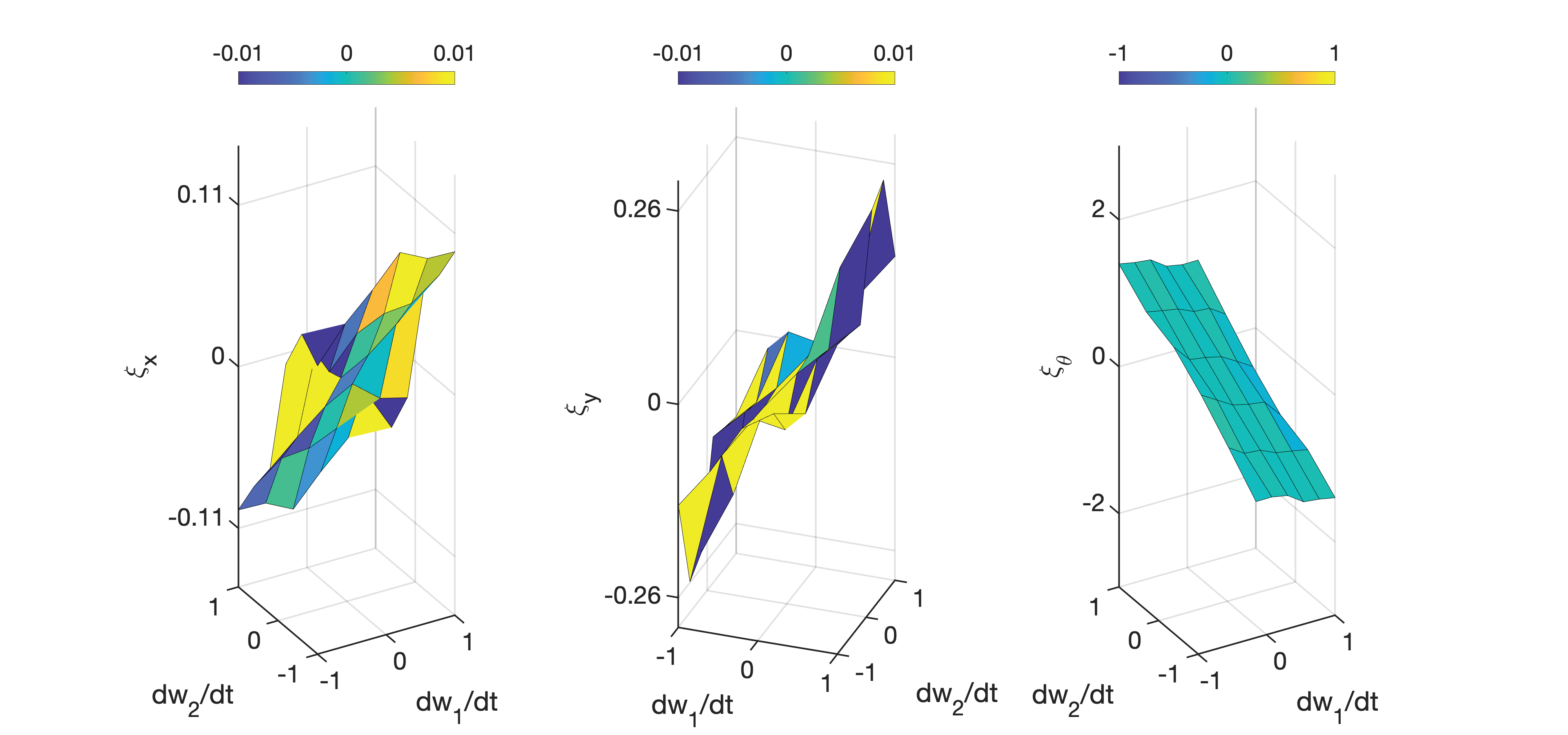}
	\caption{{\bf Testing the validity of the linearity ansatz of the local connection in a 2-dimensional reduced shape space representation of sidewinding locomotion}  The mapping between shape velocity ($\dot{w_1}$ and $\dot{w_2}$) and body velocity ($\xi_x$, $\xi_y$, and $\xi_\theta$) evaluated at shape variable ($w_1= -0.7388$, $w_2=-0.7388$) are illustrated. Color on surface denotes the deviation from actual body velocity from a linear plane fitting. The mapping evaluated at other shape variables are available online.
	}
	\label{linearity4}
\end{figure}

\begin{figure}[h]
	\centering
	\includegraphics[width=0.8\textwidth]{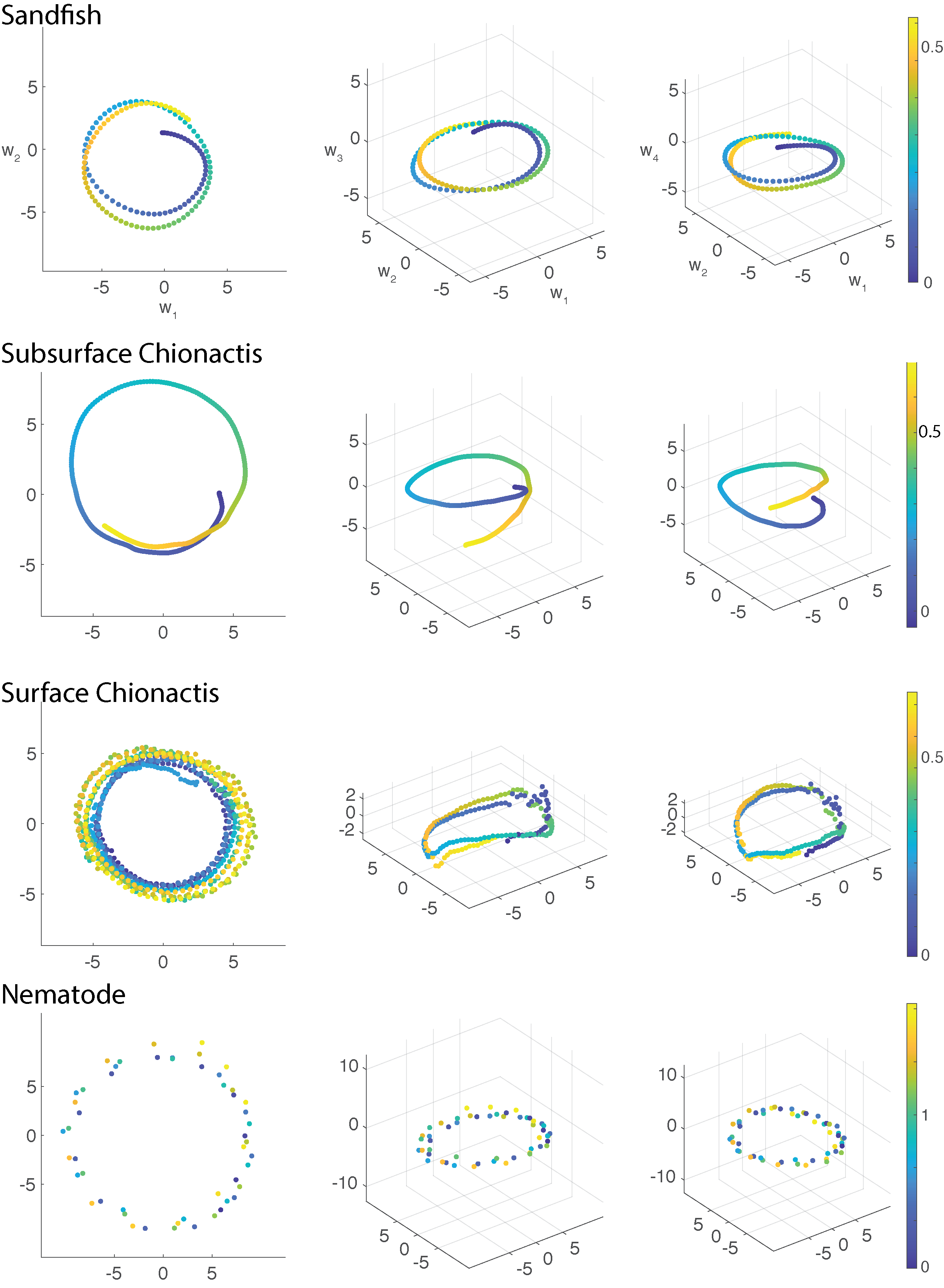}
	\caption{{\bf The trajectory of body undulation kinematics} (\textit{left}) The trajectory of body undulation in 2-dimensional shape space ($w_1$-$w_2$ space) and three-dimensional shape space (\textit{mid}: $w_1$-$w_2$-$w_3$ space; \textit{right}: $w_1$-$w_2$-$w_3$). Time is indicated by color. Color bars have the units of sec. Four species are compared: (from top to bottom) sandfish, subsurface snake, surface snake, and nematodes.}
	\label{Fig:traj}
\end{figure}

\begin{figure}[h]
	\centering
	\includegraphics[width=0.8\textwidth]{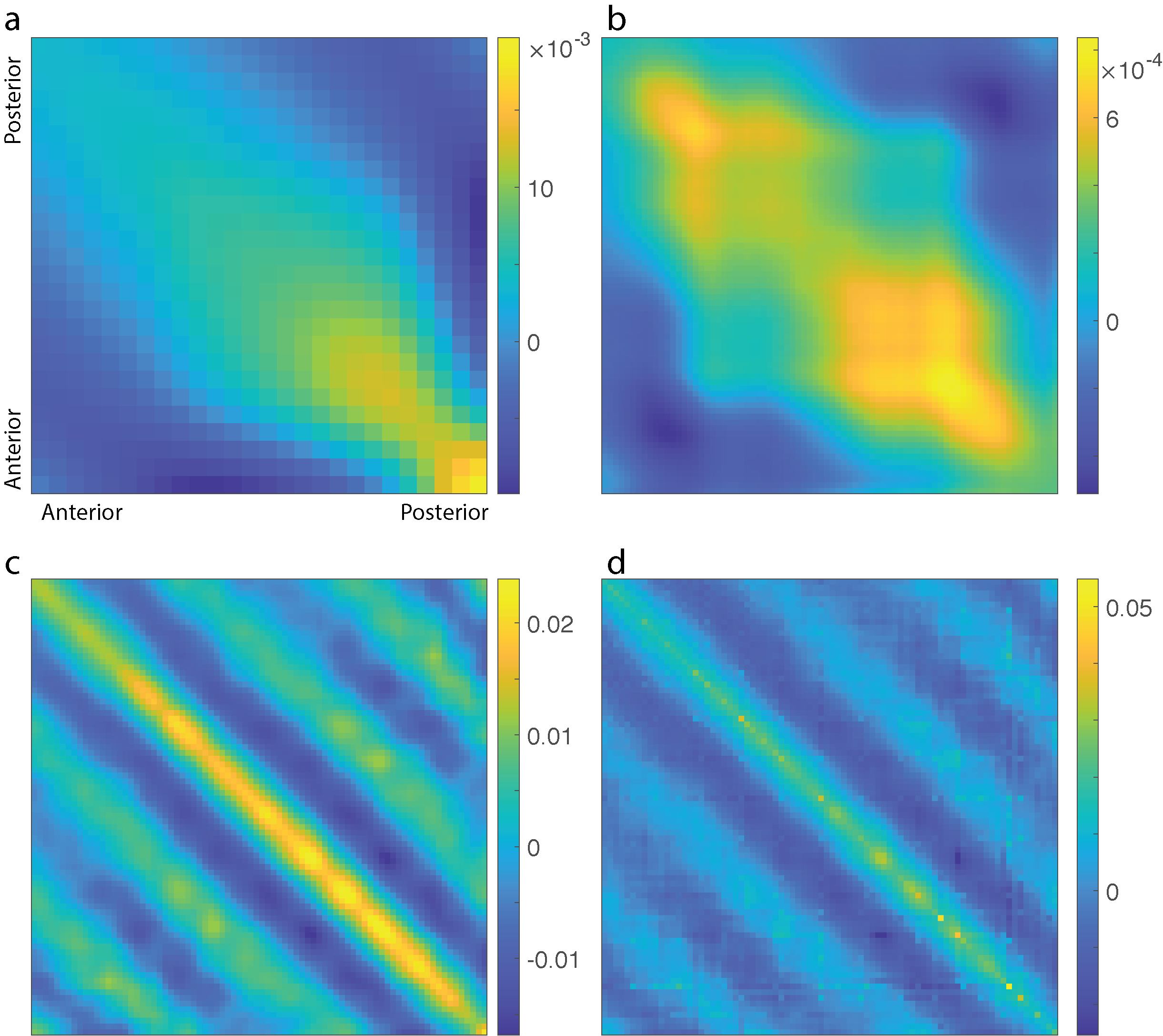}
	\caption{ Covariance Matrix of curvature profiles for (a) nematodes (b) sandfish, (c) subsurface snake, and (d) surface snake. Horizontal axis and vertical axis denote the positions on the body from anterior to posterior. Color bar denotes the covariance matrix fluctuations in the relative curvature $\kappa(t,s)\lambda$. Axes are identical for all panels.
	}
	\label{covmat}
\end{figure}

\begin{figure}[h]
	\centering
	\includegraphics[width=0.8\textwidth]{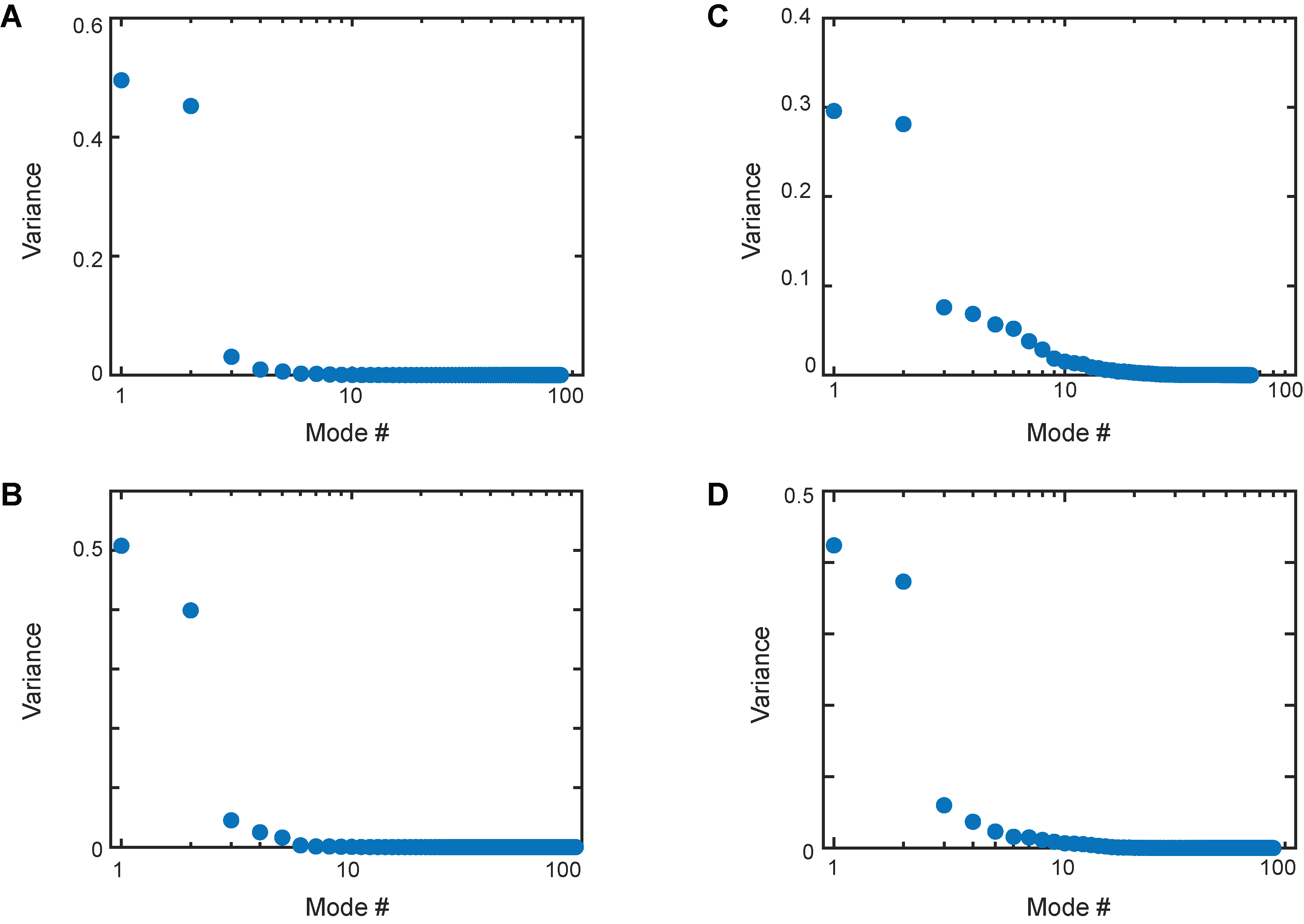}
	\caption{The fraction of the total variance explained by each principal component for (A) the nematode worm (\textit{Caenorhabditis elegans}) in S-basal buffer (B) the sandfish lizard (\textit{Scincus scincus}) $7.6$~cm below the surface of and fully immersed in $300$-$\mu$m glass beads, (C) the shovel-nosed snake (\textit{Chionactis occipitalis}) $7.6$~cm below the surface of and fully immersed in $300$-$\mu$m glass beads, and (D) \textit{Chionactis occipitalis} moving on the surface of $300$-$\mu$m glass beads. The calculation is based on the Matlab function `pca'.}
	\label{modevar}
\end{figure}

\clearpage
\section{Tables}

\begin{table}[h]
\tiny
\centering
\caption{\textit{C. elegans}, \textit{Scincus scincus}, \textit{Chionactis occipitalis} information and modeling parameters.}
\setlength{\tabcolsep}{1mm}
\begin{tabular}{|c|c|c|c|c| } 
\hline
&\textit{Caenorhabditis elegans}& \textit{Scincus scincus}&\textit{Chionactis occipitalis}& \textit{Chionactis occipitalis} \\
\hline
Substrate &Buffer Solution& $300$-$\mu$m glass beads & $300$-$\mu$m glass beads & $300$-$\mu$m glass beads \\

Imaging type; frame rate & Brightfield microscopy; 15 fps & X-Ray; 250 fps  & X-Ray; 250 fps & Visible light; 250 fps \\

RFT relation & viscous fluid; $c_\perp / c_\parallel = 1.5$ & subsurface granular & subsurface granular & surface granular  \\

RFT Penetration depth, $d$ &N/A & $7.6$~cm &$7.6$~cm & $0.8$~cm\footnote{\cite{schiebel2019mitigating} reported that the depth of the track was $d = 0.5$~cm but used $d=0.8$~cm for modeling. Differences in predicted performance were small compared to experimental variation, and the fits to experimental RFT force measurements for $d=0.8$~cm were much better than those for $d=0.5$~mm.} \\

Spatial frequency, $n$ & $0.73 \pm 0.10$  & $1.0 \pm 0.1$ & $3.5 \pm 0.7$ & $1.90 \pm  0.14$  \\

$\lambda_s$ & $1.5 \pm 0.2$~mm & $10.6 \pm 1.2$~cm & $9.1 \pm 1.9$~cm & $20.1 \pm  1.5$~cm \\

Body length, $L$ & $1.1 \pm 0.1$~mm  & $10.6 \pm 0.6$~cm &$31.8 \pm 2.4$~cm & $38.0 \pm  1.3$~cm \\

Body mass, $m$ & N/A  & $17.6\pm2.8$~g & $17.5\pm3.4$~g & $16.4\pm3.4$~g  \\

Forward-ventral skin-grain friction, $\mu$ & N/A & $0.194 \pm 0.022$ & $0.109 \pm 0.016$ & $0.109 \pm 0.016$  \\

Number of individuals, $N_i$ & $7$ & $4$ & $3$ & $9$ \\

Total number of trials, $N_t$ &16 & $29$  & $14$ & $30$ \\

Average body curvatures, $\overline{\kappa_m \lambda_s}$ &$7.9 \pm 1.1$ & $4.3\pm 1.2$;  $7.5 \pm 0.8$ & $4.1\pm 1.7$; $6.0 \pm 1.3$ & $5.1\pm 0.7$\\

Displacement [BL/cycle] & $0.20 \pm 0.06$ &$0.27\pm 0.09$  & $0.16\pm0.04$ & $0.44 \pm 0.03$ \\

Frequency [Hz] & $1.28\pm0.16$ & $2.10 \pm 0.75$  & $1.4 \pm 0.2$ & $3.9 \pm 0.4$ \\

Coasting number & 0.01 & [0.01, 0.1]  & [0.01, 0.1] & [0.3, 3] \\

References &\cite{fang2010biomechanical} & \cite{maladen2009undulatory,sharpe2015locomotor} & \cite{sharpe2015locomotor} & \cite{schiebel2019mitigating}  \\
\hline
\end{tabular}
\label{tab:param}
\end{table}

\begin{table}[h]
\small
\centering
\caption{\textit{Crotalus cerastes} information and modeling parameters.}
\setlength{\tabcolsep}{1mm}
\begin{tabular}{|c|c|c|c| } 
\hline
& \textit{Crotalus cerastes}&\textit{Crotalus cerastes}& \textit{Crotalus cerastes} \\
\hline
Substrate & Yuma sand & Smooth oak board & Roughened oak board \\

Imaging type; frame rate & Infrared; 120 fps  & Infrared; 120 fps & Infrared; 120 fps\\

RFT relation & surface granular & Coulomb friction & Coulomb friction \\

RFT Penetration depth, $d$ & $1.2$~cm & N/A & N/A \\

Spatial frequency, $n$ & $1.50 \pm 0.19$ & $1.51 \pm 0.15$ & $1.53 \pm 0.13$  \\

$\lambda_s$ & $24.4 \pm 2.6$~cm & $24.5 \pm 3.0$~cm & $24.5 \pm 2.9$~cm \\

Tracked body length & $36.2\pm 3.8$~cm & $36.9 \pm 3.7$~cm & $37.2 \pm 3.4$~cm \\

Total body length & $48.6 \pm 6$~cm & $48.6 \pm 6$~cm & $48.6 \pm 6$~cm \\
Body mass, $m$ & $98\pm18$~g & $98\pm18$~g & $98\pm18$~g\\


Number of individuals, $N_i$ & $4$ & $4$ & $4$ \\

Total number of trials, $N_t$ & $18$  & $30$ & $37$\\

Average body curvatures, $\overline{\kappa_m \lambda_s}$ & $8.2\pm 1.8$ &  $8.8 \pm 1.5$ & $8.9\pm 1.4$ \\

Displacement [BL/cycle] & $0.49\pm 0.03$  & $0.58\pm0.05$ & $0.58 \pm 0.05$ \\

Average track angle, $\phi$ & $38.1^\circ \pm 8.1^\circ$ & $29.7^\circ \pm 7.3^\circ$ & $28.6^\circ \pm 7.3^\circ$\\

Frequency [Hz] & $0.64 \pm 0.24$  & $0.67 \pm 0.19$ & $0.46 \pm 0.17$ \\

Coasting number & [0.01, 0.1] & [0.01, 0.1]  & [0.01, 0.1] \\
References & \cite{marvi2014sidewinding,astley2015modulation} & N/A & N/A \\
\hline
\end{tabular}
\label{tab:param2}
\end{table}

\subsection*{Experimental systems and analysis}
\subsubsection*{Curvature measurements}
\label{methods:KL}
Extracted animal mid-lines were interpolated and up-sampled to have $100$ evenly-spaced points along the backbone of the animal. For each trial, curvatures, $\kappa$ were measured along the body and throughout the duration of an experiment. These curvatures throughout time were non-dimensionalized by the approximate arc-length associated with one wave along the body, $\lambda_s$.  Here, average values of $\lambda_s$ were determined by measuring $\lambda_s / 2$, the arc length between two adjacent points of zero curvature.  Average values of $\lambda_s$ were used for each animal. We note that $\lambda_s$ is also given by the length of the animal, $L$, divided by the number of waves along the body, $n$.

\subsubsection*{Principal components analysis}
\label{methods:pca}
Non-dimensional curvatures, $\kappa \lambda_s$, for all individuals and all trials (within each animal species) were then combined into a single matrix. We then used principal components analysis (PCA) to diagonalize the covariance matrix of all curvatures and identify an orthogonal set of basis modes whose weighted superposition can be used to capture observed animal postures. Eigenvalues associated with each PC indicate the variance explained by each mode (and therefore the fraction of the variance explained by each mode is given by the eigenvalue divided by the sum of all eigenvalues. For each animal, we found that more than half of the variance in observed postures can be captured by the two modes with the largest eigenvalues).
We therefore choose the first two modes as our low-dimensional of animal postures.

\subsubsection*{Curvature calculations and comparisons with previous results}
\label{methods:curvComp}
For the subsurface self-propulsion of the sandfish lizard \textit{S. scincus}, average animal gaits are shown for different data sets and analysis procedures. The darker data point, originally reported in ~\cite{sharpe2015locomotor}, represents the average and standard deviation of maximum curvatures manually measured once per cycle, and includes trials originally reported in \cite{maladen2009undulatory} in which animals were not tracked.  The light blue data point only includes newer data in which animal positions were tracked through by identifying the positions of lead markers along the backbone of the animals, and includes a maximum curvature measurement for each moment in time (as shown in Fig.~2A$(iv)$). We expect that including all frames would decrease the average maximum curvature value. Also, we note that for animal kinematics in the more recent data, wires implanted for electromyography (EMG) recordings~\cite{sharpe2013environmental} may have resulted in lower curvatures. Therefore, we expect that these two data points should bound the average time-resolved maximum curvatures.  Within this curvature range, the comparison with the geometric calculations reveals displacements per cycle are quantitatively close to predictions, and that, animals often used local curvatures that result in near-maximal displacements per cycle.

For the subsurface propulsion of \textit{Chionactis occipitalis}, we note that the first two modes only capture about $60\%$ of the variance observed in the animal postures. As a result, when we project the full data set onto the first two modes, we obtain a smaller value for curvatures than was previously-reported. As we increase the number of modes within our description of postures, our measured value approaches the previously-reported value (see supplemental). 

\subsubsection*{Sidewinding track angles}
Previous studies investigating sidewinding locomotion~\cite{jayne1986kinematics,gans1992kinematic,Brain1960} did not report body curvatures, but instead reported track angles. While we did not measure track angles angles at the time of our experiments, we were able to determine track angles from kinematic data and observations reported in previous studies~\cite{astley2015modulation,jayne1986kinematics}. During sidewinding, contact patches coincide with regions of low curvature along the body, and that these contact patches leave behind straight-line shallow troughs or tracks in the sand, oriented at an angle relative to the center of mass motion. We therefore measured track angles in the following way: for each trial, we created a matrix containing curvature along the body and throughout time. We then identified diagonal bands of near-zero curvature that start at the head and are subsequently propagated down the body. The (x,y) points associated with static patches form dense straight lines when combined through time, so we fit these points to a line.  We also fit the overall center of mass movement to a line, and the track angle is given by the angle between these two lines.

\subsection*{RFT modifications for surface movement}
\subsubsection*{Estimating coasting number}

\paragraph{Viscous Stokes drag}
Let $\gamma$ be the viscosity in Stokes drag such that $F=\gamma v$, where $F$ is the reaction force and $v$ is the velocity. Let $v_0$ be the steady-state velocity. Once the locomotor stops self-deforming, its velocity will decrease exponentially: $\dot{v} = -\frac{\gamma}{m} v$. Then we have $v(t) = v_0e^{-\frac{\gamma}{m}t}$. The half-life decay time can be approximated as $\tau_{coast}= \eta m/\gamma$. where $\eta = -\log(0.5)$. The coasting distance is $v_0m/\gamma$.

Typically, an adult \textit{C. elegans} has a diameter of about 50 $\mu m$ and about 1 $mm$ long~\cite{andrews2019new}. The density of an adult \textit{C. elegans} is approximately $10^3\times kg/m^3$~\cite{reina2013shifts}. Thus the body weight for a typical \textit{C. elegans} is approximately $2\times 10^{-9} kg$.

From Stoke's law, $\gamma = 6\pi \mu R$, where $R$ is the radius of the spherical object and $\mu$ is the dynamic viscosity. In buffer, $\mu\approx10^{-3} kg/(m*s)$~\cite{sznitman2010effects}. Approximating $R$ by the radius of \textit{C. elegans}, $\gamma$ is approximately $4\times10^{-7} kg/s$. In this way, the coasting time for buffer is approximately $5$ ms. Notably, for nematode worm swimming in buffer, the Reynolds number is approximately 0.4~\cite{sznitman2010propulsive}.

\paragraph{Friction}
Let $\mu$ and $g$ be the friction coefficient and gravitational acceleration constants respectively. Let $v_0$ be the steady-state velocity. Once the locomotor stops self-deforming, its velocity will decrease linearly: $\dot{v} = -\mu g$. Then we have $v(t) = v_0-\mu g t$. The mean lifetime is then $\tau_{coast} = v_0/(2\mu g)$. The coasting distance is $0.5 v_0^2/(\mu g)$.

\subsubsection*{Continuous substrate contact}
\label{methods:cc}
To model movement at the surface, we included two contributions to the total infinitesimal force acting on each segment: the resistive granular forces that result from the snake's lateral surfaces pushing on the sides of the track~\cite{schiebel2019mitigating} (see Fig.~3A for experimentally-measured environmental stress relations for a partially-submerged intruder) and a kinetic Coulomb friction term (that opposes the local segment velocity) from the interaction of the snake's ventral surface with the sand. Note that in principle, the resistive granular forces is speed-dependent~\cite{schiebel2019mitigating}. However, in the speed range discussed in this manuscript, the speed dependence is weak and thus we model the magnitude of resistive granular forces as independent from speed. With these forces, we use our geometric phase approach to create the height function (Fig.~5C$(i)$) and integrate over circular gaits to predict forward displacements for surface movement as a function of maximum curvature along the body. \cite{schiebel2019mitigating} found that the inclusion of a kinetic Coulomb friction term did not significantly affect RFT predictions.

\subsubsection*{Changing substrate contact}
\label{methods:chc}
As with \textit{Ch. occipitalis}, we included both surface granular resistive forces and a kinetic Coulomb friction term to the net force acting on each infinitesimal segment. Given the varying environmental contact, we assigned a weighted local normal force to infinitesimal segment to ensure that the total normal force (and therefore the snake's weight) were constant throughout all calculations. A direct simulation that included these forces and solved Eq.~5 for undulatory locomotion revealed that sidewinding behavior results from this contact-dependent force balance. Given that we do not know the coefficient of friction between the \emph{Cr. cerastes} and sand, we used a range of values, see supplemental. 
To reduce clutter, we only show the curve for $\mu=0.2$.

\subsubsection*{Sidewinding shape parameterization}
 We characterized the animals’ postures using a two-step process: first, in-plane local curvatures were derived from horizontal kinematic data for movement on a $7.6$-cm-deep sandy substrate (originally reported in~\cite{astley2015modulation}); and second, the vertical contact state of each posture was set by introducing a continuously-varying curvature-dependent periodic function. PCA applied to the full data set combining in-plane curvatures throughout time for all trials revealed that nearly $80\%$ of the variation in the postural data is captured within the space spanned by the first two PCs (Fig.~8A). Projecting the full data set onto the two-dimensional space spanned by these PCs reveals that a traveling wave of lateral curvatures (passed from head to tail down the body) provides a low-dimensional description of body shape changes (Fig.~8B). Finally, we use the phase-shifted vertical wave identified in~\cite{astley2015modulation} to define an environmental contact function $c \in [0,1]$, where a body segment with $c = 1$ is in full contact with the substrate and $c=0$ corresponds to a lifted body segment. Fig.~7B depicts, for one posture, the relationship of the contact pattern with vertical and lateral waves, and Fig.~8C shows how the body-environment contact varies with throughout the two-dimensional postural space.

\subsection*{Geometric formulation}
\label{methods:form}

\subsubsection*{Swimmer geometry}\label{methods:geom}

The swimmer geometry is defined by the curvature of its backbone as a function of arclength along the body, $\kappa(s)$. For the analysis in this paper, we defined this curvature as the superposition of two sinusoidal waves,

\begin{eqnarray}
{v_1}(s) = \sin{(2\pi s/L)}, \\
{v_2}(s) = \cos{(2\pi s/L)},
\label{eq:B}
\end{eqnarray}

\noindent weighted by coefficients

\begin{equation}
    \boldsymbol{\alpha} = \begin{bmatrix} w_{1} \\ w_{2} \end{bmatrix}
\end{equation}

\noindent such that the total curvature of the body at any shape $\boldsymbol{\alpha}$ is

\begin{equation}
    \kappa(s,\boldsymbol{\alpha}) = \begin{bmatrix} v_{1}(s) & v_{2}(s) \end{bmatrix}\begin{bmatrix} w_{1} \\ w_{2} \end{bmatrix}
\end{equation}

\noindent Integrating this curvature along the body as

\begin{equation}
\begin{split}
    h(s,\boldsymbol{\alpha}) = & \begin{bmatrix} x \\ y\\ \theta \end{bmatrix}(s,\boldsymbol{\alpha}) \\ = &  \int_{0}^{s} \begin{bmatrix} \cos{\theta(\sigma,\boldsymbol{\alpha})} & -\sin {\theta(\sigma,\boldsymbol{\alpha})} & 0 \\  \sin{\theta(\sigma,\boldsymbol{\alpha})} & \phantom{-}\cos{\theta(\sigma,\boldsymbol{\alpha})} & 0 \\ 0 & 0 & 1 \end{bmatrix} \begin{bmatrix} 1 \\ 0 \\ \kappa(\sigma,\boldsymbol{\alpha}) \end{bmatrix} d\sigma
    \end{split}
\end{equation}

\noindent (which takes the curvature as the rate of change of the body tangent orientation, and rotates the unit flow along the body by the amount the tangent has rotated at each point $s$) provides a locus of points and tangent orientations along the body, relative to the position of the $s=0$ frame.

\subsubsection*{Swimmer movement and environmental forces}
\label{methods:movement}
Once the locus of body points has been found, the local body velocity\footnote{Sometimes referred to as $\xi$ in other references.}, $\fibercirc$ (in tangential/normal/rotational directions) of each point on the backbone for a given shape, shape velocity and overall system body velocity can be calculated as

\begin{equation}
    \fibercirc(\fibercircnot,s,\boldsymbol{\alpha},\boldsymbol{\dot{\alpha}}) =  \overbrace{\begin{bmatrix} Ad_{h(s,\boldsymbol{\alpha})}^{-1} &  J^{b}_{\text{internal}}(s,\boldsymbol{\alpha})\end{bmatrix}}^{J^{b}_{\text{full}}} \begin{bmatrix} \fibercircnot \\ \boldsymbol{\dot{\alpha}} \end{bmatrix} ,
\end{equation}

\noindent where the \emph{adjoint-inverse} matrix

\begin{equation}
    Ad_{h}^{-1} = \begin{bmatrix} \phantom{-}\cos{\theta} & \sin {\theta} & x\sin\theta-y\cos\theta \\  -\sin{\theta} & \cos{\theta} & x\cos\theta+y\sin\theta \\ 0 & 0 & 1 \end{bmatrix}
\end{equation}

\noindent encodes the cross-products and rotation matrices needed to map the body velocity of the body frame to the body velocity of any other point on the body when the system is rigid (not changing shape), and

\begin{equation}
    \begin{split}
J^{b}_{\text{internal}}(s,\boldsymbol{\alpha})  =  \int_{0}^{s}  & Ad_{h(\sigma,\boldsymbol{\alpha})}^{-1} \begin{bmatrix} 0 & 0 \\ 0 & 0 \\ \frac{\partial\kappa(\sigma,\boldsymbol{\alpha})}{\partial w_{1}} & \frac{\partial\kappa(\sigma,\boldsymbol{\alpha})}{\partial w_{2}} \end{bmatrix} d\sigma \\
     =  \int_{0}^{s}  & Ad_{h(\sigma,\boldsymbol{\alpha})}^{-1} \begin{bmatrix} 0 & 0 \\ 0 & 0 \\ v_{1}(\sigma) & v_{2}(\sigma) \end{bmatrix} d\sigma
    \end{split}
\end{equation}

\noindent maps the rotational velocity contributed at $\sigma$ by bending each shape mode to translational and rotational velocity of all points $s$ that are distal to that contribution (further along the backbone).

The forces acting on each segment are found from our resistive force model,

\begin{equation}
\begin{split}
    F_{\text{local}}(\fibercircnot,s,\boldsymbol{\alpha},\boldsymbol{\dot{\alpha}})  & = \begin{bmatrix} F_{\parallel}  \\ F_{\perp} \\ F_{\circlearrowleft} \end{bmatrix}_{\text{local}}\hspace{-1em}(\fibercircnot,s,\boldsymbol{\alpha},\boldsymbol{\dot{\alpha}}) \\ 
    & =  F_{\text{local}} \bigl( \fibercirc(\fibercircnot,s,\boldsymbol{\alpha},\boldsymbol{\dot{\alpha}})\bigr),
    \end{split}
\end{equation}

\noindent and then mapped to net forces and moments acting on the system at the $s=0$ frame by the transpose of the adjoint-inverse matrix,

\begin{equation}
    F_{\text{total}}(\fibercircnot,\boldsymbol{\alpha},\boldsymbol{\dot{\alpha}}) = \int_{0}^{s_{\text{max}}} Ad_{h(s,\boldsymbol{\alpha})}^{-T} F_{\text{resistive}}(s,\boldsymbol{\alpha},\boldsymbol{\dot{\alpha}})\, ds
\end{equation}

\noindent which encodes the cross-products and rotation matrices needed to move forces between frames that are physically attached to each other.

Finally, we generate a map from shape and shape velocity to body velocity by imposing the quasi-static equilibrium approximation that the drag forces dominate any acceleration forces felt by the system and solving for the $\fibercircnot$ system body velocity that minimizes the net drag force on the body for a given (shape, shape velocity) pair $(\boldsymbol{\alpha},\boldsymbol{\dot{\alpha}})$. We then approximate a local connection matrix $\mixedconn(\boldsymbol{\alpha})$ for the system by taking a linear regression of $\fibercircnot$ against $\boldsymbol{\dot{\alpha}}$ at each $\boldsymbol{\alpha}$, and populating the coefficients of $\mixedconn$ with the slopes of the regression. 

\subsubsection*{Minimum perturbation coordinates}\label{methods:mpc}
The accuracy of approximating the net displacement via the integral of constraint curvature depends on the choice of body frame used for the system (e.g., placing the body frame at the center of mass and aligning it with a principle axis, vs attaching it to the head of the system)~\cite{hatton2011geometric}. The errors in the approximation appear because noncommutative interactions between translation and rotation are nonlinear, and the Lie bracket only captures these interactions to second order.

The residual errors in the noncommutative terms scale 
with the amount of intermediate rotation that the body frame experiences during the gait. This intermediate rotation in turn scales with the amplitude of the gait and the magnitude of A, so the integral is most accurate
so the integral is most accurate if we select a body frame constructed to minimize this magnitude~\cite{hatton2015nonconservativity}. Constructing this body frame is directly equivalent to constructing the Coulomb gauge for an electromagnetic system, and, as in the case of a Coulomb gauge, this body frame can be found by first finding the local connection $\mixedconn$ in any frame convenient for calculation, and then applying a Hodge-Helmholtz decomposition to it~\cite{hatton2015nonconservativity}. For snake-like systems, the locations of such frames are similar to those of center-of-mass frames, but are defined by the drag forces acting on the system, rather than its mass distribution.

Notably, he accuracy of the approximation presented in this manuscript is optimal when considering motion in a single direction (e.g., forward motion in Fig. 5, in-place turning motion in Fig. 6, or sidewinding motion in Fig. 9). In cases where there are mixed-direction motions (e.g., differential turn in Fig. 10), the parallel parking effect can be significant. We acknowledge that to enhance the accuracy of the approximation, it is desirable to express velocities and positions in exponential terms. For detailed discussions, we refer readers to~\cite{bass2022characterizing}.

Briefly, solving Eq.~3, we have:

\begin{align}
    \begin{bmatrix}
    \Delta x (T) \\
    \Delta y (T) \\
    \Delta \theta (T) \\
    \end{bmatrix} & = \int_0^T \begin{bmatrix} 
    \cos{\big(\theta(t)\big)} & -\sin{\big(\theta(t)\big)} & 0 \\
    \sin{\big(\theta(t)\big)} & \cos{\big(\theta(t)\big)} & 0 \\
    0 & 0 & 1 \\
    \end{bmatrix}\fibercirc(t) dt
    \label{SIeq:fullRFT}
\end{align}

Let $\fibercircave$ be the average velocity over a cycle:

\begin{equation}
    \fibercircave =     \begin{bmatrix}
    \Tilde{v}_x \\
    \Tilde{v}_y \\
    \Tilde{v}_\theta \\
    \end{bmatrix}  = \frac{1}{T} \int_0^T \fibercirc(t) dt.
\end{equation}

We can simplify Eq.~\ref{SIeq:fullRFT} as:

\begin{equation}
    \begin{bmatrix}
    \Delta x (T) \\
    \Delta y (T) \\
    \Delta \theta (T) \\
    \end{bmatrix} \approx  
    \begin{bmatrix}
    \Delta \Tilde{x} (T) \\
    \Delta \Tilde{y} (T)\\
    \Delta \Tilde{\theta} (T) \\
    \end{bmatrix}= \int_0^T \begin{bmatrix} 
    \cos{\big(\theta(t)\big)} & -\sin{\big(\theta(t)\big)} & 0 \\
    \sin{\big(\theta(t)\big)} & \cos{\big(\theta(t)\big)} & 0 \\
    0 & 0 & 1 \\
    \end{bmatrix}\fibercircave dt    
    \label{SIeq:approx}
\end{equation}

Notably, Eq.~\ref{SIeq:approx} has an analytical solution~\cite{bass2022characterizing}:

\begin{equation}
    \begin{bmatrix}
    \Delta \Tilde{x} (T) \\
    \Delta \Tilde{y} (T)\\
    \Delta \Tilde{\theta} (T) \\
    \end{bmatrix} =  e^{\fibercircave T}, \label{SIeq:fullEq}
\end{equation}

\noindent where 

\begin{equation}
    e^{\fibercircave} = \frac{1}{\Tilde{v}_\theta } \begin{bmatrix} 
    \sin{(\Tilde{v}_\theta)} & \cos{(\Tilde{v}_\theta)}-1 & 0 \\
    1-\cos{(\Tilde{v}_\theta)} & \sin{(\Tilde{v}_\theta)} & 0 \\
    0 & 0 & \Tilde{v}_\theta \\
    \end{bmatrix} \begin{bmatrix}
    \Tilde{v}_x \\
    \Tilde{v}_y \\
    \Tilde{v}_\theta \\
    \end{bmatrix} \nonumber
\end{equation}

Note that the third row in Eq.~\ref{SIeq:fullEq} can a simple form: $\Delta \Tilde{\theta} (T) = \Tilde{v}_\theta T$, which justifies our analysis on turning behaviors (in-place worm omega turn in Fig. 6 and differential turn in Fig. 10). Further, in the case where $\Tilde{v}_\theta \rightarrow 0$, we have:

\begin{equation}
\lim_{\Tilde{v}_\theta \rightarrow 0}\Bigg(\frac{1}{\Tilde{v}_\theta } \begin{bmatrix} 
    \sin{(\Tilde{v}_\theta T)} & \cos{(\Tilde{v}_\theta T)}-1 & 0 \\
    1-\cos{(\Tilde{v}_\theta T)} & \sin{(\Tilde{v}_\theta T)} & 0 \\
    0 & 0 & \Tilde{v}_\theta \\
    \end{bmatrix}\Bigg) = I_3,
\end{equation}

\noindent where $I_3$ is a $3\times3$ identity matrix. In this case, Eq.~\ref{SIeq:fullEq} reduces to:

\begin{equation}
    \begin{bmatrix}
    \Delta \Tilde{x} (T) \\
    \Delta \Tilde{y} (T)\\
    \Delta \Tilde{\theta} (T) \\
    \end{bmatrix} = \fibercircave T \label{SIeq:forward}
\end{equation}

Notably, Eq.~\ref{SIeq:forward} is applicable to our analysis to forward motion in Fig. 5 and sidewinding motion in Fig. 9.

\subsubsection*{Numerical implementation}
All numerical calculations were performed in Matlab.  The built-in function fsolve was used to determine the swimming speed that balanced net forces and torques for perturbations to swimmer shape. The initial condition for each calculation was set to be the solution for the same configuration and perturbation in viscous fluid. Default tolerances were used for each calculation, and each solution was only used for subsequent calculations (including plane fits) if the exit flag returned was equal to $1$ (indicating that a solution was found and first-order optimality is small). Note that in the very rare case (only 10 cases in over 1800 numerical calculations) where there is no solution for force and torque balance under nonlinear RFT (e.g., Coulomb friction or surface granular force equations), we approximate the local connections using the linear viscous friction. Note that in straight sidewinding, the head, similar to the rest of body, will make and break contact with substrate. In this way, we extend our measured PCA body wave over the entire body. For differential turn, since the head is always off the substrate, we only analyzed the body wave dynamics (not extend the wave over the entire body) where there is active making and breaking contact.

Note that in numerical calculations, we simutaneously obtained the forward, lateral, and rotational height functions (correspond to $\xi_x, \xi_y, \xi_\theta$ respectively). We choose the height function of interest in different section to further our analysis.


\begin{thebibliography}{97}%
	\makeatletter
	\providecommand \@ifxundefined [1]{%
		\@ifx{#1\undefined}
	}%
	\providecommand \@ifnum [1]{%
		\ifnum #1\expandafter \@firstoftwo
		\else \expandafter \@secondoftwo
		\fi
	}%
	\providecommand \@ifx [1]{%
		\ifx #1\expandafter \@firstoftwo
		\else \expandafter \@secondoftwo
		\fi
	}%
	\providecommand \natexlab [1]{#1}%
	\providecommand \enquote  [1]{``#1''}%
	\providecommand \bibnamefont  [1]{#1}%
	\providecommand \bibfnamefont [1]{#1}%
	\providecommand \citenamefont [1]{#1}%
	\providecommand \href@noop [0]{\@secondoftwo}%
	\providecommand \href [0]{\begingroup \@sanitize@url \@href}%
	\providecommand \@href[1]{\@@startlink{#1}\@@href}%
	\providecommand \@@href[1]{\endgroup#1\@@endlink}%
	\providecommand \@sanitize@url [0]{\catcode `\\12\catcode `\$12\catcode
		`\&12\catcode `\#12\catcode `\^12\catcode `\_12\catcode `\%12\relax}%
	\providecommand \@@startlink[1]{}%
	\providecommand \@@endlink[0]{}%
	\providecommand \url  [0]{\begingroup\@sanitize@url \@url }%
	\providecommand \@url [1]{\endgroup\@href {#1}{\urlprefix }}%
	\providecommand \urlprefix  [0]{URL }%
	\providecommand \Eprint [0]{\href }%
	\providecommand \doibase [0]{https://doi.org/}%
	\providecommand \selectlanguage [0]{\@gobble}%
	\providecommand \bibinfo  [0]{\@secondoftwo}%
	\providecommand \bibfield  [0]{\@secondoftwo}%
	\providecommand \translation [1]{[#1]}%
	\providecommand \BibitemOpen [0]{}%
	\providecommand \bibitemStop [0]{}%
	\providecommand \bibitemNoStop [0]{.\EOS\space}%
	\providecommand \EOS [0]{\spacefactor3000\relax}%
	\providecommand \BibitemShut  [1]{\csname bibitem#1\endcsname}%
	\let\auto@bib@innerbib\@empty
	\bibitem [{\citenamefont {Alexander}(2003)}]{alexander2003principles}%
	\BibitemOpen
	\bibfield  {author} {\bibinfo {author} {\bibfnamefont {R.~M.}\ \bibnamefont
			{Alexander}},\ }\href@noop {} {\emph {\bibinfo {title} {Principles of animal
				locomotion}}}\ (\bibinfo  {publisher} {Princeton University Press},\ \bibinfo
	{year} {2003})\BibitemShut {NoStop}%
	\bibitem [{\citenamefont {Childress}\ \emph {et~al.}(2012)\citenamefont
		{Childress}, \citenamefont {Hosoi}, \citenamefont {Schultz},\ and\
		\citenamefont {Wang}}]{childress2012natural}%
	\BibitemOpen
	\bibfield  {author} {\bibinfo {author} {\bibfnamefont {S.}~\bibnamefont
			{Childress}}, \bibinfo {author} {\bibfnamefont {A.}~\bibnamefont {Hosoi}},
		\bibinfo {author} {\bibfnamefont {W.~W.}\ \bibnamefont {Schultz}},\ and\
		\bibinfo {author} {\bibfnamefont {J.}~\bibnamefont {Wang}},\ }\href@noop {}
	{\emph {\bibinfo {title} {Natural locomotion in fluids and on surfaces:
				swimming, flying, and sliding}}},\ Vol.\ \bibinfo {volume} {155}\ (\bibinfo
	{publisher} {Springer},\ \bibinfo {year} {2012})\BibitemShut {NoStop}%
	\bibitem [{\citenamefont {Gravish}\ and\ \citenamefont
		{Lauder}(2018)}]{gravish2018robotics}%
	\BibitemOpen
	\bibfield  {author} {\bibinfo {author} {\bibfnamefont {N.}~\bibnamefont
			{Gravish}}\ and\ \bibinfo {author} {\bibfnamefont {G.~V.}\ \bibnamefont
			{Lauder}},\ }\bibfield  {title} {\bibinfo {title} {Robotics-inspired
			biology},\ }\href@noop {} {\bibfield  {journal} {\bibinfo  {journal} {Journal
				of Experimental Biology}\ }\textbf {\bibinfo {volume} {221}},\ \bibinfo
		{pages} {jeb138438} (\bibinfo {year} {2018})}\BibitemShut {NoStop}%
	\bibitem [{\citenamefont {Kim}\ \emph {et~al.}(2013)\citenamefont {Kim},
		\citenamefont {Laschi},\ and\ \citenamefont {Trimmer}}]{kim2013soft}%
	\BibitemOpen
	\bibfield  {author} {\bibinfo {author} {\bibfnamefont {S.}~\bibnamefont
			{Kim}}, \bibinfo {author} {\bibfnamefont {C.}~\bibnamefont {Laschi}},\ and\
		\bibinfo {author} {\bibfnamefont {B.}~\bibnamefont {Trimmer}},\ }\bibfield
	{title} {\bibinfo {title} {Soft robotics: a bioinspired evolution in
			robotics},\ }\href@noop {} {\bibfield  {journal} {\bibinfo  {journal} {Trends
				in biotechnology}\ }\textbf {\bibinfo {volume} {31}},\ \bibinfo {pages} {287}
		(\bibinfo {year} {2013})}\BibitemShut {NoStop}%
	\bibitem [{\citenamefont {Aguilar}\ \emph {et~al.}(2016)\citenamefont
		{Aguilar}, \citenamefont {Zhang}, \citenamefont {Qian}, \citenamefont
		{Kingsbury}, \citenamefont {McInroe}, \citenamefont {Mazouchova},
		\citenamefont {Li}, \citenamefont {Maladen}, \citenamefont {Gong},
		\citenamefont {Travers}, \citenamefont {Hatton}, \citenamefont {Choset},
		\citenamefont {Umbanhowar},\ and\ \citenamefont {Goldman}}]{Aguilar:2016bq}%
	\BibitemOpen
	\bibfield  {author} {\bibinfo {author} {\bibfnamefont {J.}~\bibnamefont
			{Aguilar}}, \bibinfo {author} {\bibfnamefont {T.}~\bibnamefont {Zhang}},
		\bibinfo {author} {\bibfnamefont {F.}~\bibnamefont {Qian}}, \bibinfo {author}
		{\bibfnamefont {M.}~\bibnamefont {Kingsbury}}, \bibinfo {author}
		{\bibfnamefont {B.}~\bibnamefont {McInroe}}, \bibinfo {author} {\bibfnamefont
			{N.}~\bibnamefont {Mazouchova}}, \bibinfo {author} {\bibfnamefont
			{C.}~\bibnamefont {Li}}, \bibinfo {author} {\bibfnamefont {R.}~\bibnamefont
			{Maladen}}, \bibinfo {author} {\bibfnamefont {C.}~\bibnamefont {Gong}},
		\bibinfo {author} {\bibfnamefont {M.}~\bibnamefont {Travers}}, \bibinfo
		{author} {\bibfnamefont {R.~L.}\ \bibnamefont {Hatton}}, \bibinfo {author}
		{\bibfnamefont {H.}~\bibnamefont {Choset}}, \bibinfo {author} {\bibfnamefont
			{P.~B.}\ \bibnamefont {Umbanhowar}},\ and\ \bibinfo {author} {\bibfnamefont
			{D.~I.}\ \bibnamefont {Goldman}},\ }\bibfield  {title} {\bibinfo {title} {{A
				review on locomotion robophysics: the study of movement at the intersection
				of robotics, soft matter and dynamical systems}},\ }\href@noop {} {\bibfield
		{journal} {\bibinfo  {journal} {Reports on Progress in Physics}\ }\textbf
		{\bibinfo {volume} {79}},\ \bibinfo {pages} {1} (\bibinfo {year}
		{2016})}\BibitemShut {NoStop}%
	\bibitem [{\citenamefont {Grillner}(1985)}]{grillner1985neurobiological}%
	\BibitemOpen
	\bibfield  {author} {\bibinfo {author} {\bibfnamefont {S.}~\bibnamefont
			{Grillner}},\ }\bibfield  {title} {\bibinfo {title} {Neurobiological bases of
			rhythmic motor acts in vertebrates},\ }\href@noop {} {\bibfield  {journal}
		{\bibinfo  {journal} {Science}\ }\textbf {\bibinfo {volume} {228}},\ \bibinfo
		{pages} {143} (\bibinfo {year} {1985})}\BibitemShut {NoStop}%
	\bibitem [{\citenamefont {Loeb}(2012)}]{loeb2012optimal}%
	\BibitemOpen
	\bibfield  {author} {\bibinfo {author} {\bibfnamefont {G.~E.}\ \bibnamefont
			{Loeb}},\ }\bibfield  {title} {\bibinfo {title} {Optimal isn’t good
			enough},\ }\href@noop {} {\bibfield  {journal} {\bibinfo  {journal}
			{Biological cybernetics}\ }\textbf {\bibinfo {volume} {106}},\ \bibinfo
		{pages} {757} (\bibinfo {year} {2012})}\BibitemShut {NoStop}%
	\bibitem [{\citenamefont {Holmes}\ \emph {et~al.}(2006)\citenamefont {Holmes},
		\citenamefont {Full}, \citenamefont {Koditschek},\ and\ \citenamefont
		{Guckenheimer}}]{Holmes:2006ku}%
	\BibitemOpen
	\bibfield  {author} {\bibinfo {author} {\bibfnamefont {P.}~\bibnamefont
			{Holmes}}, \bibinfo {author} {\bibfnamefont {R.~J.}\ \bibnamefont {Full}},
		\bibinfo {author} {\bibfnamefont {D.}~\bibnamefont {Koditschek}},\ and\
		\bibinfo {author} {\bibfnamefont {J.}~\bibnamefont {Guckenheimer}},\
	}\bibfield  {title} {\bibinfo {title} {{The Dynamics of Legged Locomotion:
				Models, Analyses, and Challenges}},\ }\href@noop {} {\bibfield  {journal}
		{\bibinfo  {journal} {SIAM Review}\ }\textbf {\bibinfo {volume} {48}},\
		\bibinfo {pages} {207} (\bibinfo {year} {2006})}\BibitemShut {NoStop}%
	\bibitem [{\citenamefont {Schilling}\ \emph {et~al.}(2013)\citenamefont
		{Schilling}, \citenamefont {Hoinville}, \citenamefont {Schmitz},\ and\
		\citenamefont {Cruse}}]{schilling2013walknet}%
	\BibitemOpen
	\bibfield  {author} {\bibinfo {author} {\bibfnamefont {M.}~\bibnamefont
			{Schilling}}, \bibinfo {author} {\bibfnamefont {T.}~\bibnamefont
			{Hoinville}}, \bibinfo {author} {\bibfnamefont {J.}~\bibnamefont {Schmitz}},\
		and\ \bibinfo {author} {\bibfnamefont {H.}~\bibnamefont {Cruse}},\ }\bibfield
	{title} {\bibinfo {title} {Walknet, a bio-inspired controller for hexapod
			walking},\ }\href@noop {} {\bibfield  {journal} {\bibinfo  {journal}
			{Biological cybernetics}\ }\textbf {\bibinfo {volume} {107}},\ \bibinfo
		{pages} {397} (\bibinfo {year} {2013})}\BibitemShut {NoStop}%
	\bibitem [{\citenamefont {Boyle}\ \emph {et~al.}(2012)\citenamefont {Boyle},
		\citenamefont {Berri},\ and\ \citenamefont {Cohen}}]{boyle2012gait}%
	\BibitemOpen
	\bibfield  {author} {\bibinfo {author} {\bibfnamefont {J.~H.}\ \bibnamefont
			{Boyle}}, \bibinfo {author} {\bibfnamefont {S.}~\bibnamefont {Berri}},\ and\
		\bibinfo {author} {\bibfnamefont {N.}~\bibnamefont {Cohen}},\ }\bibfield
	{title} {\bibinfo {title} {Gait modulation in c. elegans: an integrated
			neuromechanical model},\ }\href@noop {} {\bibfield  {journal} {\bibinfo
			{journal} {Frontiers in computational neuroscience}\ }\textbf {\bibinfo
			{volume} {6}},\ \bibinfo {pages} {10} (\bibinfo {year} {2012})}\BibitemShut
	{NoStop}%
	\bibitem [{\citenamefont {Johnson}\ \emph {et~al.}(2020)\citenamefont
		{Johnson}, \citenamefont {Lewis},\ and\ \citenamefont
		{Guy}}]{johnson2020neuromechanical}%
	\BibitemOpen
	\bibfield  {author} {\bibinfo {author} {\bibfnamefont {C.~L.}\ \bibnamefont
			{Johnson}}, \bibinfo {author} {\bibfnamefont {T.~J.}\ \bibnamefont {Lewis}},\
		and\ \bibinfo {author} {\bibfnamefont {R.~D.}\ \bibnamefont {Guy}},\
	}\bibfield  {title} {\bibinfo {title} {Neuromechanical mechanisms of gait
			adaptation in \textit{C. Elegans}: Relative roles of neural and mechanical
			coupling},\ }\href@noop {} {\bibfield  {journal} {\bibinfo  {journal} {arXiv
				preprint arXiv:2006.10122}\ }\textbf {\bibinfo {volume} {XX}} (\bibinfo
		{year} {2020})}\BibitemShut {NoStop}%
	\bibitem [{\citenamefont {McMillen}\ \emph {et~al.}(2008)\citenamefont
		{McMillen}, \citenamefont {Williams},\ and\ \citenamefont
		{Holmes}}]{mcmillen2008nonlinear}%
	\BibitemOpen
	\bibfield  {author} {\bibinfo {author} {\bibfnamefont {T.}~\bibnamefont
			{McMillen}}, \bibinfo {author} {\bibfnamefont {T.}~\bibnamefont {Williams}},\
		and\ \bibinfo {author} {\bibfnamefont {P.}~\bibnamefont {Holmes}},\
	}\bibfield  {title} {\bibinfo {title} {Nonlinear muscles, passive
			viscoelasticity and body taper conspire to create neuromechanical phase lags
			in anguilliform swimmers},\ }\href@noop {} {\bibfield  {journal} {\bibinfo
			{journal} {PLoS computational biology}\ }\textbf {\bibinfo {volume} {4}},\
		\bibinfo {pages} {e1000157} (\bibinfo {year} {2008})}\BibitemShut {NoStop}%
	\bibitem [{\citenamefont {Full}\ and\ \citenamefont
		{Koditschek}(1999)}]{full1999templates}%
	\BibitemOpen
	\bibfield  {author} {\bibinfo {author} {\bibfnamefont {R.~J.}\ \bibnamefont
			{Full}}\ and\ \bibinfo {author} {\bibfnamefont {D.~E.}\ \bibnamefont
			{Koditschek}},\ }\bibfield  {title} {\bibinfo {title} {Templates and anchors:
			neuromechanical hypotheses of legged locomotion on land},\ }\href@noop {}
	{\bibfield  {journal} {\bibinfo  {journal} {Journal of Experimental Biology}\
		}\textbf {\bibinfo {volume} {202}},\ \bibinfo {pages} {3325} (\bibinfo {year}
		{1999})}\BibitemShut {NoStop}%
	\bibitem [{\citenamefont {Roth}\ \emph {et~al.}(2014)\citenamefont {Roth},
		\citenamefont {Sponberg},\ and\ \citenamefont {Cowan}}]{roth2014comparative}%
	\BibitemOpen
	\bibfield  {author} {\bibinfo {author} {\bibfnamefont {E.}~\bibnamefont
			{Roth}}, \bibinfo {author} {\bibfnamefont {S.}~\bibnamefont {Sponberg}},\
		and\ \bibinfo {author} {\bibfnamefont {N.}~\bibnamefont {Cowan}},\ }\bibfield
	{title} {\bibinfo {title} {A comparative approach to closed-loop
			computation},\ }\href@noop {} {\bibfield  {journal} {\bibinfo  {journal}
			{Current opinion in neurobiology}\ }\textbf {\bibinfo {volume} {25}},\
		\bibinfo {pages} {54} (\bibinfo {year} {2014})}\BibitemShut {NoStop}%
	\bibitem [{\citenamefont {Cowan}\ \emph {et~al.}(2014)\citenamefont {Cowan},
		\citenamefont {Ankarali}, \citenamefont {Dyhr}, \citenamefont {Madhav},
		\citenamefont {Roth}, \citenamefont {Sefati}, \citenamefont {Sponberg},
		\citenamefont {Stamper}, \citenamefont {Fortune},\ and\ \citenamefont
		{Daniel}}]{cowan2014feedback}%
	\BibitemOpen
	\bibfield  {author} {\bibinfo {author} {\bibfnamefont {N.~J.}\ \bibnamefont
			{Cowan}}, \bibinfo {author} {\bibfnamefont {M.~M.}\ \bibnamefont {Ankarali}},
		\bibinfo {author} {\bibfnamefont {J.~P.}\ \bibnamefont {Dyhr}}, \bibinfo
		{author} {\bibfnamefont {M.~S.}\ \bibnamefont {Madhav}}, \bibinfo {author}
		{\bibfnamefont {E.}~\bibnamefont {Roth}}, \bibinfo {author} {\bibfnamefont
			{S.}~\bibnamefont {Sefati}}, \bibinfo {author} {\bibfnamefont
			{S.}~\bibnamefont {Sponberg}}, \bibinfo {author} {\bibfnamefont {S.~A.}\
			\bibnamefont {Stamper}}, \bibinfo {author} {\bibfnamefont {E.~S.}\
			\bibnamefont {Fortune}},\ and\ \bibinfo {author} {\bibfnamefont {T.~L.}\
			\bibnamefont {Daniel}},\ }\bibfield  {title} {\bibinfo {title} {Feedback
			control as a framework for understanding tradeoffs in biology},\ }\href@noop
	{} {\bibfield  {journal} {\bibinfo  {journal} {Integrative and Comparative
				Biology}\ }\textbf {\bibinfo {volume} {54}},\ \bibinfo {pages} {223}
		(\bibinfo {year} {2014})}\BibitemShut {NoStop}%
	\bibitem [{\citenamefont {Sponberg}(2017)}]{sponberg2017emergent}%
	\BibitemOpen
	\bibfield  {author} {\bibinfo {author} {\bibfnamefont {S.}~\bibnamefont
			{Sponberg}},\ }\bibfield  {title} {\bibinfo {title} {The emergent physics of
			animal locomotion},\ }\href@noop {} {\bibfield  {journal} {\bibinfo
			{journal} {Physics Today}\ }\textbf {\bibinfo {volume} {70}},\ \bibinfo
		{pages} {34} (\bibinfo {year} {2017})}\BibitemShut {NoStop}%
	\bibitem [{\citenamefont {Srinivasan}\ and\ \citenamefont
		{Ruina}(2006)}]{srinivasan2006computer}%
	\BibitemOpen
	\bibfield  {author} {\bibinfo {author} {\bibfnamefont {M.}~\bibnamefont
			{Srinivasan}}\ and\ \bibinfo {author} {\bibfnamefont {A.}~\bibnamefont
			{Ruina}},\ }\bibfield  {title} {\bibinfo {title} {Computer optimization of a
			minimal biped model discovers walking and running},\ }\href@noop {}
	{\bibfield  {journal} {\bibinfo  {journal} {Nature}\ }\textbf {\bibinfo
			{volume} {439}},\ \bibinfo {pages} {72} (\bibinfo {year} {2006})}\BibitemShut
	{NoStop}%
	\bibitem [{\citenamefont {Ding}\ \emph {et~al.}(2013)\citenamefont {Ding},
		\citenamefont {Sharpe}, \citenamefont {Wiesenfeld},\ and\ \citenamefont
		{Goldman}}]{ding2013emergence}%
	\BibitemOpen
	\bibfield  {author} {\bibinfo {author} {\bibfnamefont {Y.}~\bibnamefont
			{Ding}}, \bibinfo {author} {\bibfnamefont {S.~S.}\ \bibnamefont {Sharpe}},
		\bibinfo {author} {\bibfnamefont {K.}~\bibnamefont {Wiesenfeld}},\ and\
		\bibinfo {author} {\bibfnamefont {D.~I.}\ \bibnamefont {Goldman}},\
	}\bibfield  {title} {\bibinfo {title} {Emergence of the advancing
			neuromechanical phase in a resistive force dominated medium},\ }\href@noop {}
	{\bibfield  {journal} {\bibinfo  {journal} {Proceedings of the National
				Academy of Sciences}\ }\textbf {\bibinfo {volume} {110}},\ \bibinfo {pages}
		{10123} (\bibinfo {year} {2013})}\BibitemShut {NoStop}%
	\bibitem [{\citenamefont {Bhounsule}\ \emph {et~al.}(2012)\citenamefont
		{Bhounsule}, \citenamefont {Cortell},\ and\ \citenamefont
		{Ruina}}]{bhounsule2012design}%
	\BibitemOpen
	\bibfield  {author} {\bibinfo {author} {\bibfnamefont {P.~A.}\ \bibnamefont
			{Bhounsule}}, \bibinfo {author} {\bibfnamefont {J.}~\bibnamefont {Cortell}},\
		and\ \bibinfo {author} {\bibfnamefont {A.}~\bibnamefont {Ruina}},\ }\bibfield
	{title} {\bibinfo {title} {Design and control of ranger: an
			energy-efficient, dynamic walking robot},\ }in\ \href@noop {} {\emph
		{\bibinfo {booktitle} {Adaptive Mobile Robotics}}}\ (\bibinfo  {publisher}
	{World Scientific},\ \bibinfo {year} {2012})\ pp.\ \bibinfo {pages}
	{441--448}\BibitemShut {NoStop}%
	\bibitem [{\citenamefont {Seok}\ \emph {et~al.}(2014)\citenamefont {Seok},
		\citenamefont {Wang}, \citenamefont {Chuah}, \citenamefont {Hyun},
		\citenamefont {Lee}, \citenamefont {Otten}, \citenamefont {Lang},\ and\
		\citenamefont {Kim}}]{seok2014design}%
	\BibitemOpen
	\bibfield  {author} {\bibinfo {author} {\bibfnamefont {S.}~\bibnamefont
			{Seok}}, \bibinfo {author} {\bibfnamefont {A.}~\bibnamefont {Wang}}, \bibinfo
		{author} {\bibfnamefont {M.~Y.}\ \bibnamefont {Chuah}}, \bibinfo {author}
		{\bibfnamefont {D.~J.}\ \bibnamefont {Hyun}}, \bibinfo {author}
		{\bibfnamefont {J.}~\bibnamefont {Lee}}, \bibinfo {author} {\bibfnamefont
			{D.~M.}\ \bibnamefont {Otten}}, \bibinfo {author} {\bibfnamefont {J.~H.}\
			\bibnamefont {Lang}},\ and\ \bibinfo {author} {\bibfnamefont
			{S.}~\bibnamefont {Kim}},\ }\bibfield  {title} {\bibinfo {title} {Design
			principles for energy-efficient legged locomotion and implementation on the
			mit cheetah robot},\ }\href@noop {} {\bibfield  {journal} {\bibinfo
			{journal} {Ieee/asme transactions on mechatronics}\ }\textbf {\bibinfo
			{volume} {20}},\ \bibinfo {pages} {1117} (\bibinfo {year}
		{2014})}\BibitemShut {NoStop}%
	\bibitem [{\citenamefont {Astley}\ \emph {et~al.}(2015)\citenamefont {Astley},
		\citenamefont {Gong}, \citenamefont {Dai}, \citenamefont {Travers},
		\citenamefont {Serrano}, \citenamefont {Vela}, \citenamefont {Choset},
		\citenamefont {Mendelson}, \citenamefont {Hu},\ and\ \citenamefont
		{Goldman}}]{astley2015modulation}%
	\BibitemOpen
	\bibfield  {author} {\bibinfo {author} {\bibfnamefont {H.~C.}\ \bibnamefont
			{Astley}}, \bibinfo {author} {\bibfnamefont {C.}~\bibnamefont {Gong}},
		\bibinfo {author} {\bibfnamefont {J.}~\bibnamefont {Dai}}, \bibinfo {author}
		{\bibfnamefont {M.}~\bibnamefont {Travers}}, \bibinfo {author} {\bibfnamefont
			{M.~M.}\ \bibnamefont {Serrano}}, \bibinfo {author} {\bibfnamefont {P.~A.}\
			\bibnamefont {Vela}}, \bibinfo {author} {\bibfnamefont {H.}~\bibnamefont
			{Choset}}, \bibinfo {author} {\bibfnamefont {J.~R.}\ \bibnamefont
			{Mendelson}}, \bibinfo {author} {\bibfnamefont {D.~L.}\ \bibnamefont {Hu}},\
		and\ \bibinfo {author} {\bibfnamefont {D.~I.}\ \bibnamefont {Goldman}},\
	}\bibfield  {title} {\bibinfo {title} {Modulation of orthogonal body waves
			enables high maneuverability in sidewinding locomotion},\ }\href@noop {}
	{\bibfield  {journal} {\bibinfo  {journal} {Proceedings of the National
				Academy of Sciences}\ }\textbf {\bibinfo {volume} {112}},\ \bibinfo {pages}
		{6200} (\bibinfo {year} {2015})}\BibitemShut {NoStop}%
	\bibitem [{\citenamefont {Chong}\ \emph {et~al.}(2022)\citenamefont {Chong},
		\citenamefont {Wang}, \citenamefont {Erickson}, \citenamefont {Bergmann},\
		and\ \citenamefont {Goldman}}]{chong2022coordinating}%
	\BibitemOpen
	\bibfield  {author} {\bibinfo {author} {\bibfnamefont {B.}~\bibnamefont
			{Chong}}, \bibinfo {author} {\bibfnamefont {T.}~\bibnamefont {Wang}},
		\bibinfo {author} {\bibfnamefont {E.}~\bibnamefont {Erickson}}, \bibinfo
		{author} {\bibfnamefont {P.~J.}\ \bibnamefont {Bergmann}},\ and\ \bibinfo
		{author} {\bibfnamefont {D.~I.}\ \bibnamefont {Goldman}},\ }\bibfield
	{title} {\bibinfo {title} {Coordinating tiny limbs and long bodies: geometric
			mechanics of diverse undulatory lizard locomotion (accepted)},\ }\href@noop
	{} {\bibfield  {journal} {\bibinfo  {journal} {Proceedings of the National
				Academy of Sciences}\ }\textbf {\bibinfo {volume} {TBD}} (\bibinfo {year}
		{2022})}\BibitemShut {NoStop}%
	\bibitem [{\citenamefont {Ijspeert}\ \emph {et~al.}(2007)\citenamefont
		{Ijspeert}, \citenamefont {Crespi}, \citenamefont {Ryczko},\ and\
		\citenamefont {Cabelguen}}]{ijspeert2007swimming}%
	\BibitemOpen
	\bibfield  {author} {\bibinfo {author} {\bibfnamefont {A.~J.}\ \bibnamefont
			{Ijspeert}}, \bibinfo {author} {\bibfnamefont {A.}~\bibnamefont {Crespi}},
		\bibinfo {author} {\bibfnamefont {D.}~\bibnamefont {Ryczko}},\ and\ \bibinfo
		{author} {\bibfnamefont {J.-M.}\ \bibnamefont {Cabelguen}},\ }\bibfield
	{title} {\bibinfo {title} {From swimming to walking with a salamander robot
			driven by a spinal cord model},\ }\href@noop {} {\bibfield  {journal}
		{\bibinfo  {journal} {science}\ }\textbf {\bibinfo {volume} {315}},\ \bibinfo
		{pages} {1416} (\bibinfo {year} {2007})}\BibitemShut {NoStop}%
	\bibitem [{\citenamefont {Berman}(2018)}]{berman2018measuring}%
	\BibitemOpen
	\bibfield  {author} {\bibinfo {author} {\bibfnamefont {G.~J.}\ \bibnamefont
			{Berman}},\ }\bibfield  {title} {\bibinfo {title} {Measuring behavior across
			scales},\ }\href@noop {} {\bibfield  {journal} {\bibinfo  {journal} {BMC
				biology}\ }\textbf {\bibinfo {volume} {16}},\ \bibinfo {pages} {23} (\bibinfo
		{year} {2018})}\BibitemShut {NoStop}%
	\bibitem [{\citenamefont {Shapere}\ and\ \citenamefont
		{Wilczek}(1987)}]{shapere1987self}%
	\BibitemOpen
	\bibfield  {author} {\bibinfo {author} {\bibfnamefont {A.}~\bibnamefont
			{Shapere}}\ and\ \bibinfo {author} {\bibfnamefont {F.}~\bibnamefont
			{Wilczek}},\ }\bibfield  {title} {\bibinfo {title} {Self-propulsion at low
			reynolds number},\ }\href@noop {} {\bibfield  {journal} {\bibinfo  {journal}
			{Physical Review Letters}\ }\textbf {\bibinfo {volume} {58}},\ \bibinfo
		{pages} {2051} (\bibinfo {year} {1987})}\BibitemShut {NoStop}%
	\bibitem [{\citenamefont {Berry}(1990)}]{berry1990anticipations}%
	\BibitemOpen
	\bibfield  {author} {\bibinfo {author} {\bibfnamefont {M.}~\bibnamefont
			{Berry}},\ }\bibfield  {title} {\bibinfo {title} {Anticipations of the
			geometric phase},\ }\href@noop {} {\bibfield  {journal} {\bibinfo  {journal}
			{Physics Today}\ }\textbf {\bibinfo {volume} {43}},\ \bibinfo {pages} {34}
		(\bibinfo {year} {1990})}\BibitemShut {NoStop}%
	\bibitem [{\citenamefont {Berry}(1988)}]{berry1988geometric}%
	\BibitemOpen
	\bibfield  {author} {\bibinfo {author} {\bibfnamefont {M.}~\bibnamefont
			{Berry}},\ }\bibfield  {title} {\bibinfo {title} {The geometric phase},\
	}\href@noop {} {\bibfield  {journal} {\bibinfo  {journal} {Scientific
				American}\ }\textbf {\bibinfo {volume} {259}},\ \bibinfo {pages} {46}
		(\bibinfo {year} {1988})}\BibitemShut {NoStop}%
	\bibitem [{\citenamefont {Montgomery}(1990)}]{montgomery1990}%
	\BibitemOpen
	\bibfield  {author} {\bibinfo {author} {\bibfnamefont {R.}~\bibnamefont
			{Montgomery}},\ }\bibfield  {title} {\bibinfo {title} {Isoholonomic problems
			and some applications},\ }\href
	{https://projecteuclid.org:443/euclid.cmp/1104180539} {\bibfield  {journal}
		{\bibinfo  {journal} {Comm. Math. Phys.}\ }\textbf {\bibinfo {volume}
			{128}},\ \bibinfo {pages} {565} (\bibinfo {year} {1990})}\BibitemShut
	{NoStop}%
	\bibitem [{\citenamefont {Marsden}\ \emph {et~al.}(1990)\citenamefont
		{Marsden}, \citenamefont {Montgomery},\ and\ \citenamefont
		{Rațiu}}]{marsden1990reduction}%
	\BibitemOpen
	\bibfield  {author} {\bibinfo {author} {\bibfnamefont {J.~E.}\ \bibnamefont
			{Marsden}}, \bibinfo {author} {\bibfnamefont {R.}~\bibnamefont
			{Montgomery}},\ and\ \bibinfo {author} {\bibfnamefont {T.~S.}\ \bibnamefont
			{Rațiu}},\ }\href@noop {} {\emph {\bibinfo {title} {Reduction, symmetry, and
				phases in mechanics}}},\ Vol.\ \bibinfo {volume} {436}\ (\bibinfo
	{publisher} {American Mathematical Soc.},\ \bibinfo {year}
	{1990})\BibitemShut {NoStop}%
	\bibitem [{\citenamefont {{Krishnaprasad}}\ and\ \citenamefont
		{{Tsakiris}}(1994)}]{krishnaprasad1994}%
	\BibitemOpen
	\bibfield  {author} {\bibinfo {author} {\bibfnamefont {P.~S.}\ \bibnamefont
			{{Krishnaprasad}}}\ and\ \bibinfo {author} {\bibfnamefont {D.~P.}\
			\bibnamefont {{Tsakiris}}},\ }\bibfield  {title} {\bibinfo {title} {G-snakes:
			nonholonomic kinematic chains on lie groups},\ }in\ \href@noop {} {\emph
		{\bibinfo {booktitle} {Proceedings of 1994 33rd IEEE Conference on Decision
				and Control}}},\ Vol.~\bibinfo {volume} {3}\ (\bibinfo {year} {1994})\ pp.\
	\bibinfo {pages} {2955--2960 vol.3}\BibitemShut {NoStop}%
	\bibitem [{\citenamefont {Kelly}\ and\ \citenamefont
		{Murray}(1995)}]{kelly1995geometric}%
	\BibitemOpen
	\bibfield  {author} {\bibinfo {author} {\bibfnamefont {S.~D.}\ \bibnamefont
			{Kelly}}\ and\ \bibinfo {author} {\bibfnamefont {R.~M.}\ \bibnamefont
			{Murray}},\ }\bibfield  {title} {\bibinfo {title} {Geometric phases and
			robotic locomotion},\ }\href@noop {} {\bibfield  {journal} {\bibinfo
			{journal} {Journal of Robotic Systems}\ }\textbf {\bibinfo {volume} {12}},\
		\bibinfo {pages} {417} (\bibinfo {year} {1995})}\BibitemShut {NoStop}%
	\bibitem [{\citenamefont {Ostrowski}\ and\ \citenamefont
		{Burdick}(1998)}]{ostrowski1998geometric}%
	\BibitemOpen
	\bibfield  {author} {\bibinfo {author} {\bibfnamefont {J.}~\bibnamefont
			{Ostrowski}}\ and\ \bibinfo {author} {\bibfnamefont {J.}~\bibnamefont
			{Burdick}},\ }\bibfield  {title} {\bibinfo {title} {The geometric mechanics
			of undulatory robotic locomotion},\ }\href@noop {} {\bibfield  {journal}
		{\bibinfo  {journal} {The international journal of robotics research}\
		}\textbf {\bibinfo {volume} {17}},\ \bibinfo {pages} {683} (\bibinfo {year}
		{1998})}\BibitemShut {NoStop}%
	\bibitem [{\citenamefont {{Lewis}}(2000)}]{lewis2000}%
	\BibitemOpen
	\bibfield  {author} {\bibinfo {author} {\bibfnamefont {A.~D.}\ \bibnamefont
			{{Lewis}}},\ }\bibfield  {title} {\bibinfo {title} {Simple mechanical control
			systems with constraints},\ }\href@noop {} {\bibfield  {journal} {\bibinfo
			{journal} {IEEE Transactions on Automatic Control}\ }\textbf {\bibinfo
			{volume} {45}},\ \bibinfo {pages} {1420} (\bibinfo {year}
		{2000})}\BibitemShut {NoStop}%
	\bibitem [{\citenamefont {Bloch}(2003)}]{Bloch2003}%
	\BibitemOpen
	\bibfield  {author} {\bibinfo {author} {\bibfnamefont {A.~M.}\ \bibnamefont
			{Bloch}},\ }\bibinfo {title} {Nonholonomic mechanics},\ in\ \href
	{https://doi.org/10.1007/b97376_5} {\emph {\bibinfo {booktitle} {Nonholonomic
				Mechanics and Control}}}\ (\bibinfo  {publisher} {Springer New York},\
	\bibinfo {address} {New York, NY},\ \bibinfo {year} {2003})\ Chap.~\bibinfo
	{chapter} {5}, pp.\ \bibinfo {pages} {207--276}\BibitemShut {NoStop}%
	\bibitem [{\citenamefont {Melli}\ \emph {et~al.}(2006)\citenamefont {Melli},
		\citenamefont {Rowley},\ and\ \citenamefont {Rufat}}]{melli2006}%
	\BibitemOpen
	\bibfield  {author} {\bibinfo {author} {\bibfnamefont {J.~B.}\ \bibnamefont
			{Melli}}, \bibinfo {author} {\bibfnamefont {C.~W.}\ \bibnamefont {Rowley}},\
		and\ \bibinfo {author} {\bibfnamefont {D.~S.}\ \bibnamefont {Rufat}},\
	}\bibfield  {title} {\bibinfo {title} {Motion planning for an articulated
			body in a perfect planar fluid},\ }\href {https://doi.org/10.1137/060649884}
	{\bibfield  {journal} {\bibinfo  {journal} {SIAM Journal on Applied Dynamical
				Systems}\ }\textbf {\bibinfo {volume} {5}},\ \bibinfo {pages} {650} (\bibinfo
		{year} {2006})},\ \Eprint
	{https://arxiv.org/abs/https://doi.org/10.1137/060649884}
	{https://doi.org/10.1137/060649884} \BibitemShut {NoStop}%
	\bibitem [{\citenamefont {Berry}(1993)}]{berry1993geometric}%
	\BibitemOpen
	\bibfield  {author} {\bibinfo {author} {\bibfnamefont {M.~V.}\ \bibnamefont
			{Berry}},\ }\href@noop {} {\emph {\bibinfo {title} {Geometric phases}}},\
	\bibinfo {type} {Tech. Rep.}\ (\bibinfo  {institution} {CERN-VIDEO-C-144-A},\
	\bibinfo {year} {1993})\BibitemShut {NoStop}%
	\bibitem [{\citenamefont {Hatton}\ and\ \citenamefont
		{Choset}(2015{\natexlab{a}})}]{hatton2015EPJ}%
	\BibitemOpen
	\bibfield  {author} {\bibinfo {author} {\bibfnamefont {R.}~\bibnamefont
			{Hatton}}\ and\ \bibinfo {author} {\bibfnamefont {H.}~\bibnamefont
			{Choset}},\ }\bibfield  {title} {\bibinfo {title} {Nonconservativity and
			noncommutativity in locomotion},\ }\href@noop {} {\bibfield  {journal}
		{\bibinfo  {journal} {The European Physical Journal Special Topics}\ }\textbf
		{\bibinfo {volume} {224}},\ \bibinfo {pages} {3141} (\bibinfo {year}
		{2015}{\natexlab{a}})}\BibitemShut {NoStop}%
	\bibitem [{\citenamefont {Hatton}\ \emph {et~al.}(2013)\citenamefont {Hatton},
		\citenamefont {Ding}, \citenamefont {Choset},\ and\ \citenamefont
		{Goldman}}]{hatton2013geometric}%
	\BibitemOpen
	\bibfield  {author} {\bibinfo {author} {\bibfnamefont {R.~L.}\ \bibnamefont
			{Hatton}}, \bibinfo {author} {\bibfnamefont {Y.}~\bibnamefont {Ding}},
		\bibinfo {author} {\bibfnamefont {H.}~\bibnamefont {Choset}},\ and\ \bibinfo
		{author} {\bibfnamefont {D.~I.}\ \bibnamefont {Goldman}},\ }\bibfield
	{title} {\bibinfo {title} {Geometric visualization of self-propulsion in a
			complex medium},\ }\href@noop {} {\bibfield  {journal} {\bibinfo  {journal}
			{Physical review letters}\ }\textbf {\bibinfo {volume} {110}},\ \bibinfo
		{pages} {078101} (\bibinfo {year} {2013})}\BibitemShut {NoStop}%
	\bibitem [{\citenamefont {Purcell}(1977)}]{purcell1977life}%
	\BibitemOpen
	\bibfield  {author} {\bibinfo {author} {\bibfnamefont {E.~M.}\ \bibnamefont
			{Purcell}},\ }\bibfield  {title} {\bibinfo {title} {Life at low reynolds
			number},\ }\href@noop {} {\bibfield  {journal} {\bibinfo  {journal} {American
				journal of physics}\ }\textbf {\bibinfo {volume} {45}},\ \bibinfo {pages} {3}
		(\bibinfo {year} {1977})}\BibitemShut {NoStop}%
	\bibitem [{\citenamefont {Becker}\ \emph {et~al.}(2003)\citenamefont {Becker},
		\citenamefont {Koehler},\ and\ \citenamefont
		{Stone}}]{becker_koehler_stone_2003}%
	\BibitemOpen
	\bibfield  {author} {\bibinfo {author} {\bibfnamefont {L.~E.}\ \bibnamefont
			{Becker}}, \bibinfo {author} {\bibfnamefont {S.~A.}\ \bibnamefont
			{Koehler}},\ and\ \bibinfo {author} {\bibfnamefont {H.~A.}\ \bibnamefont
			{Stone}},\ }\bibfield  {title} {\bibinfo {title} {On self-propulsion of
			micro-machines at low reynolds number: Purcell's three-link swimmer},\ }\href
	{https://doi.org/10.1017/S0022112003005184} {\bibfield  {journal} {\bibinfo
			{journal} {Journal of Fluid Mechanics}\ }\textbf {\bibinfo {volume} {490}},\
		\bibinfo {pages} {15} (\bibinfo {year} {2003})}\BibitemShut {NoStop}%
	\bibitem [{\citenamefont {Tam}\ and\ \citenamefont
		{Hosoi}(2007)}]{tam2007optimal}%
	\BibitemOpen
	\bibfield  {author} {\bibinfo {author} {\bibfnamefont {D.}~\bibnamefont
			{Tam}}\ and\ \bibinfo {author} {\bibfnamefont {A.~E.}\ \bibnamefont
			{Hosoi}},\ }\bibfield  {title} {\bibinfo {title} {Optimal stroke patterns for
			purcell’s three-link swimmer},\ }\href@noop {} {\bibfield  {journal}
		{\bibinfo  {journal} {Physical Review Letters}\ }\textbf {\bibinfo {volume}
			{98}},\ \bibinfo {pages} {068105} (\bibinfo {year} {2007})}\BibitemShut
	{NoStop}%
	\bibitem [{\citenamefont {Wuhr}\ and\ \citenamefont
		{Staley}(1995)}]{wuhr1995satellite}%
	\BibitemOpen
	\bibfield  {author} {\bibinfo {author} {\bibfnamefont {M.}~\bibnamefont
			{Wuhr}}\ and\ \bibinfo {author} {\bibfnamefont {D.~A.}\ \bibnamefont
			{Staley}},\ }\bibfield  {title} {\bibinfo {title} {Satellite reorientation
			maneuvers by momentum transfer (aka reorienting without using thrusters,
			torquers or nutation dampers)},\ }in\ \href@noop {} {\emph {\bibinfo
			{booktitle} {Space OPS 2004 Conference}}}\ (\bibinfo {year} {1995})\ p.\
	\bibinfo {pages} {421}\BibitemShut {NoStop}%
	\bibitem [{\citenamefont {Dai}\ \emph {et~al.}(2016)\citenamefont {Dai},
		\citenamefont {Faraji}, \citenamefont {Gong}, \citenamefont {Hatton},
		\citenamefont {Goldman},\ and\ \citenamefont {Choset}}]{dai2016rss}%
	\BibitemOpen
	\bibfield  {author} {\bibinfo {author} {\bibfnamefont {J.}~\bibnamefont
			{Dai}}, \bibinfo {author} {\bibfnamefont {H.}~\bibnamefont {Faraji}},
		\bibinfo {author} {\bibfnamefont {C.}~\bibnamefont {Gong}}, \bibinfo {author}
		{\bibfnamefont {R.~L.}\ \bibnamefont {Hatton}}, \bibinfo {author}
		{\bibfnamefont {D.~I.}\ \bibnamefont {Goldman}},\ and\ \bibinfo {author}
		{\bibfnamefont {H.}~\bibnamefont {Choset}},\ }\bibfield  {title} {\bibinfo
		{title} {Geometric swimming on a granular surface.},\ }in\ \href@noop {}
	{\emph {\bibinfo {booktitle} {Robotics: Science and Systems}}}\ (\bibinfo
	{year} {2016})\BibitemShut {NoStop}%
	\bibitem [{\citenamefont {Chong}\ \emph
		{et~al.}(2021{\natexlab{a}})\citenamefont {Chong}, \citenamefont {Wang},
		\citenamefont {Lin}, \citenamefont {Li}, \citenamefont {Blekherman},
		\citenamefont {Choset},\ and\ \citenamefont {Goldman}}]{chongmoving}%
	\BibitemOpen
	\bibfield  {author} {\bibinfo {author} {\bibfnamefont {B.}~\bibnamefont
			{Chong}}, \bibinfo {author} {\bibfnamefont {T.}~\bibnamefont {Wang}},
		\bibinfo {author} {\bibfnamefont {B.}~\bibnamefont {Lin}}, \bibinfo {author}
		{\bibfnamefont {S.}~\bibnamefont {Li}}, \bibinfo {author} {\bibfnamefont
			{G.}~\bibnamefont {Blekherman}}, \bibinfo {author} {\bibfnamefont
			{H.}~\bibnamefont {Choset}},\ and\ \bibinfo {author} {\bibfnamefont {D.~I.}\
			\bibnamefont {Goldman}},\ }\bibfield  {title} {\bibinfo {title} {Moving
			sidewinding forward: optimizing contact patterns for limbless robots via
			geometric mechanics},\ }in\ \href@noop {} {\emph {\bibinfo {booktitle}
			{Robotics: Science and Systems}}}\ (\bibinfo {year} {2021})\BibitemShut
	{NoStop}%
	\bibitem [{\citenamefont {Chong}\ \emph
		{et~al.}(2021{\natexlab{b}})\citenamefont {Chong}, \citenamefont
		{Ozkan-Aydin}, \citenamefont {Gong}, \citenamefont {Sartoretti},
		\citenamefont {Wu}, \citenamefont {Rieser}, \citenamefont {Xing},
		\citenamefont {Rankin}, \citenamefont {Michel}, \citenamefont {Nicieza},
		\citenamefont {Hutchinson}, \citenamefont {Goldman},\ and\ \citenamefont
		{Choset}}]{Chong2019Coordination}%
	\BibitemOpen
	\bibfield  {author} {\bibinfo {author} {\bibfnamefont {B.}~\bibnamefont
			{Chong}}, \bibinfo {author} {\bibfnamefont {Y.}~\bibnamefont {Ozkan-Aydin}},
		\bibinfo {author} {\bibfnamefont {C.}~\bibnamefont {Gong}}, \bibinfo {author}
		{\bibfnamefont {G.}~\bibnamefont {Sartoretti}}, \bibinfo {author}
		{\bibfnamefont {Y.}~\bibnamefont {Wu}}, \bibinfo {author} {\bibfnamefont
			{J.~M.}\ \bibnamefont {Rieser}}, \bibinfo {author} {\bibfnamefont
			{H.}~\bibnamefont {Xing}}, \bibinfo {author} {\bibfnamefont {J.}~\bibnamefont
			{Rankin}}, \bibinfo {author} {\bibfnamefont {K.}~\bibnamefont {Michel}},
		\bibinfo {author} {\bibfnamefont {A.}~\bibnamefont {Nicieza}}, \bibinfo
		{author} {\bibfnamefont {J.}~\bibnamefont {Hutchinson}}, \bibinfo {author}
		{\bibfnamefont {D.~I.}\ \bibnamefont {Goldman}},\ and\ \bibinfo {author}
		{\bibfnamefont {H.}~\bibnamefont {Choset}},\ }\bibfield  {title} {\bibinfo
		{title} {{Coordination of lateral body bending and leg movements for sprawled
				posture quadrupedal locomotion}},\ }\href@noop {} {\bibfield  {journal}
		{\bibinfo  {journal} {International Journal of Robotics Research}\ }\textbf
		{\bibinfo {volume} {40}},\ \bibinfo {pages} {747} (\bibinfo {year}
		{2021}{\natexlab{b}})}\BibitemShut {NoStop}%
	\bibitem [{\citenamefont {Lauga}\ and\ \citenamefont
		{Powers}(2009)}]{lauga2009hydrodynamics}%
	\BibitemOpen
	\bibfield  {author} {\bibinfo {author} {\bibfnamefont {E.}~\bibnamefont
			{Lauga}}\ and\ \bibinfo {author} {\bibfnamefont {T.~R.}\ \bibnamefont
			{Powers}},\ }\bibfield  {title} {\bibinfo {title} {The hydrodynamics of
			swimming microorganisms},\ }\href@noop {} {\bibfield  {journal} {\bibinfo
			{journal} {Reports on Progress in Physics}\ }\textbf {\bibinfo {volume}
			{72}},\ \bibinfo {pages} {096601} (\bibinfo {year} {2009})}\BibitemShut
	{NoStop}%
	\bibitem [{\citenamefont {Gray}\ and\ \citenamefont
		{Hancock}(1955)}]{gray1955propulsion}%
	\BibitemOpen
	\bibfield  {author} {\bibinfo {author} {\bibfnamefont {J.}~\bibnamefont
			{Gray}}\ and\ \bibinfo {author} {\bibfnamefont {G.}~\bibnamefont {Hancock}},\
	}\bibfield  {title} {\bibinfo {title} {The propulsion of sea-urchin
			spermatozoa},\ }\href@noop {} {\bibfield  {journal} {\bibinfo  {journal}
			{Journal of Experimental Biology}\ }\textbf {\bibinfo {volume} {32}},\
		\bibinfo {pages} {802} (\bibinfo {year} {1955})}\BibitemShut {NoStop}%
	\bibitem [{\citenamefont {Hu}\ \emph {et~al.}(2009)\citenamefont {Hu},
		\citenamefont {Nirody}, \citenamefont {Scott},\ and\ \citenamefont
		{Shelley}}]{hu2009mechanics}%
	\BibitemOpen
	\bibfield  {author} {\bibinfo {author} {\bibfnamefont {D.~L.}\ \bibnamefont
			{Hu}}, \bibinfo {author} {\bibfnamefont {J.}~\bibnamefont {Nirody}}, \bibinfo
		{author} {\bibfnamefont {T.}~\bibnamefont {Scott}},\ and\ \bibinfo {author}
		{\bibfnamefont {M.~J.}\ \bibnamefont {Shelley}},\ }\bibfield  {title}
	{\bibinfo {title} {The mechanics of slithering locomotion},\ }\href@noop {}
	{\bibfield  {journal} {\bibinfo  {journal} {Proceedings of the National
				Academy of Sciences}\ }\textbf {\bibinfo {volume} {106}},\ \bibinfo {pages}
		{10081} (\bibinfo {year} {2009})}\BibitemShut {NoStop}%
	\bibitem [{\citenamefont {Maladen}\ \emph {et~al.}(2009)\citenamefont
		{Maladen}, \citenamefont {Ding}, \citenamefont {Li},\ and\ \citenamefont
		{Goldman}}]{maladen2009undulatory}%
	\BibitemOpen
	\bibfield  {author} {\bibinfo {author} {\bibfnamefont {R.~D.}\ \bibnamefont
			{Maladen}}, \bibinfo {author} {\bibfnamefont {Y.}~\bibnamefont {Ding}},
		\bibinfo {author} {\bibfnamefont {C.}~\bibnamefont {Li}},\ and\ \bibinfo
		{author} {\bibfnamefont {D.~I.}\ \bibnamefont {Goldman}},\ }\bibfield
	{title} {\bibinfo {title} {Undulatory swimming in sand: subsurface locomotion
			of the sandfish lizard},\ }\href@noop {} {\bibfield  {journal} {\bibinfo
			{journal} {science}\ }\textbf {\bibinfo {volume} {325}},\ \bibinfo {pages}
		{314} (\bibinfo {year} {2009})}\BibitemShut {NoStop}%
	\bibitem [{\citenamefont {Sharpe}\ \emph {et~al.}(2015)\citenamefont {Sharpe},
		\citenamefont {Koehler}, \citenamefont {Kuckuk}, \citenamefont {Serrano},
		\citenamefont {Vela}, \citenamefont {Mendelson},\ and\ \citenamefont
		{Goldman}}]{sharpe2015locomotor}%
	\BibitemOpen
	\bibfield  {author} {\bibinfo {author} {\bibfnamefont {S.~S.}\ \bibnamefont
			{Sharpe}}, \bibinfo {author} {\bibfnamefont {S.~A.}\ \bibnamefont {Koehler}},
		\bibinfo {author} {\bibfnamefont {R.~M.}\ \bibnamefont {Kuckuk}}, \bibinfo
		{author} {\bibfnamefont {M.}~\bibnamefont {Serrano}}, \bibinfo {author}
		{\bibfnamefont {P.~A.}\ \bibnamefont {Vela}}, \bibinfo {author}
		{\bibfnamefont {J.}~\bibnamefont {Mendelson}},\ and\ \bibinfo {author}
		{\bibfnamefont {D.~I.}\ \bibnamefont {Goldman}},\ }\bibfield  {title}
	{\bibinfo {title} {Locomotor benefits of being a slender and slick sand
			swimmer},\ }\href@noop {} {\bibfield  {journal} {\bibinfo  {journal} {Journal
				of Experimental Biology}\ }\textbf {\bibinfo {volume} {218}},\ \bibinfo
		{pages} {440} (\bibinfo {year} {2015})}\BibitemShut {NoStop}%
	\bibitem [{\citenamefont {Zhang}\ and\ \citenamefont
		{Goldman}(2014)}]{zhang2014effectiveness}%
	\BibitemOpen
	\bibfield  {author} {\bibinfo {author} {\bibfnamefont {T.}~\bibnamefont
			{Zhang}}\ and\ \bibinfo {author} {\bibfnamefont {D.~I.}\ \bibnamefont
			{Goldman}},\ }\bibfield  {title} {\bibinfo {title} {The effectiveness of
			resistive force theory in granular locomotion},\ }\href@noop {} {\bibfield
		{journal} {\bibinfo  {journal} {Physics of Fluids}\ }\textbf {\bibinfo
			{volume} {26}},\ \bibinfo {pages} {101308} (\bibinfo {year}
		{2014})}\BibitemShut {NoStop}%
	\bibitem [{\citenamefont {Stephens}\ \emph {et~al.}(2008)\citenamefont
		{Stephens}, \citenamefont {Johnson-Kerner}, \citenamefont {Bialek},\ and\
		\citenamefont {Ryu}}]{stephens2008dimensionality}%
	\BibitemOpen
	\bibfield  {author} {\bibinfo {author} {\bibfnamefont {G.~J.}\ \bibnamefont
			{Stephens}}, \bibinfo {author} {\bibfnamefont {B.}~\bibnamefont
			{Johnson-Kerner}}, \bibinfo {author} {\bibfnamefont {W.}~\bibnamefont
			{Bialek}},\ and\ \bibinfo {author} {\bibfnamefont {W.~S.}\ \bibnamefont
			{Ryu}},\ }\bibfield  {title} {\bibinfo {title} {Dimensionality and dynamics
			in the behavior of c. elegans},\ }\href@noop {} {\bibfield  {journal}
		{\bibinfo  {journal} {PLoS computational biology}\ }\textbf {\bibinfo
			{volume} {4}},\ \bibinfo {pages} {e1000028} (\bibinfo {year}
		{2008})}\BibitemShut {NoStop}%
	\bibitem [{\citenamefont {Astley}\ \emph
		{et~al.}(2020{\natexlab{a}})\citenamefont {Astley}, \citenamefont
		{Mendelson}, \citenamefont {Dai}, \citenamefont {Gong}, \citenamefont
		{Chong}, \citenamefont {Rieser}, \citenamefont {Schiebel}, \citenamefont
		{Sharpe}, \citenamefont {Hatton}, \citenamefont {Choset} \emph
		{et~al.}}]{astley2020surprising}%
	\BibitemOpen
	\bibfield  {author} {\bibinfo {author} {\bibfnamefont {H.~C.}\ \bibnamefont
			{Astley}}, \bibinfo {author} {\bibfnamefont {J.~R.}\ \bibnamefont
			{Mendelson}}, \bibinfo {author} {\bibfnamefont {J.}~\bibnamefont {Dai}},
		\bibinfo {author} {\bibfnamefont {C.}~\bibnamefont {Gong}}, \bibinfo {author}
		{\bibfnamefont {B.}~\bibnamefont {Chong}}, \bibinfo {author} {\bibfnamefont
			{J.~M.}\ \bibnamefont {Rieser}}, \bibinfo {author} {\bibfnamefont {P.~E.}\
			\bibnamefont {Schiebel}}, \bibinfo {author} {\bibfnamefont {S.~S.}\
			\bibnamefont {Sharpe}}, \bibinfo {author} {\bibfnamefont {R.~L.}\
			\bibnamefont {Hatton}}, \bibinfo {author} {\bibfnamefont {H.}~\bibnamefont
			{Choset}}, \emph {et~al.},\ }\bibfield  {title} {\bibinfo {title} {Surprising
			simplicities and syntheses in limbless self-propulsion in sand},\ }\href@noop
	{} {\bibfield  {journal} {\bibinfo  {journal} {Journal of Experimental
				Biology}\ }\textbf {\bibinfo {volume} {223}} (\bibinfo {year}
		{2020}{\natexlab{a}})}\BibitemShut {NoStop}%
	\bibitem [{\citenamefont {Fang-Yen}\ \emph {et~al.}(2010)\citenamefont
		{Fang-Yen}, \citenamefont {Wyart}, \citenamefont {Xie}, \citenamefont
		{Kawai}, \citenamefont {Kodger}, \citenamefont {Chen}, \citenamefont {Wen},\
		and\ \citenamefont {Samuel}}]{fang2010biomechanical}%
	\BibitemOpen
	\bibfield  {author} {\bibinfo {author} {\bibfnamefont {C.}~\bibnamefont
			{Fang-Yen}}, \bibinfo {author} {\bibfnamefont {M.}~\bibnamefont {Wyart}},
		\bibinfo {author} {\bibfnamefont {J.}~\bibnamefont {Xie}}, \bibinfo {author}
		{\bibfnamefont {R.}~\bibnamefont {Kawai}}, \bibinfo {author} {\bibfnamefont
			{T.}~\bibnamefont {Kodger}}, \bibinfo {author} {\bibfnamefont
			{S.}~\bibnamefont {Chen}}, \bibinfo {author} {\bibfnamefont {Q.}~\bibnamefont
			{Wen}},\ and\ \bibinfo {author} {\bibfnamefont {A.~D.}\ \bibnamefont
			{Samuel}},\ }\bibfield  {title} {\bibinfo {title} {Biomechanical analysis of
			gait adaptation in the nematode caenorhabditis elegans},\ }\href@noop {}
	{\bibfield  {journal} {\bibinfo  {journal} {Proceedings of the National
				Academy of Sciences}\ }\textbf {\bibinfo {volume} {107}},\ \bibinfo {pages}
		{20323} (\bibinfo {year} {2010})}\BibitemShut {NoStop}%
	\bibitem [{\citenamefont {Schiebel}\ \emph {et~al.}(2020)\citenamefont
		{Schiebel}, \citenamefont {Astley}, \citenamefont {Rieser}, \citenamefont
		{Agarwal}, \citenamefont {Hubicki}, \citenamefont {Hubbard}, \citenamefont
		{Diaz}, \citenamefont {Mendelson~III}, \citenamefont {Kamrin},\ and\
		\citenamefont {Goldman}}]{schiebel2019mitigating}%
	\BibitemOpen
	\bibfield  {author} {\bibinfo {author} {\bibfnamefont {P.~E.}\ \bibnamefont
			{Schiebel}}, \bibinfo {author} {\bibfnamefont {H.~C.}\ \bibnamefont
			{Astley}}, \bibinfo {author} {\bibfnamefont {J.~M.}\ \bibnamefont {Rieser}},
		\bibinfo {author} {\bibfnamefont {S.}~\bibnamefont {Agarwal}}, \bibinfo
		{author} {\bibfnamefont {C.}~\bibnamefont {Hubicki}}, \bibinfo {author}
		{\bibfnamefont {A.~M.}\ \bibnamefont {Hubbard}}, \bibinfo {author}
		{\bibfnamefont {K.}~\bibnamefont {Diaz}}, \bibinfo {author} {\bibfnamefont
			{J.~R.}\ \bibnamefont {Mendelson~III}}, \bibinfo {author} {\bibfnamefont
			{K.}~\bibnamefont {Kamrin}},\ and\ \bibinfo {author} {\bibfnamefont {D.~I.}\
			\bibnamefont {Goldman}},\ }\bibfield  {title} {\bibinfo {title} {Mitigating
			memory effects during undulatory locomotion on hysteretic materials},\
	}\href@noop {} {\bibfield  {journal} {\bibinfo  {journal} {eLife}\ }\textbf
		{\bibinfo {volume} {9}},\ \bibinfo {pages} {e51412} (\bibinfo {year}
		{2020})}\BibitemShut {NoStop}%
	\bibitem [{\citenamefont {Schiebel}\ \emph {et~al.}(2019)\citenamefont
		{Schiebel}, \citenamefont {Rieser}, \citenamefont {Hubbard}, \citenamefont
		{Chen}, \citenamefont {Rocklin},\ and\ \citenamefont
		{Goldman}}]{schiebel2019mechanical}%
	\BibitemOpen
	\bibfield  {author} {\bibinfo {author} {\bibfnamefont {P.~E.}\ \bibnamefont
			{Schiebel}}, \bibinfo {author} {\bibfnamefont {J.~M.}\ \bibnamefont
			{Rieser}}, \bibinfo {author} {\bibfnamefont {A.~M.}\ \bibnamefont {Hubbard}},
		\bibinfo {author} {\bibfnamefont {L.}~\bibnamefont {Chen}}, \bibinfo {author}
		{\bibfnamefont {D.~Z.}\ \bibnamefont {Rocklin}},\ and\ \bibinfo {author}
		{\bibfnamefont {D.~I.}\ \bibnamefont {Goldman}},\ }\bibfield  {title}
	{\bibinfo {title} {Mechanical diffraction reveals the role of passive
			dynamics in a slithering snake},\ }\href@noop {} {\bibfield  {journal}
		{\bibinfo  {journal} {Proceedings of the National Academy of Sciences}\
		}\textbf {\bibinfo {volume} {116}},\ \bibinfo {pages} {4798} (\bibinfo {year}
		{2019})}\BibitemShut {NoStop}%
	\bibitem [{\citenamefont {Hirose}(1993)}]{Hirose:1993}%
	\BibitemOpen
	\bibfield  {author} {\bibinfo {author} {\bibfnamefont {S.}~\bibnamefont
			{Hirose}},\ }\href@noop {} {\emph {\bibinfo {title} {Biologically Inspired
				Robots}}}\ (\bibinfo  {publisher} {Oxford University Press},\ \bibinfo {year}
	{1993})\BibitemShut {NoStop}%
	\bibitem [{\citenamefont {Maladen}\ \emph {et~al.}(2011)\citenamefont
		{Maladen}, \citenamefont {Ding}, \citenamefont {Umbanhowar}, \citenamefont
		{Kamor},\ and\ \citenamefont {Goldman}}]{Maladen:2011es}%
	\BibitemOpen
	\bibfield  {author} {\bibinfo {author} {\bibfnamefont {R.~D.}\ \bibnamefont
			{Maladen}}, \bibinfo {author} {\bibfnamefont {Y.}~\bibnamefont {Ding}},
		\bibinfo {author} {\bibfnamefont {P.~B.}\ \bibnamefont {Umbanhowar}},
		\bibinfo {author} {\bibfnamefont {A.}~\bibnamefont {Kamor}},\ and\ \bibinfo
		{author} {\bibfnamefont {D.~I.}\ \bibnamefont {Goldman}},\ }\bibfield
	{title} {\bibinfo {title} {{Mechanical models of sandfish locomotion reveal
				principles of high performance subsurface sand-swimming}},\ }\href@noop {}
	{\bibfield  {journal} {\bibinfo  {journal} {Journal of The Royal Society
				Interface}\ }\textbf {\bibinfo {volume} {8}},\ \bibinfo {pages} {1332}
		(\bibinfo {year} {2011})}\BibitemShut {NoStop}%
	\bibitem [{\citenamefont {Berri}\ \emph {et~al.}(2009)\citenamefont {Berri},
		\citenamefont {Boyle}, \citenamefont {Tassieri}, \citenamefont {Hope},\ and\
		\citenamefont {Cohen}}]{berri2009forward}%
	\BibitemOpen
	\bibfield  {author} {\bibinfo {author} {\bibfnamefont {S.}~\bibnamefont
			{Berri}}, \bibinfo {author} {\bibfnamefont {J.~H.}\ \bibnamefont {Boyle}},
		\bibinfo {author} {\bibfnamefont {M.}~\bibnamefont {Tassieri}}, \bibinfo
		{author} {\bibfnamefont {I.~A.}\ \bibnamefont {Hope}},\ and\ \bibinfo
		{author} {\bibfnamefont {N.}~\bibnamefont {Cohen}},\ }\bibfield  {title}
	{\bibinfo {title} {Forward locomotion of the nematode c. elegans is achieved
			through modulation of a single gait},\ }\href@noop {} {\bibfield  {journal}
		{\bibinfo  {journal} {HFSP journal}\ }\textbf {\bibinfo {volume} {3}},\
		\bibinfo {pages} {186} (\bibinfo {year} {2009})}\BibitemShut {NoStop}%
	\bibitem [{\citenamefont {Sznitman}\ \emph
		{et~al.}(2010{\natexlab{a}})\citenamefont {Sznitman}, \citenamefont {Shen},
		\citenamefont {Sznitman},\ and\ \citenamefont
		{Arratia}}]{sznitman2010propulsive}%
	\BibitemOpen
	\bibfield  {author} {\bibinfo {author} {\bibfnamefont {J.}~\bibnamefont
			{Sznitman}}, \bibinfo {author} {\bibfnamefont {X.}~\bibnamefont {Shen}},
		\bibinfo {author} {\bibfnamefont {R.}~\bibnamefont {Sznitman}},\ and\
		\bibinfo {author} {\bibfnamefont {P.~E.}\ \bibnamefont {Arratia}},\
	}\bibfield  {title} {\bibinfo {title} {Propulsive force measurements and flow
			behavior of undulatory swimmers at low reynolds number},\ }\href
	{https://doi.org/10.1063/1.3529236} {\bibfield  {journal} {\bibinfo
			{journal} {Physics of Fluids}\ }\textbf {\bibinfo {volume} {22}},\ \bibinfo
		{pages} {121901} (\bibinfo {year} {2010}{\natexlab{a}})},\ \Eprint
	{https://arxiv.org/abs/https://doi.org/10.1063/1.3529236}
	{https://doi.org/10.1063/1.3529236} \BibitemShut {NoStop}%
	\bibitem [{\citenamefont {Rodenborn}\ \emph {et~al.}(2013)\citenamefont
		{Rodenborn}, \citenamefont {Chen}, \citenamefont {Swinney}, \citenamefont
		{Liu},\ and\ \citenamefont {Zhang}}]{RodenbornE338}%
	\BibitemOpen
	\bibfield  {author} {\bibinfo {author} {\bibfnamefont {B.}~\bibnamefont
			{Rodenborn}}, \bibinfo {author} {\bibfnamefont {C.-H.}\ \bibnamefont {Chen}},
		\bibinfo {author} {\bibfnamefont {H.~L.}\ \bibnamefont {Swinney}}, \bibinfo
		{author} {\bibfnamefont {B.}~\bibnamefont {Liu}},\ and\ \bibinfo {author}
		{\bibfnamefont {H.~P.}\ \bibnamefont {Zhang}},\ }\bibfield  {title} {\bibinfo
		{title} {Propulsion of microorganisms by a helical flagellum},\ }\href
	{https://doi.org/10.1073/pnas.1219831110} {\bibfield  {journal} {\bibinfo
			{journal} {Proceedings of the National Academy of Sciences}\ }\textbf
		{\bibinfo {volume} {110}},\ \bibinfo {pages} {E338} (\bibinfo {year}
		{2013})},\ \Eprint
	{https://arxiv.org/abs/https://www.pnas.org/content/110/5/E338.full.pdf}
	{https://www.pnas.org/content/110/5/E338.full.pdf} \BibitemShut {NoStop}%
	\bibitem [{\citenamefont {Berg}(1993)}]{berg1993random}%
	\BibitemOpen
	\bibfield  {author} {\bibinfo {author} {\bibfnamefont {H.~C.}\ \bibnamefont
			{Berg}},\ }\href@noop {} {\emph {\bibinfo {title} {Random walks in
				biology}}}\ (\bibinfo  {publisher} {Princeton University Press},\ \bibinfo
	{year} {1993})\BibitemShut {NoStop}%
	\bibitem [{\citenamefont {Backholm}\ \emph {et~al.}(2015)\citenamefont
		{Backholm}, \citenamefont {Kasper}, \citenamefont {Schulman}, \citenamefont
		{Ryu},\ and\ \citenamefont {Dalnoki-Veress}}]{Backholm2015effects}%
	\BibitemOpen
	\bibfield  {author} {\bibinfo {author} {\bibfnamefont {M.}~\bibnamefont
			{Backholm}}, \bibinfo {author} {\bibfnamefont {A.~K.~S.}\ \bibnamefont
			{Kasper}}, \bibinfo {author} {\bibfnamefont {R.~D.}\ \bibnamefont
			{Schulman}}, \bibinfo {author} {\bibfnamefont {W.~S.}\ \bibnamefont {Ryu}},\
		and\ \bibinfo {author} {\bibfnamefont {K.}~\bibnamefont {Dalnoki-Veress}},\
	}\bibfield  {title} {\bibinfo {title} {The effects of viscosity on the
			undulatory swimming dynamics of c. elegans},\ }\href
	{https://doi.org/10.1063/1.4931795} {\bibfield  {journal} {\bibinfo
			{journal} {Physics of Fluids}\ }\textbf {\bibinfo {volume} {27}},\ \bibinfo
		{pages} {091901} (\bibinfo {year} {2015})},\ \Eprint
	{https://arxiv.org/abs/https://aip.scitation.org/doi/pdf/10.1063/1.4931795}
	{https://aip.scitation.org/doi/pdf/10.1063/1.4931795} \BibitemShut {NoStop}%
	\bibitem [{\citenamefont {Shammas}\ \emph {et~al.}(2007)\citenamefont
		{Shammas}, \citenamefont {Choset},\ and\ \citenamefont
		{Rizzi}}]{shammas2007geometric}%
	\BibitemOpen
	\bibfield  {author} {\bibinfo {author} {\bibfnamefont {E.~A.}\ \bibnamefont
			{Shammas}}, \bibinfo {author} {\bibfnamefont {H.}~\bibnamefont {Choset}},\
		and\ \bibinfo {author} {\bibfnamefont {A.~A.}\ \bibnamefont {Rizzi}},\
	}\bibfield  {title} {\bibinfo {title} {Geometric motion planning analysis for
			two classes of underactuated mechanical systems},\ }\href@noop {} {\bibfield
		{journal} {\bibinfo  {journal} {The International Journal of Robotics
				Research}\ }\textbf {\bibinfo {volume} {26}},\ \bibinfo {pages} {1043}
		(\bibinfo {year} {2007})}\BibitemShut {NoStop}%
	\bibitem [{\citenamefont {Hatton}\ and\ \citenamefont
		{Choset}(2011)}]{hatton2011geometric}%
	\BibitemOpen
	\bibfield  {author} {\bibinfo {author} {\bibfnamefont {R.~L.}\ \bibnamefont
			{Hatton}}\ and\ \bibinfo {author} {\bibfnamefont {H.}~\bibnamefont
			{Choset}},\ }\bibfield  {title} {\bibinfo {title} {Geometric motion planning:
			The local connection, stokes’ theorem, and the importance of coordinate
			choice},\ }\href@noop {} {\bibfield  {journal} {\bibinfo  {journal} {The
				International Journal of Robotics Research}\ }\textbf {\bibinfo {volume}
			{30}},\ \bibinfo {pages} {988} (\bibinfo {year} {2011})}\BibitemShut
	{NoStop}%
	\bibitem [{\citenamefont {Sharpe}\ \emph {et~al.}(2013)\citenamefont {Sharpe},
		\citenamefont {Ding},\ and\ \citenamefont
		{Goldman}}]{sharpe2013environmental}%
	\BibitemOpen
	\bibfield  {author} {\bibinfo {author} {\bibfnamefont {S.~S.}\ \bibnamefont
			{Sharpe}}, \bibinfo {author} {\bibfnamefont {Y.}~\bibnamefont {Ding}},\ and\
		\bibinfo {author} {\bibfnamefont {D.~I.}\ \bibnamefont {Goldman}},\
	}\bibfield  {title} {\bibinfo {title} {Environmental interaction influences
			muscle activation strategy during sand-swimming in the sandfish lizard
			scincus scincus},\ }\href@noop {} {\bibfield  {journal} {\bibinfo  {journal}
			{Journal of Experimental Biology}\ }\textbf {\bibinfo {volume} {216}},\
		\bibinfo {pages} {260} (\bibinfo {year} {2013})}\BibitemShut {NoStop}%
	\bibitem [{\citenamefont {Hatton}\ \emph {et~al.}(2017)\citenamefont {Hatton},
		\citenamefont {Dear},\ and\ \citenamefont {Choset}}]{hatton2017cartography}%
	\BibitemOpen
	\bibfield  {author} {\bibinfo {author} {\bibfnamefont {R.~L.}\ \bibnamefont
			{Hatton}}, \bibinfo {author} {\bibfnamefont {T.}~\bibnamefont {Dear}},\ and\
		\bibinfo {author} {\bibfnamefont {H.}~\bibnamefont {Choset}},\ }\bibfield
	{title} {\bibinfo {title} {Kinematic cartography and the efficiency of
			viscous swimming},\ }\href@noop {} {\bibfield  {journal} {\bibinfo  {journal}
			{IEEE Transactions on Robotics}\ }\textbf {\bibinfo {volume} {33}},\ \bibinfo
		{pages} {523} (\bibinfo {year} {2017})}\BibitemShut {NoStop}%
	\bibitem [{\citenamefont {Chevallier}\ \emph {et~al.}(2008)\citenamefont
		{Chevallier}, \citenamefont {Ijspeert}, \citenamefont {Ryczko}, \citenamefont
		{Nagy},\ and\ \citenamefont {Cabelguen}}]{chevallier2008organisation}%
	\BibitemOpen
	\bibfield  {author} {\bibinfo {author} {\bibfnamefont {S.}~\bibnamefont
			{Chevallier}}, \bibinfo {author} {\bibfnamefont {A.~J.}\ \bibnamefont
			{Ijspeert}}, \bibinfo {author} {\bibfnamefont {D.}~\bibnamefont {Ryczko}},
		\bibinfo {author} {\bibfnamefont {F.}~\bibnamefont {Nagy}},\ and\ \bibinfo
		{author} {\bibfnamefont {J.-M.}\ \bibnamefont {Cabelguen}},\ }\bibfield
	{title} {\bibinfo {title} {Organisation of the spinal central pattern
			generators for locomotion in the salamander: biology and modelling},\
	}\href@noop {} {\bibfield  {journal} {\bibinfo  {journal} {Brain research
				reviews}\ }\textbf {\bibinfo {volume} {57}},\ \bibinfo {pages} {147}
		(\bibinfo {year} {2008})}\BibitemShut {NoStop}%
	\bibitem [{\citenamefont {Wang}\ \emph {et~al.}(2020)\citenamefont {Wang},
		\citenamefont {Chong}, \citenamefont {Diaz}, \citenamefont {Whitman},
		\citenamefont {Lu}, \citenamefont {Travers}, \citenamefont {Goldman},\ and\
		\citenamefont {Choset}}]{wang2020omega}%
	\BibitemOpen
	\bibfield  {author} {\bibinfo {author} {\bibfnamefont {T.}~\bibnamefont
			{Wang}}, \bibinfo {author} {\bibfnamefont {B.}~\bibnamefont {Chong}},
		\bibinfo {author} {\bibfnamefont {K.}~\bibnamefont {Diaz}}, \bibinfo {author}
		{\bibfnamefont {J.}~\bibnamefont {Whitman}}, \bibinfo {author} {\bibfnamefont
			{H.}~\bibnamefont {Lu}}, \bibinfo {author} {\bibfnamefont {M.}~\bibnamefont
			{Travers}}, \bibinfo {author} {\bibfnamefont {D.~I.}\ \bibnamefont
			{Goldman}},\ and\ \bibinfo {author} {\bibfnamefont {H.}~\bibnamefont
			{Choset}},\ }\bibfield  {title} {\bibinfo {title} {The omega turn: A
			biologically-inspired turning strategy for elongated limbless robots},\ }in\
	\href@noop {} {\emph {\bibinfo {booktitle} {2020 IEEE/RSJ International
				Conference on Intelligent Robots and Systems (IROS)}}}\ (\bibinfo
	{organization} {IEEE},\ \bibinfo {year} {2020})\ pp.\ \bibinfo {pages}
	{7766--7771}\BibitemShut {NoStop}%
	\bibitem [{\citenamefont {Wang}\ \emph {et~al.}(2022)\citenamefont {Wang},
		\citenamefont {Chong}, \citenamefont {Deng}, \citenamefont {Fu},
		\citenamefont {Choset},\ and\ \citenamefont {Goldman}}]{wang2022generalized}%
	\BibitemOpen
	\bibfield  {author} {\bibinfo {author} {\bibfnamefont {T.}~\bibnamefont
			{Wang}}, \bibinfo {author} {\bibfnamefont {B.}~\bibnamefont {Chong}},
		\bibinfo {author} {\bibfnamefont {Y.}~\bibnamefont {Deng}}, \bibinfo {author}
		{\bibfnamefont {R.}~\bibnamefont {Fu}}, \bibinfo {author} {\bibfnamefont
			{H.}~\bibnamefont {Choset}},\ and\ \bibinfo {author} {\bibfnamefont {D.~I.}\
			\bibnamefont {Goldman}},\ }\bibfield  {title} {\bibinfo {title} {Generalized
			omega turn gait enables agile limbless robot turning in complex
			environments},\ }in\ \href@noop {} {\emph {\bibinfo {booktitle} {2022
				International Conference on Robotics and Automation (ICRA)}}}\ (\bibinfo
	{organization} {IEEE},\ \bibinfo {year} {2022})\ pp.\ \bibinfo {pages}
	{01--07}\BibitemShut {NoStop}%
	\bibitem [{\citenamefont {Srivastava}\ \emph {et~al.}(2009)\citenamefont
		{Srivastava}, \citenamefont {Clark},\ and\ \citenamefont
		{Samuel}}]{srivastava2009temporal}%
	\BibitemOpen
	\bibfield  {author} {\bibinfo {author} {\bibfnamefont {N.}~\bibnamefont
			{Srivastava}}, \bibinfo {author} {\bibfnamefont {D.~A.}\ \bibnamefont
			{Clark}},\ and\ \bibinfo {author} {\bibfnamefont {A.~D.}\ \bibnamefont
			{Samuel}},\ }\bibfield  {title} {\bibinfo {title} {Temporal analysis of
			stochastic turning behavior of swimming c. elegans},\ }\href@noop {}
	{\bibfield  {journal} {\bibinfo  {journal} {Journal of neurophysiology}\
		}\textbf {\bibinfo {volume} {102}},\ \bibinfo {pages} {1172} (\bibinfo {year}
		{2009})}\BibitemShut {NoStop}%
	\bibitem [{\citenamefont {Hildebrand}(1965)}]{hildebrand1965symmetrical}%
	\BibitemOpen
	\bibfield  {author} {\bibinfo {author} {\bibfnamefont {M.}~\bibnamefont
			{Hildebrand}},\ }\bibfield  {title} {\bibinfo {title} {Symmetrical gaits of
			horses},\ }\href@noop {} {\bibfield  {journal} {\bibinfo  {journal}
			{science}\ }\textbf {\bibinfo {volume} {150}},\ \bibinfo {pages} {701}
		(\bibinfo {year} {1965})}\BibitemShut {NoStop}%
	\bibitem [{\citenamefont {Jayne}(1986)}]{jayne1986kinematics}%
	\BibitemOpen
	\bibfield  {author} {\bibinfo {author} {\bibfnamefont {B.~C.}\ \bibnamefont
			{Jayne}},\ }\bibfield  {title} {\bibinfo {title} {Kinematics of terrestrial
			snake locomotion},\ }\href@noop {} {\bibfield  {journal} {\bibinfo  {journal}
			{Copeia}\ }\textbf {\bibinfo {volume} {1986}},\ \bibinfo {pages} {915}
		(\bibinfo {year} {1986})}\BibitemShut {NoStop}%
	\bibitem [{\citenamefont {McInroe}\ \emph {et~al.}(2016)\citenamefont
		{McInroe}, \citenamefont {Astley}, \citenamefont {Gong}, \citenamefont
		{Kawano}, \citenamefont {Schiebel}, \citenamefont {Rieser}, \citenamefont
		{Choset}, \citenamefont {Blob},\ and\ \citenamefont
		{Goldman}}]{mcinroe2016tail}%
	\BibitemOpen
	\bibfield  {author} {\bibinfo {author} {\bibfnamefont {B.}~\bibnamefont
			{McInroe}}, \bibinfo {author} {\bibfnamefont {H.~C.}\ \bibnamefont {Astley}},
		\bibinfo {author} {\bibfnamefont {C.}~\bibnamefont {Gong}}, \bibinfo {author}
		{\bibfnamefont {S.~M.}\ \bibnamefont {Kawano}}, \bibinfo {author}
		{\bibfnamefont {P.~E.}\ \bibnamefont {Schiebel}}, \bibinfo {author}
		{\bibfnamefont {J.~M.}\ \bibnamefont {Rieser}}, \bibinfo {author}
		{\bibfnamefont {H.}~\bibnamefont {Choset}}, \bibinfo {author} {\bibfnamefont
			{R.~W.}\ \bibnamefont {Blob}},\ and\ \bibinfo {author} {\bibfnamefont
			{D.~I.}\ \bibnamefont {Goldman}},\ }\bibfield  {title} {\bibinfo {title}
		{Tail use improves performance on soft substrates in models of early
			vertebrate land locomotors},\ }\href@noop {} {\bibfield  {journal} {\bibinfo
			{journal} {Science}\ }\textbf {\bibinfo {volume} {353}},\ \bibinfo {pages}
		{154} (\bibinfo {year} {2016})}\BibitemShut {NoStop}%
	\bibitem [{\citenamefont {Chong}\ \emph
		{et~al.}(2021{\natexlab{c}})\citenamefont {Chong}, \citenamefont {Wang},
		\citenamefont {Rieser}, \citenamefont {Lin}, \citenamefont {Kaba},
		\citenamefont {Blekherman}, \citenamefont {Choset},\ and\ \citenamefont
		{Goldman}}]{chong2021frequency}%
	\BibitemOpen
	\bibfield  {author} {\bibinfo {author} {\bibfnamefont {B.}~\bibnamefont
			{Chong}}, \bibinfo {author} {\bibfnamefont {T.}~\bibnamefont {Wang}},
		\bibinfo {author} {\bibfnamefont {J.~M.}\ \bibnamefont {Rieser}}, \bibinfo
		{author} {\bibfnamefont {B.}~\bibnamefont {Lin}}, \bibinfo {author}
		{\bibfnamefont {A.}~\bibnamefont {Kaba}}, \bibinfo {author} {\bibfnamefont
			{G.}~\bibnamefont {Blekherman}}, \bibinfo {author} {\bibfnamefont
			{H.}~\bibnamefont {Choset}},\ and\ \bibinfo {author} {\bibfnamefont {D.~I.}\
			\bibnamefont {Goldman}},\ }\bibfield  {title} {\bibinfo {title} {Frequency
			modulation of body waves to improve performance of sidewinding robots},\
	}\href@noop {} {\bibfield  {journal} {\bibinfo  {journal} {The International
				Journal of Robotics Research}\ }\textbf {\bibinfo {volume} {40}},\ \bibinfo
		{pages} {1547} (\bibinfo {year} {2021}{\natexlab{c}})}\BibitemShut {NoStop}%
	\bibitem [{\citenamefont {Chong}\ \emph
		{et~al.}(2021{\natexlab{d}})\citenamefont {Chong}, \citenamefont {Aydin},
		\citenamefont {Gong}, \citenamefont {Sartoretti}, \citenamefont {Wu},
		\citenamefont {Rieser}, \citenamefont {Xing}, \citenamefont {Schiebel},
		\citenamefont {Rankin}, \citenamefont {Michel} \emph
		{et~al.}}]{chong2021coordination}%
	\BibitemOpen
	\bibfield  {author} {\bibinfo {author} {\bibfnamefont {B.}~\bibnamefont
			{Chong}}, \bibinfo {author} {\bibfnamefont {Y.~O.}\ \bibnamefont {Aydin}},
		\bibinfo {author} {\bibfnamefont {C.}~\bibnamefont {Gong}}, \bibinfo {author}
		{\bibfnamefont {G.}~\bibnamefont {Sartoretti}}, \bibinfo {author}
		{\bibfnamefont {Y.}~\bibnamefont {Wu}}, \bibinfo {author} {\bibfnamefont
			{J.~M.}\ \bibnamefont {Rieser}}, \bibinfo {author} {\bibfnamefont
			{H.}~\bibnamefont {Xing}}, \bibinfo {author} {\bibfnamefont {P.~E.}\
			\bibnamefont {Schiebel}}, \bibinfo {author} {\bibfnamefont {J.~W.}\
			\bibnamefont {Rankin}}, \bibinfo {author} {\bibfnamefont {K.~B.}\
			\bibnamefont {Michel}}, \emph {et~al.},\ }\bibfield  {title} {\bibinfo
		{title} {Coordination of lateral body bending and leg movements for sprawled
			posture quadrupedal locomotion},\ }\href@noop {} {\bibfield  {journal}
		{\bibinfo  {journal} {The International Journal of Robotics Research}\
		}\textbf {\bibinfo {volume} {40}},\ \bibinfo {pages} {747} (\bibinfo {year}
		{2021}{\natexlab{d}})}\BibitemShut {NoStop}%
	\bibitem [{\citenamefont {Secor}(1994)}]{Secor1994}%
	\BibitemOpen
	\bibfield  {author} {\bibinfo {author} {\bibfnamefont {S.~M.}\ \bibnamefont
			{Secor}},\ }\bibfield  {title} {\bibinfo {title} {{Ecological significance of
				movements and activity range for the sidewinder, Crotalus cerastes}},\
	}\href@noop {} {\bibfield  {journal} {\bibinfo  {journal} {Copeia}\ }\textbf
		{\bibinfo {volume} {1994}},\ \bibinfo {pages} {631} (\bibinfo {year}
		{1994})}\BibitemShut {NoStop}%
	\bibitem [{\citenamefont {Astley}\ \emph
		{et~al.}(2020{\natexlab{b}})\citenamefont {Astley}, \citenamefont {Rieser},
		\citenamefont {Kaba}, \citenamefont {Paez}, \citenamefont {Tomkinson},
		\citenamefont {Mendelson},\ and\ \citenamefont {Goldman}}]{Astley2020}%
	\BibitemOpen
	\bibfield  {author} {\bibinfo {author} {\bibfnamefont {H.~C.}\ \bibnamefont
			{Astley}}, \bibinfo {author} {\bibfnamefont {J.~M.}\ \bibnamefont {Rieser}},
		\bibinfo {author} {\bibfnamefont {A.}~\bibnamefont {Kaba}}, \bibinfo {author}
		{\bibfnamefont {V.~M.}\ \bibnamefont {Paez}}, \bibinfo {author}
		{\bibfnamefont {I.}~\bibnamefont {Tomkinson}}, \bibinfo {author}
		{\bibfnamefont {J.~R.}\ \bibnamefont {Mendelson}},\ and\ \bibinfo {author}
		{\bibfnamefont {D.~I.}\ \bibnamefont {Goldman}},\ }\bibfield  {title}
	{\bibinfo {title} {Side-impact collision: mechanics of obstacle negotiation
			in sidewinding snakes},\ }\href@noop {} {\bibfield  {journal} {\bibinfo
			{journal} {Bioinspiration \& Biomimetics}\ }\textbf {\bibinfo {volume}
			{15}},\ \bibinfo {pages} {065005} (\bibinfo {year}
		{2020}{\natexlab{b}})}\BibitemShut {NoStop}%
	\bibitem [{\citenamefont {Mosauer}(1930)}]{Mosauer:1930}%
	\BibitemOpen
	\bibfield  {author} {\bibinfo {author} {\bibfnamefont {W.}~\bibnamefont
			{Mosauer}},\ }\bibfield  {title} {\bibinfo {title} {A note on the sidewinding
			locomotion of snakes},\ }\href@noop {} {\bibfield  {journal} {\bibinfo
			{journal} {The American Naturalist}\ }\textbf {\bibinfo {volume} {64}},\
		\bibinfo {pages} {179} (\bibinfo {year} {1930})}\BibitemShut {NoStop}%
	\bibitem [{\citenamefont {Burdick}\ \emph {et~al.}(1993)\citenamefont
		{Burdick}, \citenamefont {Radford},\ and\ \citenamefont
		{Chirikjian}}]{Burdick:1993}%
	\BibitemOpen
	\bibfield  {author} {\bibinfo {author} {\bibfnamefont {J.~W.}\ \bibnamefont
			{Burdick}}, \bibinfo {author} {\bibfnamefont {J.~E.}\ \bibnamefont
			{Radford}},\ and\ \bibinfo {author} {\bibfnamefont {G.~S.}\ \bibnamefont
			{Chirikjian}},\ }\bibfield  {title} {\bibinfo {title} {A ``sidewinding"
			locomotion gait for hyper-redundant robots},\ }\href@noop {} {\bibfield
		{journal} {\bibinfo  {journal} {Robotics and Automation}\ }\textbf {\bibinfo
			{volume} {3}},\ \bibinfo {pages} {101} (\bibinfo {year} {1993})}\BibitemShut
	{NoStop}%
	\bibitem [{\citenamefont {Hatton}\ and\ \citenamefont
		{Choset}(2010)}]{Hatton:2010ICRA:Sidewinding}%
	\BibitemOpen
	\bibfield  {author} {\bibinfo {author} {\bibfnamefont {R.~L.}\ \bibnamefont
			{Hatton}}\ and\ \bibinfo {author} {\bibfnamefont {H.}~\bibnamefont
			{Choset}},\ }\bibfield  {title} {\bibinfo {title} {Sidewinding on slopes},\
	}in\ \href@noop {} {\emph {\bibinfo {booktitle} {{Proceedings of the IEEE
					International Conference on Robotics and Automation}}}}\ (\bibinfo {address}
	{Anchorage, AK USA},\ \bibinfo {year} {2010})\ pp.\ \bibinfo {pages}
	{691--696}\BibitemShut {NoStop}%
	\bibitem [{\citenamefont {Marvi}\ \emph {et~al.}(2014)\citenamefont {Marvi},
		\citenamefont {Gong}, \citenamefont {Gravish}, \citenamefont {Astley},
		\citenamefont {Travers}, \citenamefont {Hatton}, \citenamefont {Mendelson},
		\citenamefont {Choset}, \citenamefont {Hu},\ and\ \citenamefont
		{Goldman}}]{marvi2014sidewinding}%
	\BibitemOpen
	\bibfield  {author} {\bibinfo {author} {\bibfnamefont {H.}~\bibnamefont
			{Marvi}}, \bibinfo {author} {\bibfnamefont {C.}~\bibnamefont {Gong}},
		\bibinfo {author} {\bibfnamefont {N.}~\bibnamefont {Gravish}}, \bibinfo
		{author} {\bibfnamefont {H.}~\bibnamefont {Astley}}, \bibinfo {author}
		{\bibfnamefont {M.}~\bibnamefont {Travers}}, \bibinfo {author} {\bibfnamefont
			{R.~L.}\ \bibnamefont {Hatton}}, \bibinfo {author} {\bibfnamefont {J.~R.}\
			\bibnamefont {Mendelson}}, \bibinfo {author} {\bibfnamefont {H.}~\bibnamefont
			{Choset}}, \bibinfo {author} {\bibfnamefont {D.~L.}\ \bibnamefont {Hu}},\
		and\ \bibinfo {author} {\bibfnamefont {D.~I.}\ \bibnamefont {Goldman}},\
	}\bibfield  {title} {\bibinfo {title} {Sidewinding with minimal slip: Snake
			and robot ascent of sandy slopes},\ }\href@noop {} {\bibfield  {journal}
		{\bibinfo  {journal} {Science}\ }\textbf {\bibinfo {volume} {346}},\ \bibinfo
		{pages} {224} (\bibinfo {year} {2014})}\BibitemShut {NoStop}%
	\bibitem [{\citenamefont {Rieser}\ \emph {et~al.}(2021)\citenamefont {Rieser},
		\citenamefont {Tingle}, \citenamefont {Goldman}, \citenamefont {Mendelson}
		\emph {et~al.}}]{rieser2021functional}%
	\BibitemOpen
	\bibfield  {author} {\bibinfo {author} {\bibfnamefont {J.~M.}\ \bibnamefont
			{Rieser}}, \bibinfo {author} {\bibfnamefont {J.~L.}\ \bibnamefont {Tingle}},
		\bibinfo {author} {\bibfnamefont {D.~I.}\ \bibnamefont {Goldman}}, \bibinfo
		{author} {\bibfnamefont {J.~R.}\ \bibnamefont {Mendelson}}, \emph {et~al.},\
	}\bibfield  {title} {\bibinfo {title} {Functional consequences of
			convergently evolved microscopic skin features on snake locomotion},\
	}\href@noop {} {\bibfield  {journal} {\bibinfo  {journal} {Proceedings of the
				National Academy of Sciences}\ }\textbf {\bibinfo {volume} {118}} (\bibinfo
		{year} {2021})}\BibitemShut {NoStop}%
	\bibitem [{\citenamefont {Loeb}\ and\ \citenamefont
		{Gans}(1986)}]{loeb1986electromyography}%
	\BibitemOpen
	\bibfield  {author} {\bibinfo {author} {\bibfnamefont {G.~E.}\ \bibnamefont
			{Loeb}}\ and\ \bibinfo {author} {\bibfnamefont {C.}~\bibnamefont {Gans}},\
	}\href@noop {} {\emph {\bibinfo {title} {Electromyography for
				experimentalists}}}\ (\bibinfo  {publisher} {University of Chicago press},\
	\bibinfo {year} {1986})\BibitemShut {NoStop}%
	\bibitem [{\citenamefont {Johnson}\ \emph {et~al.}(2021)\citenamefont
		{Johnson}, \citenamefont {Lewis},\ and\ \citenamefont
		{Guy}}]{johnson2021neuromechanical}%
	\BibitemOpen
	\bibfield  {author} {\bibinfo {author} {\bibfnamefont {C.~L.}\ \bibnamefont
			{Johnson}}, \bibinfo {author} {\bibfnamefont {T.~J.}\ \bibnamefont {Lewis}},\
		and\ \bibinfo {author} {\bibfnamefont {R.}~\bibnamefont {Guy}},\ }\bibfield
	{title} {\bibinfo {title} {Neuromechanical mechanisms of gait adaptation in
			c. elegans: Relative roles of neural and mechanical coupling},\ }\href@noop
	{} {\bibfield  {journal} {\bibinfo  {journal} {SIAM Journal on Applied
				Dynamical Systems}\ }\textbf {\bibinfo {volume} {20}},\ \bibinfo {pages}
		{1022} (\bibinfo {year} {2021})}\BibitemShut {NoStop}%
	\bibitem [{\citenamefont {Gart}\ \emph {et~al.}(2019)\citenamefont {Gart},
		\citenamefont {Mitchel},\ and\ \citenamefont {Li}}]{gart2019snakes}%
	\BibitemOpen
	\bibfield  {author} {\bibinfo {author} {\bibfnamefont {S.~W.}\ \bibnamefont
			{Gart}}, \bibinfo {author} {\bibfnamefont {T.~W.}\ \bibnamefont {Mitchel}},\
		and\ \bibinfo {author} {\bibfnamefont {C.}~\bibnamefont {Li}},\ }\bibfield
	{title} {\bibinfo {title} {Snakes partition their body to traverse large
			steps stably},\ }\href@noop {} {\bibfield  {journal} {\bibinfo  {journal}
			{Journal of Experimental Biology}\ }\textbf {\bibinfo {volume} {222}},\
		\bibinfo {pages} {jeb185991} (\bibinfo {year} {2019})}\BibitemShut {NoStop}%
	\bibitem [{\citenamefont {Astley}\ and\ \citenamefont
		{Jayne}(2009)}]{astley2009arboreal}%
	\BibitemOpen
	\bibfield  {author} {\bibinfo {author} {\bibfnamefont {H.~C.}\ \bibnamefont
			{Astley}}\ and\ \bibinfo {author} {\bibfnamefont {B.~C.}\ \bibnamefont
			{Jayne}},\ }\bibfield  {title} {\bibinfo {title} {Arboreal habitat structure
			affects the performance and modes of locomotion of corn snakes (elaphe
			guttata)},\ }\href@noop {} {\bibfield  {journal} {\bibinfo  {journal}
			{Journal of Experimental Zoology Part A: Ecological Genetics and Physiology}\
		}\textbf {\bibinfo {volume} {311}},\ \bibinfo {pages} {207} (\bibinfo {year}
		{2009})}\BibitemShut {NoStop}%
	\bibitem [{\citenamefont {Ramasamy}\ and\ \citenamefont
		{Hatton}(2016)}]{ramasamy2016soap}%
	\BibitemOpen
	\bibfield  {author} {\bibinfo {author} {\bibfnamefont {S.}~\bibnamefont
			{Ramasamy}}\ and\ \bibinfo {author} {\bibfnamefont {R.~L.}\ \bibnamefont
			{Hatton}},\ }\bibfield  {title} {\bibinfo {title} {Soap-bubble optimization
			of gaits},\ }in\ \href@noop {} {\emph {\bibinfo {booktitle} {2016 IEEE 55th
				Conference on Decision and Control (CDC)}}}\ (\bibinfo {organization}
	{IEEE},\ \bibinfo {year} {2016})\ pp.\ \bibinfo {pages}
	{1056--1062}\BibitemShut {NoStop}%
	\bibitem [{\citenamefont {Chong}\ \emph {et~al.}(2019)\citenamefont {Chong},
		\citenamefont {Ozkan~Aydin}, \citenamefont {Sartoretti}, \citenamefont
		{Rieser}, \citenamefont {Gong}, \citenamefont {Xing}, \citenamefont
		{Choset},\ and\ \citenamefont {Goldman}}]{chong2019hierarchical}%
	\BibitemOpen
	\bibfield  {author} {\bibinfo {author} {\bibfnamefont {B.}~\bibnamefont
			{Chong}}, \bibinfo {author} {\bibfnamefont {Y.}~\bibnamefont {Ozkan~Aydin}},
		\bibinfo {author} {\bibfnamefont {G.}~\bibnamefont {Sartoretti}}, \bibinfo
		{author} {\bibfnamefont {J.~M.}\ \bibnamefont {Rieser}}, \bibinfo {author}
		{\bibfnamefont {C.}~\bibnamefont {Gong}}, \bibinfo {author} {\bibfnamefont
			{H.}~\bibnamefont {Xing}}, \bibinfo {author} {\bibfnamefont {H.}~\bibnamefont
			{Choset}},\ and\ \bibinfo {author} {\bibfnamefont {D.~I.}\ \bibnamefont
			{Goldman}},\ }\bibfield  {title} {\bibinfo {title} {A hierarchical geometric
			framework to design locomotive gaits for highly articulated robots},\ }in\
	\href@noop {} {\emph {\bibinfo {booktitle} {Robotics: science and systems}}}\
	(\bibinfo {year} {2019})\BibitemShut {NoStop}%
	\bibitem [{\citenamefont {Mathis}\ \emph {et~al.}(2018)\citenamefont {Mathis},
		\citenamefont {Mamidanna}, \citenamefont {Cury}, \citenamefont {Abe},
		\citenamefont {Murthy}, \citenamefont {Mathis},\ and\ \citenamefont
		{Bethge}}]{mathis2018deeplabcut}%
	\BibitemOpen
	\bibfield  {author} {\bibinfo {author} {\bibfnamefont {A.}~\bibnamefont
			{Mathis}}, \bibinfo {author} {\bibfnamefont {P.}~\bibnamefont {Mamidanna}},
		\bibinfo {author} {\bibfnamefont {K.~M.}\ \bibnamefont {Cury}}, \bibinfo
		{author} {\bibfnamefont {T.}~\bibnamefont {Abe}}, \bibinfo {author}
		{\bibfnamefont {V.~N.}\ \bibnamefont {Murthy}}, \bibinfo {author}
		{\bibfnamefont {M.~W.}\ \bibnamefont {Mathis}},\ and\ \bibinfo {author}
		{\bibfnamefont {M.}~\bibnamefont {Bethge}},\ }\bibfield  {title} {\bibinfo
		{title} {Deeplabcut: markerless pose estimation of user-defined body parts
			with deep learning},\ }\href@noop {} {\bibfield  {journal} {\bibinfo
			{journal} {Nature neuroscience}\ }\textbf {\bibinfo {volume} {21}},\ \bibinfo
		{pages} {1281} (\bibinfo {year} {2018})}\BibitemShut {NoStop}%
	\bibitem [{\citenamefont {Gans}\ \emph {et~al.}(1992)\citenamefont {Gans} \emph
		{et~al.}}]{gans1992kinematic}%
	\BibitemOpen
	\bibfield  {author} {\bibinfo {author} {\bibfnamefont {C.}~\bibnamefont
			{Gans}} \emph {et~al.},\ }\bibfield  {title} {\bibinfo {title} {Kinematic
			description of the sidewinding locomotion of four vipers},\ }\href@noop {}
	{\bibfield  {journal} {\bibinfo  {journal} {Israel Journal of Ecology and
				Evolution}\ }\textbf {\bibinfo {volume} {38}},\ \bibinfo {pages} {9}
		(\bibinfo {year} {1992})}\BibitemShut {NoStop}%
	\bibitem [{\citenamefont {Brain}(1960)}]{Brain1960}%
	\BibitemOpen
	\bibfield  {author} {\bibinfo {author} {\bibfnamefont {C.~K.}\ \bibnamefont
			{Brain}},\ }\bibfield  {title} {\bibinfo {title} {{Observations on the
				locomotion of the south west African adder, Bitis peringueyi (Boulenger),
				with speculations on the origin of sidewinding}},\ }\href@noop {} {\bibfield
		{journal} {\bibinfo  {journal} {Annals of the Transvaal Museum}\ }\textbf
		{\bibinfo {volume} {24}},\ \bibinfo {pages} {19} (\bibinfo {year}
		{1960})}\BibitemShut {NoStop}%
	\bibitem [{\citenamefont {Andrews}(2019)}]{andrews2019new}%
	\BibitemOpen
	\bibfield  {author} {\bibinfo {author} {\bibfnamefont {D.~G.}\ \bibnamefont
			{Andrews}},\ }\bibfield  {title} {\bibinfo {title} {A new method for
			measuring the size of nematodes using image processing},\ }\href@noop {}
	{\bibfield  {journal} {\bibinfo  {journal} {Biology Methods and Protocols}\
		}\textbf {\bibinfo {volume} {4}},\ \bibinfo {pages} {bpz020} (\bibinfo {year}
		{2019})}\BibitemShut {NoStop}%
	\bibitem [{\citenamefont {Reina}\ \emph {et~al.}(2013)\citenamefont {Reina},
		\citenamefont {Subramaniam}, \citenamefont {Laromaine}, \citenamefont
		{Samuel},\ and\ \citenamefont {Whitesides}}]{reina2013shifts}%
	\BibitemOpen
	\bibfield  {author} {\bibinfo {author} {\bibfnamefont {A.}~\bibnamefont
			{Reina}}, \bibinfo {author} {\bibfnamefont {A.~B.}\ \bibnamefont
			{Subramaniam}}, \bibinfo {author} {\bibfnamefont {A.}~\bibnamefont
			{Laromaine}}, \bibinfo {author} {\bibfnamefont {A.~D.}\ \bibnamefont
			{Samuel}},\ and\ \bibinfo {author} {\bibfnamefont {G.~M.}\ \bibnamefont
			{Whitesides}},\ }\bibfield  {title} {\bibinfo {title} {Shifts in the
			distribution of mass densities is a signature of caloric restriction in
			caenorhabditis elegans},\ }\href@noop {} {\bibfield  {journal} {\bibinfo
			{journal} {PloS one}\ }\textbf {\bibinfo {volume} {8}},\ \bibinfo {pages}
		{e69651} (\bibinfo {year} {2013})}\BibitemShut {NoStop}%
	\bibitem [{\citenamefont {Sznitman}\ \emph
		{et~al.}(2010{\natexlab{b}})\citenamefont {Sznitman}, \citenamefont {Shen},
		\citenamefont {Purohit},\ and\ \citenamefont
		{Arratia}}]{sznitman2010effects}%
	\BibitemOpen
	\bibfield  {author} {\bibinfo {author} {\bibfnamefont {J.}~\bibnamefont
			{Sznitman}}, \bibinfo {author} {\bibfnamefont {X.}~\bibnamefont {Shen}},
		\bibinfo {author} {\bibfnamefont {P.~K.}\ \bibnamefont {Purohit}},\ and\
		\bibinfo {author} {\bibfnamefont {P.~E.}\ \bibnamefont {Arratia}},\
	}\bibfield  {title} {\bibinfo {title} {The effects of fluid viscosity on the
			kinematics and material properties of c. elegans swimming at low reynolds
			number},\ }\href@noop {} {\bibfield  {journal} {\bibinfo  {journal}
			{Experimental Mechanics}\ }\textbf {\bibinfo {volume} {50}},\ \bibinfo
		{pages} {1303} (\bibinfo {year} {2010}{\natexlab{b}})}\BibitemShut {NoStop}%
	\bibitem [{\citenamefont {Hatton}\ and\ \citenamefont
		{Choset}(2015{\natexlab{b}})}]{hatton2015nonconservativity}%
	\BibitemOpen
	\bibfield  {author} {\bibinfo {author} {\bibfnamefont {R.~L.}\ \bibnamefont
			{Hatton}}\ and\ \bibinfo {author} {\bibfnamefont {H.}~\bibnamefont
			{Choset}},\ }\bibfield  {title} {\bibinfo {title} {Nonconservativity and
			noncommutativity in locomotion},\ }\href@noop {} {\bibfield  {journal}
		{\bibinfo  {journal} {The European Physical Journal Special Topics}\ }\textbf
		{\bibinfo {volume} {224}},\ \bibinfo {pages} {3141} (\bibinfo {year}
		{2015}{\natexlab{b}})}\BibitemShut {NoStop}%
	\bibitem [{\citenamefont {Bass}\ \emph {et~al.}(2022)\citenamefont {Bass},
		\citenamefont {Ramasamy},\ and\ \citenamefont
		{Hatton}}]{bass2022characterizing}%
	\BibitemOpen
	\bibfield  {author} {\bibinfo {author} {\bibfnamefont {C.}~\bibnamefont
			{Bass}}, \bibinfo {author} {\bibfnamefont {S.}~\bibnamefont {Ramasamy}},\
		and\ \bibinfo {author} {\bibfnamefont {R.~L.}\ \bibnamefont {Hatton}},\
	}\bibfield  {title} {\bibinfo {title} {Characterizing error in noncommutative
			geometric gait analysis},\ }in\ \href@noop {} {\emph {\bibinfo {booktitle}
			{2022 International Conference on Robotics and Automation (ICRA)}}}\
	(\bibinfo {organization} {IEEE},\ \bibinfo {year} {2022})\ pp.\ \bibinfo
	{pages} {9845--9851}\BibitemShut {NoStop}%
\end{thebibliography}

\end{document}